\documentclass[
reprint,
superscriptaddress,
amsmath,
amssymb,
aps,
prb,
floatfix,
]{revtex4-2}
\usepackage{graphicx}
\usepackage{hyperref}
\usepackage{xcolor}
\usepackage{multirow}

\begin{document}

\title{Emergent B2 chemical orderings in the AlTiVNb and AlTiCrMo refractory high-entropy superalloys studied via first-principles theory and atomistic modelling}

\author{Christopher D. Woodgate}
\email{christopher.woodgate@bristol.ac.uk}
\affiliation{H.H. Wills Physics Laboratory, University of Bristol, Royal Fort, Bristol, BS8 1TL, United Kingdom}
\author{Hubert J. Naguszewski}
\affiliation{Department of Physics, University of Warwick, Coventry, CV4 7AL, United Kingdom}
\author{David Redka}
\affiliation{Department of Applied Sciences and Mechatronics, Munich University of Applied Sciences HM, Munich, Germany}
\affiliation{New Technologies Research Center, University of West Bohemia, Pilsen, Czech Republic}
\author{Ján Minár}
\affiliation{New Technologies Research Center, University of West Bohemia, Pilsen, Czech Republic}
\author{David Quigley}
\affiliation{Department of Physics, University of Warwick, Coventry, CV4 7AL, United Kingdom}
\author{Julie B. Staunton}
\affiliation{Department of Physics, University of Warwick, Coventry, CV4 7AL, United Kingdom}

\begin{abstract}
We study the thermodynamics and phase stability of the AlTiVNb and AlTiCrMo refractory high-entropy superalloys using a combination of \textit{ab initio} electronic structure theory---namely a concentration wave analysis---and atomistic Monte Carlo simulations. 
Our multiscale approach is suitable both for examining atomic short-range order in the solid solution, as well as for studying the emergence of long-range crystallographic order with decreasing temperature. 
In both alloys considered in this work, in alignment with experimental observations, we predict a B2 (CsCl) chemical ordering emerging at high temperatures, which is driven primarily by Al and Ti, with other elements expressing weaker site preferences. 
The predicted B2 ordering temperature for AlTiVNb is higher than that for AlTiCrMo. 
These chemical orderings are discussed in terms of the alloys' electronic structure, with hybridisation between the $sp$ states of Al and the $d$ states of the transition metals understood to play an important role. 
Within our modelling, the chemically ordered B2 phases for both alloys have an increased predicted residual resistivity compared to the A2 (disordered bcc) phases. These increased resistivity values are understood to originate in a reduction in the electronic density of states at the Fermi level, in conjunction with qualitative changes to the alloys' smeared-out Fermi surfaces. 
These results highlight the close connections between composition, structure, and physical properties in this technologically relevant class of materials.
\end{abstract}

\date{March 17, 2025}

\maketitle

\section{Introduction}
\label{sec:introduction}
Since first reported by Yeh \textit{et~al.}~\cite{yeh_nanostructured_2004} and Cantor \textit{et~al.}~\cite{cantor_microstructural_2004} in 2004, so-called `high-entropy', `complex concentrated', or `multi-principle element' alloys---those systems containing four or more alloying elements combined in near-equal ratios---have attracted significant and sustained attention in the field of materials science~\cite{miracle_critical_2017, george_high-entropy_2019}. The large contribution to the free energy of such multicomponent alloys made by the configurational entropy (or `entropy of mixing') is understood to stabilise single-phase solid solutions containing combinations of elements which do not readily form binary alloys. High-entropy alloys (HEAs) have frequently been shown to exhibit superior physical properties for applications when compared to traditional binary/ternary alloys. Enhanced properties of HEAs as compared to traditional alloys include improved radiation resistance~\cite{ayyagari_low_2018, el-atwani_outstanding_2019, el_atwani_quinary_2023}, fracture resistance~\cite{gludovatz_fracture-resistant_2014, gludovatz_exceptional_2016, liu_exceptional_2022}, and excellent structural properties at elevated temperatures~\cite{praveen_highentropy_2018}. In addition, some high-entropy alloys  have been shown to exhibit a range of interesting intrinsic physical phenomena such as superconductivity~\cite{kozelj_discovery_2014}, quantum critical behaviour~\cite{sales_quantum_2016}, and extreme Fermi surface smearing~\cite{robarts_extreme_2020}.

A group of HEAs which is of interest for elevated temperature applications, particularly in the nuclear and aerospace industries, is the family of refractory high-entropy alloys~\cite{senkov_development_2018}, first reported by Senkov \textit{et al.}~\cite{senkov_refractory_2010, senkov_mechanical_2011} in 2010. Based on refractory elements such as V, Nb, Mo, Ta, and W, these alloys have high melting temperatures on account of the base elements used in their compositions, and also possess excellent mechanical properties~\cite{lee_lattice_2018, maresca_mechanistic_2020}. One avenue of research in the area of refractory HEAs has been exploration of the addition of Al as an alloying element~\cite{senkov_effect_2014}. While refractory HEAs without Al present typically form disordered solid solutions with an underlying bcc lattice, the addition of Al is understood to promote formation of chemically ordered precipitates with the B2 crystal structure, with consequent impact on a variety of physical properties~\cite{senkov_microstructure_2014, jensen_characterization_2016}. Based on analogy with Ni-based superalloys, these Al-containing refractory HEAs are conventionally referred to as refractory high-entropy superalloys (RSAs)~\cite{miracle_refractory_2020}.

\begin{figure}[t]
    \centering
    \includegraphics[width=\linewidth]{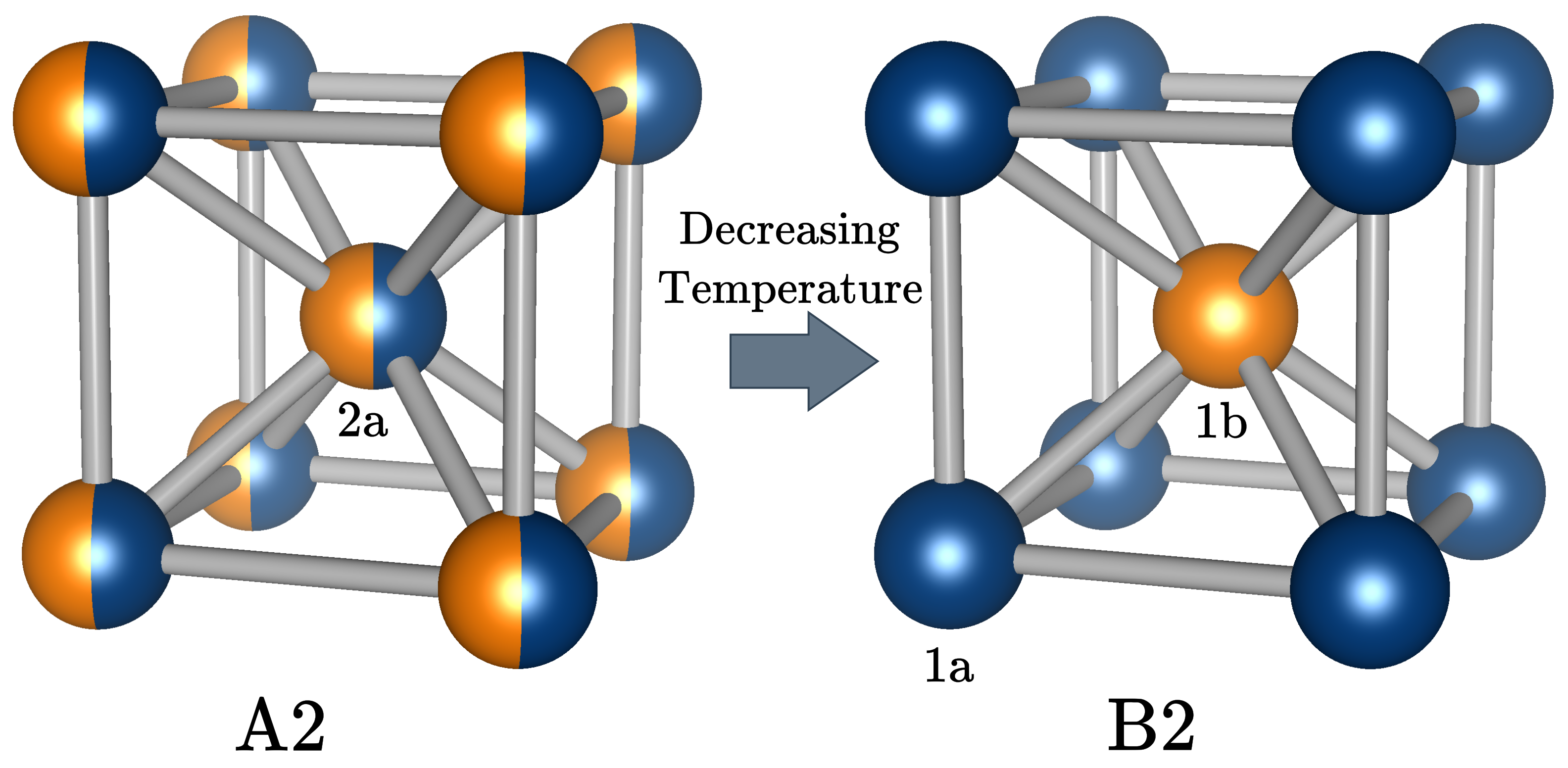}
    \caption{Illustrations of A2 and B2 (CsCl) structures for an equiatomic binary $AB$ alloy. Non-equivalent lattice sites are given their Wyckoff labels. In this context, the A2 structure is a bcc lattice where there is substitutional disorder on all lattice sites---denoted by partially coloured spheres. The B2 structure is a chemically ordered structure imposed on the bcc lattice and consists of two interpenetrating simple cubic sublattices. Crystallographic orderings such as this are anticipated to emerge as an alloy is gradually cooled from high temperature. Images generated using {\it VESTA}~\cite{momma_vesta_2011}.}
    \label{fig:A2_B2}
\end{figure}

Detection of such B2 chemical orderings in RSAs can be experimentally challenging, however. For example, the AlTiVNb RSA was first characterised as having a disordered bcc structure~\cite{stepanov_structure_2015}, before later results showed that this alloy in fact forms a single B2 phase across a wide temperature range~\cite{yurchenko_effect_2018}. Further, in the AlTiCrMo RSA, it has been shown that X-ray diffraction patterns alone fail to distinguish between the A2 (disordered bcc) and B2 crystal structures~\cite{chen_crystallographic_2019}. In addition, given a composition containing three or more separate elements at near equal ratios, as occurs in RSAs, a B2 chemical ordering---illustrated in Fig.~\ref{fig:A2_B2}---is not unambiguously defined by the system's stoichiometry. (For a schematic illustration of this problem, we refer the reader to panels (c), (d), and (e) of Fig.~\ref{fig:concentration_waves}.) Consequently, it is desirable to understand whether different elements in a given composition have stronger/weaker preferences for sitting on different sublattices in the ordered structure. It is also useful to understand the temperature at which such B2 ordered structures are likely to emerge, to help guide materials processing when seeking to promote/inhibit formation of such chemically ordered phases. Finally, as many HEAs are expected to become eventually metastable with decreasing temperature, it is important to simulate the phase stability of the alloy below any initial chemical ordering temperature, to understand whether there are likely to be additional sublattice orderings and/or eventual phase decomposition. Deeper understanding of these aspects can facilitate improved understanding of the behaviour of existing materials, as well as potentially suggesting new compositions which could be investigated for applications.

Theory and simulation have an important role to play in addressing the aforementioned issues. Modelling materials at the atomic and sub-atomic length scales can provide insights into aspects of materials' behaviour which are not always experimentally accessible. In addition, first principles calculations of materials' electronic structure can shed light on the physical origins of experimentally observed phenomena such as chemical disorder/order transitions in alloys. Studying the phase stability of HEAs, however, presents a number of challenges from the perspective of computational materials modelling~\cite{widom_modeling_2018, ferrari_frontiers_2020, ferrari_simulating_2023}. As the size of the configuration space for a given alloy grows combinatorially with the number of elements present in the composition, a large number of configurations must be sampled for there to be confidence that results are well-converged and representative of the thermodynamic phases. In addition, given the huge range of HEA compositions being continually reported, it is undesirable to use computationally intensive methodologies which produce results specific to a single composition. 

Despite these challenges, there are a number of well-established techniques for modelling the thermodynamics and phase stability of HEAs, which use a range of sampling techniques and/or free energy calculations to explore the configuration space of a given alloy composition. These techniques can be distinguished from one another by consideration of the constructions of the description of disordered and partially ordered phases, and by consideration of the means by which energies (and energy differences between different structures) are obtained. In the context of Monte Carlo and/or molecular dynamics calculations performed on supercells, it is possible to take energies directly from DFT calculations~\cite{widom_hybrid_2014, tamm_atomic-scale_2015, widom_first-principles_2024}, to use a range of semi-empirical and machine-learned interatomic potentials~\cite{kostiuchenko_impact_2019, rosenbrock_machine-learned_2021, kormann_b2_2021, zhou_thermodynamics_2022, zhang_roadmap_2025}, or to apply lattice-based models such as cluster expansions~\cite{fernandez-caballero_short-range_2017, sobieraj_chemical_2020, kim_interaction_2023, vazquez_deciphering_2024}. In the context of calculations working with partial lattice site occupancies, there are also a range of techniques based on effective medium theories such as the coherent potential approximation (CPA)~\cite{singh_atomic_2015, kormann_long-ranged_2017, singh_ta-nb-mo-w_2018}. There are also approaches available based on direct free energy calculations~\cite{feng_first-principles_2017}, or semi-empirical approaches based on thermodynamic databases such as CALPHAD~\cite{li_calphad-aided_2023}.

Our own methodology~\cite{khan_statistical_2016, woodgate_compositional_2022, woodgate_short-range_2023, woodgate_modelling_2024} for assessing the thermodynamics and phase stability of multicomponent alloys is based on a perturbative analysis of the change in free energy of a disordered alloy due to applied, inhomogeneous, atomic-scale chemical fluctuations described within a \textit{concentration wave} formalism. Our approach combines electronic structure calculations, the aforementioned perturbative chemical stability analysis, and atomistic Monte Carlo simulations using a real-space, pairwise form of the alloy internal energy (pair potential) recovered from the \textit{ab initio} data. In a demonstration of this multiscale modelling approach, in this work, we study the thermodynamics and phase stability of the AlTiVNb and AlTiCrMo RSAs, both of which are known experimentally to crystallographically order into the B2 structure~\cite{yurchenko_effect_2018, chen_crystallographic_2019}. For both alloys, we predict the chemical disorder/order transition temperature, as well as which elements preferentially sit on which sublattice. Our atomistic Monte Carlo simulations facilitate further exploration of the configuration space, and demonstrate the emergence of additional sublattice atom-atom correlations in both systems at low temperatures. We are able to shed light on physical origins of these chemical ordering tendencies by considering details of the electronic structure of the alloys in both disordered (A2) and ordered (B2) phases. Finally, to demonstrate the impact such chemical orderings can have on materials' physical properties, we calculate the differences in lattice parameter, bulk modulus and residual resistivity between A2 and B2 phases for both of the considered RSAs.

The rest of this paper is structured as follows. In Section~\ref{sec:methodology}, we outline the modelling approach employed in this study to examine the thermodynamics and phase stability of the selected RSAs, discussing in detail the concentration wave formalism, as well as outlining our Monte Carlo simulations and residual resistivity calculations. Then, in Section~\ref{sec:results}, we present results for the two considered alloy compositions, comparing and contrasting how the different elements present in each of the two compositions alter the predicted chemical ordering tendencies. In-depth discussion of the electronic structure of disordered and partially ordered compositional phases facilitates understanding of the underlying physical mechanisms driving these experimentally observed chemical orderings. Finally, in Section~\ref{sec:conclusions}, we summarise our results, venture an outlook on their implications, and suggest potential future avenues of exploration.

\section{Methods}
\label{sec:methodology}

\subsection{Internal energy of the solid solution: the coherent potential approximation (CPA)}
\label{sec:cpa}

\begin{figure*}[t]
    \centering
    \includegraphics[width=\linewidth]{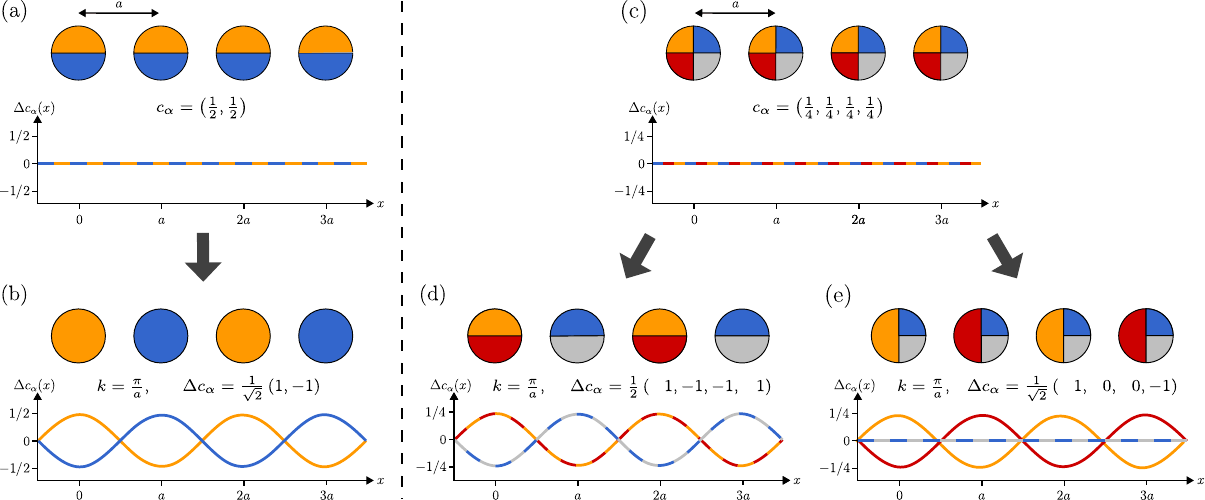}
    \caption{Schematic illustrations of \textit{concentration waves} modulating partial lattice site occupancies of two toy, 1-dimensional alloys. The colours orange, blue, grey, and red denote chemical species $A$, $B$, $C$, and $D$, respectively. An equiatomic $AB$ binary composition with homogeneous partial lattice site occupancies (a) can be transformed to a chemically ordered structure (b) by application of a concentration wave with wave vector $k = \frac{\pi}{a}$ and a normalised chemical polarisation (or `eigenvector') of $\Delta c_\alpha = \frac{1}{\sqrt{2}} (1, -1)$. For the quarternary $ABCD$ composition (c) the wave vector alone does not uniquely determine the chemically ordered structure---both (d) and (e) are example chemical orderings described by $k = \frac{\pi}{a}$. However, the two structures may be distinguished once the chemical polarisation, $\Delta c_\alpha$, of the applied concentration wave is considered.}
    \label{fig:concentration_waves}
\end{figure*}

In a substitutional alloy with a fixed underlying crystal lattice, with the set of lattice positions denoted by $\{\mathbf{R}_i\}$, an arrangement of atoms (a `configuration') can be uniquely specified by a set of site occupancies, $\{\xi_{i\alpha}\}$, where $\xi_{i\alpha}=1$ if site $i$ is occupied by an atom of chemical species $\alpha$, and $\xi_{i\alpha}=0$ otherwise. Each lattice site is constrained to be occupied by one (and only one) atom, which leads to the condition 
\begin{equation}
    \sum_\alpha \xi_{i\alpha} = 1.
    \label{eq:occupancy_condition}
\end{equation}
It is then natural to consider the average value of these site occupancies, which leads to the definition of the site-wise concentrations,
\begin{equation}
    c_{i\alpha} := \langle \xi_{i\alpha} \rangle,
    \label{eq:site-wise_concentrations}
\end{equation}
where $\langle \cdot \rangle$ denotes an average taken with respect to the relevant thermodynamic ensemble. Note that, by construction, we have that $0\leq c_{i\alpha} \leq 1$. These site-wise concentrations represent long-range order parameters classifying potential chemically ordered phases.

In the limit of high-temperature, where the free energy landscape is dominated by entropic contributions and the alloy is maximally disordered, the site-wise concentrations become spatially homogeneous, {\it i.e.} any atom can occupy any lattice site with probability equal to its average concentration in the alloy. This is equivalent to the statement that
\begin{equation}
    \lim_{T \to \infty} c_{i\alpha} = c_\alpha,
    \label{eq:total_concentrations}
\end{equation}
where $c_\alpha$ is the overall (total) concentration of species $\alpha$. 

The ensemble average of the internal energy of this disordered phase can be described \textit{ab initio} via application of the coherent potential approxiation (CPA)~\cite{soven_coherent-potential_1967, gyorffy_coherent-potential_1972, stocks_complete_1978} within the Korringa--Kohn--Rostoker (KKR) formulation~\cite{korringa_calculation_1947, kohn_solution_1954, ebert_calculating_2011} of density functional theory (DFT)~\cite{hohenberg_inhomogeneous_1964, kohn_self-consistent_1965, martin_electronic_2004}. The KKR method uses multiple scattering theory (MST) to construct the single-particle Green's function for the Kohn--Sham equations, while the CPA seeks to construct an effective medium of electronic scatterers whose overall scattering properties approximate those of the disordered solid solution~\cite{faulkner_multiple_2018}. We emphasise here that the CPA has been established as a powerful and efficient technique for describing the electronic structure and physical properties of high-entropy alloys. It has previously been shown to accurately capture details of their electronic structure~\cite{redka_interplay_2024}, the nature of their heavily-smeared Fermi surfaces~\cite{robarts_extreme_2020}, aspects of their magnetism~\cite{billington_bulk_2020, bista_fast_2025}, and also a range of transport~\cite{samolyuk_temperature_2018, mu_uncovering_2019} and structural~\cite{tian_structural_2013, tian_impact_2017, huang_elasticity_2018} properties.

\subsection{Alloy thermodynamics and phase stability: Concentration wave analysis}
\label{sec:concentration_wave_analysis}

\subsubsection{Describing chemical fluctuations: concentration waves}

The KKR-CPA is not limited to studies of systems where all lattice sites have the same set of partial occupancies. It can also be applied to structures with multiple sublattices, each with different associated partial lattice site occupancies. For a particular (partially) ordered structure and a given set of (partial) sublattice occupancies, it is therefore possible to compute the total DFT energy of the system, $E_\textrm{int}(\{c_{i\alpha}\})$. However, when assessing which chemically ordered structures are most energetically favourable, direct evaluation of the energy of all potential chemically ordered (and segregated) structures is both laborious and computationally expensive, requiring evaluation of a large number of (partially) ordered supercells. A more elegant and efficient approach is to consider the change of the internal energy of the system in response to an infinitesimal chemical perturbation applied to the homogeneous solid solution~\cite{khan_statistical_2016, woodgate_compositional_2022, woodgate_short-range_2023, woodgate_modelling_2024}.

These perturbations are naturally expressed using the language of concentration waves. In this approach, we write an inhomogeneous set of site-wise concentrations (Eq.~\ref{eq:site-wise_concentrations}) as a perturbation to the homogeneous site-wise concentrations (Eq.~\ref{eq:total_concentrations}),
\begin{equation}
    c_{i\alpha} = c_\alpha + \Delta c_{i\alpha},
    \label{eq:perturbation}
\end{equation}
where $\Delta c_{i\alpha}$ denotes a spatially inhomogeneous chemical fluctuation. Owing to the translational symmetry of the underlying crystal lattice, it is convenient to express these chemical fluctuations in reciprocal space,
\begin{equation}
c_{i\alpha} = c_\alpha + \sum_{\mathbf{k}} e^{i \mathbf{k} \cdot \mathbf{R}_i} \Delta c_\alpha(\mathbf{k}),
\end{equation}
where $\Delta c_\alpha (\mathbf{k})$ represents a static \textit{concentration wave} with wavevector $\mathbf{k}$ and chemical polarisation $\Delta c_\alpha$. (For most crystallographically ordered structures, the sum over $\mathbf{k}$ typically runs over a handful of high-symmetry $\mathbf{k}$-vectors.) 

This multicomponent concentration wave formalism provides a concise and elegant way of describing a range of chemically ordered structures, and has its roots in pioneering work on concentration waves in binary alloys by Khachaturyan~\cite{khachaturyan_ordering_1978} and Győrffy and Stocks~\cite{gyorffy_concentration_1983}. Some one-dimensional examples of concentration waves and associated chemical polarisations are provided in Fig.~\ref{fig:concentration_waves}. In three dimensions, when the underlying lattice is bcc, the wavevector(s) associated with a B2 chemical ordering shown in Fig.~\ref{fig:A2_B2} is $\mathbf{k}_\textrm{ord} = (0,0, \frac{2\pi}{a})$ and equivalent, while the associated chemical polarisation, $\Delta c_\alpha((0,0,\frac{2\pi}{a}))$ then describes which chemical species are moving to which sublattice.

\subsubsection{Energetic cost of chemical fluctuations: the $S^{(2)}$ theory for multicomponent alloys}

To assess the energetic cost of such chemical fluctuations, and the temperatures at which they may emerge, we begin by writing down an approximate expression for the grand potential, or Landau free energy, $\Omega$. In general, this takes the form
\begin{equation}
    \Omega = U - TS -\mu N,
\end{equation}
where $U$ is the internal energy, $T$ the temperature, $S$ the entropy, $\mu$ the chemical potential(s), and $N$ the particle number(s). For the description of the alloy considered in this paper, the free energy is approximated via
\begin{equation}
    \Omega^{(1)} = \langle E_\text{int} \rangle [\{c_{i\alpha}\}] - \beta^{-1} \sum_{i \alpha} c_{i\alpha} \ln{c_{i\alpha}} - \sum_{i \alpha} \nu_{i \alpha} c_{i\alpha}.
    \label{eq:alloy_free_energy}
\end{equation}
In the above expression, the first term represents the average internal energy as obtained within the CPA, the second term represents the so-called entropy of mixing, and the third term represents the contribution to the free energy made by the sitewise chemical potentials, $\nu_{i\alpha}$, which, in principle, can vary for each chemical species and lattice site. We then expand this free energy about the homogeneous reference state ({\it i.e.} the disordered solid solution) in terms of the applied inhomogeneous fluctuation $\{\Delta c_{i\alpha}\}$. This is a Landau-type series expansion and is written
\begin{align}
    \Omega^{(1)}[\{c_{i\alpha}\}] &= \Omega^{(1)}[\{c_{\alpha}\}] + \sum_{i\alpha} \frac{\partial \Omega^{(1)}}{\partial c_{i\alpha}} \Big\vert_{\{c_{\alpha}\}} \Delta c_{i\alpha} \nonumber \\ 
    &+ \frac{1}{2} \sum_{i\alpha; j\alpha'} \frac{\partial^2 \Omega^{(1)}}{\partial c_{i\alpha} \partial c_{j\alpha'}} \Big\vert_{\{c_{\alpha}\}} \Delta c_{i\alpha}\Delta c_{j\alpha'} + \dots.
\label{eq:landau}   
\end{align}
In the full linear response theory, the site-wise chemical potentials, $\nu_{i \alpha}$, of Eq.~\ref{eq:alloy_free_energy} serve as Lagrange multipliers to conserve the overall concentrations of each chemical species. However, the variation of these multipliers is understood to be irrelevant to the underlying physics, so terms involving these derivatives are dropped~\cite{khan_statistical_2016, woodgate_compositional_2022, woodgate_short-range_2023, woodgate_modelling_2024}. In addition, combined with the translational symmetry of the disordered solid solution, the requirement that overall concentrations of each chemical species be conserved ensures that the first-order term of Eq.~\ref{eq:landau} is zero. Keeping terms to second-order, the change in energy due to an applied chemical perturbation is written
\begin{equation}
    \delta \Omega^{(1)} = \frac{1}{2} \sum_{i\alpha; j\alpha'} \Delta c_{i\alpha} [\beta^{-1} \, C_{\alpha\alpha'}^{-1} - S^{(2)}_{i\alpha, j\alpha'}] \Delta c_{j\alpha'}.
\label{eq:chemical_stability_real}
\end{equation}
The first term in square brackets, $C_{\alpha \alpha'}^{-1} = \frac{\delta_{\alpha \alpha'}}{c_\alpha}$, is a diagonal, positive definite matrix is associated with entropic contributions to the free energy, while the second term, $-\frac{\partial^2 \langle \Omega_\text{el} \rangle_0}{\partial c_{i\alpha} \partial c_{j\alpha'}} \equiv S^{(2)}_{i\alpha;j\alpha'}$ is the second-order concentration derivative of the average energy of the disordered alloy as evaluated within the CPA. 

Evaluation of $S^{(2)}_{i\alpha;j\alpha'}$ amounts to self-consistently solving a ring of coupled equations in terms of various CPA-relevant quantities, carefully incorporating the rearrangement of the electrons due to the applied chemical perturbation. This set of coupled equations are defined in Ref.~\cite{khan_statistical_2016}, and their solutions first examined and discussed in Ref.~\cite{woodgate_compositional_2022}. It should be emphasised that the present scheme is a rigorous generalisation of an earlier theory used to examine the phase stability of binary alloys in a similar manner~\cite{staunton_compositional_1994, johnson_first-principles_1994, clark_van_1995}.

Within the outlined concentration wave formalism, $S^{(2)}_{i\alpha;j\alpha'}$ is evaluated in reciprocal space, and therefore the change in free energy of Eq.~\ref{eq:chemical_stability_real} is written accordingly as:
\begin{equation}
    \delta \Omega^{(1)} = \frac{1}{2} \sum_{\bf k} \sum_{\alpha, \alpha'} \Delta c_\alpha({\bf k}) [\beta^{-1} C^{-1}_{\alpha \alpha'} -S^{(2)}_{\alpha \alpha'}({\bf k})] \Delta c_{\alpha'}({\bf k}).
\label{eq:chemical_stability_reciprocal}
\end{equation}
The matrix in square brackets, $[\beta^{-1} C^{-1}_{\alpha \alpha'} -S^{(2)}_{\alpha \alpha'}({\bf k})]$, referred to as the chemical stability matrix, is directly related to an estimate of the atomic short-range order (ASRO)~\cite{khan_statistical_2016}. This matrix can be thought of as analogous to a Hessian matrix of second derivatives evaluated at a stationary point of the free energy landscape. When searching for a disorder-order transition, we start from the high temperature solid solution, where all eigenvalues of this matrix are positive and the system is stable to applied chemical perturbations. We then progressively lower the temperature and look for the point at which the lowest lying eigenvalue of this matrix passes through zero for any $\mathbf{k}$-vector in the irreducible Brillouin zone. When this eigenvalue passes through zero at some temperature $T_\text{ord}$ and wavevector $\mathbf{k}_\text{ord}$, the solid solution is unstable to that applied chemical perturbation and we infer the presence of a disorder-order transition with chemical polarisation $\Delta c_\alpha$ given by the associated eigenvector. In this fashion we can predict both dominant ASRO and also the temperature at which the solid solution becomes unstable and a chemically ordered phase emerges.

\subsubsection{Pairwise form of the alloy internal energy: the Bragg-Williams Hamiltonian}
\label{sec:pairwise_energy}

The above concentration wave analysis and associated linear response theory provides information about the initial inferred chemical ordering in the system, but it is also possible to go further and map the derivatives of the internal energy of the disordered alloy, $S^{(2)}_{\alpha\alpha'}(\mathbf{k})$, to a real-space effective pair interaction describing the internal energy of a configuration with discrete lattice site occupancies. When an appropriate sampling technique is applied to this model, the phase behaviour can be studied in detail and equilibrated, lattice-based configurations extracted for illustration and further study.

The real space model and associated atom-atom effective pair interactions are for the Bragg--Williams Hamiltonian~\cite{bragg_effect_1934, bragg_effect_1935}, which can be thought of as a multicomponent Lenz--Ising Hamiltonian~\cite{brush_history_1967}, taking the form
\begin{equation}
    H(\{\xi_{i\alpha}\}) = \frac{1}{2}\sum_{i \alpha; j\alpha'} V_{i\alpha; j\alpha'} \xi_{i \alpha} \xi_{j \alpha'},
    \label{eq:b-w1}
\end{equation}
where $V_{i\alpha; j\alpha'}$ denotes the effective pair interaction between an atom of chemical species $\alpha$ on lattice site $i$ and an atom of chemical species $\alpha'$ on lattice site $j$. Assuming interactions are homogeneous and isotropic, we can express this as a sum of interactions over first-nearest neighbours, second-nearest neighbours, \textit{etc.}, writing $V^{(n)}_{\alpha \alpha'}$ to denote the interaction between species $\alpha$ and $\alpha'$ on coordination shell $n$. The effective pairwise interactions, are recovered from $S^{(2)}_{\alpha\alpha'}(\mathbf{k})$ by means of a inverse Fourier transform. The mapping from reciprocal-space to real-space and fixing of the gauge degree of freedom on the $V_{i\alpha; j\alpha'}$ is specified in earlier works~\cite{khan_statistical_2016, woodgate_compositional_2022, woodgate_short-range_2023, woodgate_modelling_2024}.

\subsection{Monte Carlo simulations}
\label{sec:atomistic}

Using the pairwise Hamiltonian of Eq.~\ref{eq:b-w1}, with atom-atom effective pair interactions recovered from the \textit{ab initio} data, we can employ lattice-based Monte Carlo simulations to further explore the alloy configuration space. Such simulations facilitate validation of the concentration wave analysis outlined above, and also allow us to search for further phase transitions occurring below any initial chemical ordering temperature. In this work, these atomistic Monte Carlo simulations consist of two different Markov chain random walks, both making use of Metropolis--Kawasaki dynamics \cite{kawasaki_diffusion_1966}, to explore the configuration space of the AlTiVNb and AlTiCrMo RSAs. Kawasaki dynamics ensure conservation of the overall concentration of each chemical species in the alloy by only performing swaps of pairs of atoms in the simulation cell. The sampling methods employed are Metropolis--Hastings Monte Carlo \cite{metropolis_equation_1953} and Wang-Landau sampling \cite{wang_efficient_2001}. We outline the details of these two sampling algorithms below.

\subsubsection{Metropolis--Hastings Algorithm}

The Metropolis--Hastings algorithm allows for a system to follow a chain of states to an equilibrium ensemble in a indeterminate but finite time \cite{metropolis_equation_1953, landau_guide_2014}. Phase equilibria are obtained by performing swap moves with a probability which depends on the energy difference between the initial and subsequent states and the simulation temperature. 

For relaxation models such as the ones in this paper the time-dependent behaviour obeys
\begin{equation}
    \frac{\partial P_n(t)}{\partial t} = -\sum_{n\neq m}\left[P_n(t)W_{n\rightarrow m}-P_m(t)W_{m\rightarrow n}\right]
\end{equation}
where $P_n(t)$ is the probability of the system being in a state $n$ at a time $t$, $m$ is the final state and $W_{n\rightarrow m}$ is the transition rate $n\rightarrow m$. When the system is in equilibrium,  $\frac{\partial P_n(t)}{\partial t}=0$, and we obtain the expression
\begin{equation}
    P_n(t)W_{n\rightarrow m}=P_m(t)W_{m\rightarrow n}
\end{equation}
which is known as detailed balance. If the dynamics satisfy this equation then the simulation is able to reach equilibrium. To calculate the acceptance rate (attempt rate multiplied by acceptance rate), the only quantity needed is the energy difference between the initial and proposed state, $\Delta E = E_m - E_n$, resulting in the the Metropolis acceptance probability
\begin{equation}
    W_{n\rightarrow m} = 
    \begin{cases}
        \; \text{exp}\left(-\Delta E/k_BT\right), &\quad \Delta E > 0\\
        \; 1, &\quad \Delta E \leq 0,
    \end{cases}
\label{eq:metropolis_transition_probability}
\end{equation}
where $T$ denotes the simulation temperature, and $k_B$ is the usual Boltzmann constant. In practice, for the alloy simulations conducted in this work, the Metropolis--Hastings algorithm allows us to obtain sample atomic configurations from an equilibrated ensemble at a range of temperatures which are suitable for visualisation and further study. At a given temperature, we initialise a supercell containing the correct overall concentration of each chemical species, where the initial atomic site occupancies are randomly generated. Two lattice sites (not necessarily nearest neighbours) are then selected at random and the energy difference obtained by swapping their atomic occupancies is computed. The swap is accepted according to the transition probability of Eq.~\ref{eq:metropolis_transition_probability}. This process is repeated until the internal energy of the simulation has reached a stable equilibrium by monitoring how the energy fluctuates about a stable energy average across a set number of previous states. Phase equilibria obtained using this method, allows for visualisation of alloy ordering and contextualise the results of Wang-Landau sampling method, the details of which follow.

\subsubsection{Wang-Landau Sampling}
\label{sec:wl_section}

The Wang-Landau sampling method~\cite{wang_efficient_2001, landau_guide_2014} is a flat energy histogram method that has a wide applicability due to its ability to obtain the density of states in energy from which information on thermodynamic quantities at any temperature can be obtained. In this context, the simulation density of states can be defined via the classical partition function. The conventional definition of the partition function with discretely labelled configurations $i$ is rewritten as
\begin{equation}
    Z = \sum_i e^{-E_i/k_BT} \approx  \sum_{E_j} g\left(E_j\right) e^{-E_j/k_BT},
\end{equation}
where $E$ is the energy, $k_B$ is the Boltzmann factor, $T$ is the temperature and $g(E)$ is the density of states. For the case of $E_j$, the energy represents a discretised energy bin within the energy histogram which is treated as a single energy macrostate. Wang-Landau sampling obtains an estimate for $g(E)$ across a chosen energy range during a Monte Carlo simulation. The density of states can begin with a simple estimate such as $g(E) \equiv 1$ and is improved throughout the course of the simulation. Atoms are swapped using the same method as in the Metropolis--Hastings case according to the probability
\begin{equation}
    P\left(E_1\rightarrow E_2\right) = \min \left( 1, \frac{g(E_1)}{g(E_2)} \right)
\end{equation}
where $E_1$ is the initial energy and $E_2$ is the energy of the proposed swap. After each proposed swap, the density of states is updated according to
\begin{equation}
    g(E_j) \rightarrow g(E_j)f_k,
\end{equation}
where $E$ is the energy of the resultant state, $f_k$ is a modification factor initially ($k=0$) greater than 1, and $k$ is the current iteration index of the Wang-Landau sampling algorithm. A histogram of the energies visited is maintained, $H(E_j)$, as is a measure of the `flatness', $F$ of the histogram,
\begin{equation}
    F = \frac{\min(H(E))}{\frac{1}{N}\sum_j^N H(E_j)}.
\end{equation}
Once $F$ is below a given tolerance, sampling is interrupted and $f$ is reduced for the next sampling iteration, \textit{e.g.} $f_{i+1}=\sqrt{f_i}$. Then the histogram entries are set to zero and the sampling continues until the flatness tolerance is achieved again. This process continues until the modification factor, $f$, is sufficiently close to unity, \textit{e.g.} $f<\text{exp}(10^{-6})$, and it is decided that $g(E)$ is known to an acceptable tolerance.

Once the density of states for the simulation has been obtained, the energy probability distribution at a given temperature can be found by using
\begin{equation}
    P(E_j, T) = \frac{g(E_j) e^{-E_j/k_BT}}{Z}
    \label{eq:prob_dist}
\end{equation}
from which we can extract a variety of of system properties. One such property is the specific heat capacity, $C$, as a function of temperature, $T$, recovered via~\cite{landau_guide_2014}
\begin{equation}
    C(T) = \frac{\langle E^2\rangle - \langle E\rangle^2}{k_BT^2},
\end{equation}
where $E$ is the energy of a simulation at a particular temperature, and the relevant ensemble averages are taken using the energy probability distribution from Eq.~\ref{eq:prob_dist}.

\subsubsection{Warren-Cowley atomic short-range order (ASRO) parameters}

To quantify the temperature-dependent atomic short range order (ASRO) in our Monte Carlo simulations, we use the Warren-Cowley ASRO parameters \cite{cowley_approximate_1950, norman_x-ray_1951, cowley_short-range_1965} adapted to the multicomponent setting, defined as
\begin{equation}
    \alpha^{pq}_n=1-\frac{P^{pq}_n}{c_q}
\end{equation}
where $n$ refers to the $n$\textsuperscript{th} coordination shell, $P^{pq}_n$ is the conditional probability of an atom of type $q$ neighbouring an atom of type $p$ on shell $n$, and $c_q$ is the total concentration of atom type $q$. When $\alpha^{pq}_n > 0$, $p$-$q$ pairs are disfavoured on shell $n$ and, when $\alpha^{pq}_n < 0$ they are favoured. The value $\alpha^{pq}_n = 0$ corresponds to the ideal, maximally disordered solid solution.

These ASRO parameters are evaluated across the sampled energy range (in practice, by evaluating the average ASRO for configurations drawn from each `bin' of the energy histogram) and then reconfigured to be written as a function of temperature
\begin{equation}
    \alpha^{pq}_n(T) = \sum_{E_j} \alpha^{pq}_n(E_j) \cdot P(E_j, T),
\end{equation}
where $P(E, T)$ is the energy probability distribution at a given temperature.

In combination, the specific heat capacity and Warren-Cowley ASRO parameters facilitate a detailed description of emergent phase transitions in a multicomponent alloy. The specific heat capacity data allows us to accurately identify the temperature at which a phase transition occurs, while the ASRO parameters help to quantify the nature of the transition in terms of atom-atom correlations.

\subsection{Residual resistivity calculations}
\label{sec:resistivity_theory}

A fundamental quantity of solid state physics is the electrical resistivity of a material, $\rho$. At its most basic level, this quantity allows one to distinguish between metals, semiconductors, and insulators. In metals and alloys, the electrical resistivity is directly connected to the electronic mean free path of states at the Fermi level, $\lambda_e(E_F)$. In a crystalline, metallic material in the limit $T\rightarrow0$~K, Bloch states are eigenvalues of the electronic Hamiltonian and the electronic mean free path diverges, $\lambda_e(E_F) \rightarrow \infty$, resulting in a vanishing resistivity, $\rho \rightarrow 0$. However, in a disordered, substitutional alloy, translational symmetry is violated, Bloch's theorem does not apply, and the electronic mean free path is finite even at $T=0$~K. This results in a finite \textit{residual} resistivity, $\rho_0$, defined as the limiting value of the resistivity as $T\rightarrow0$, which can provide insight into the effects of chemical disorder on the electronic structure of a material \cite{mu_uncovering_2019}.

The KKR-CPA naturally captures the broken translational symmetry of a substitutionally disordered alloy, and provides a means by which to evaluate the Green's function $G$ for such a system~\cite{ebert_calculating_2011}. From this Green's function, one can then use the linear response Kubo--Greenwood formula~\cite{kubo_statistical-mechanical_1957, greenwood_boltzmann_1958} applied to the KKR-CPA effective medium~\cite{butler_theory_1985} to evaluate the resistivity at both zero and finite temperatures, where the latter can also incorporate the effects of thermally induced ionic displacements within an `alloy analogy' model~\cite{ebert_calculating_2011}. The Kubo--Greenwood formula for the conductivity tensor, $\sigma_{\mu\nu}$, implemented in \textit{SPR-KKR}~\cite{ebert_munich_nodate,ebert_calculating_2011} is given by
\begin{equation}
    \sigma_{\mu\nu} = \frac{\hbar}{\pi V} \text{Tr} \langle j_\mu \text{Im} G^+ (E_F) j_\nu \text{Im} G^+ (E_F) \rangle_\mathrm{c},
\label{eq:kubo_greenwood}
\end{equation}
where $V$ is the simulation cell volume, $G^+(E_F)$ is the retarded Green's function at the Fermi level, and $j_\mu$ the current density operator with Cartesian coordinate indices $\mu$ and $\nu$. Angled brackets, $\langle \cdot \rangle_\mathrm{c}$, indicate the CPA average over substitutional disorder. From the conductivity tensor, it is a simple matter to recover the resistivity of a material. In this study, we employ this approach to evaluate the influence of (partial) chemical order on the electronic transport properties of the examined alloys.

\begin{figure*}[ht!]
    \centering
    \includegraphics[width=\textwidth]{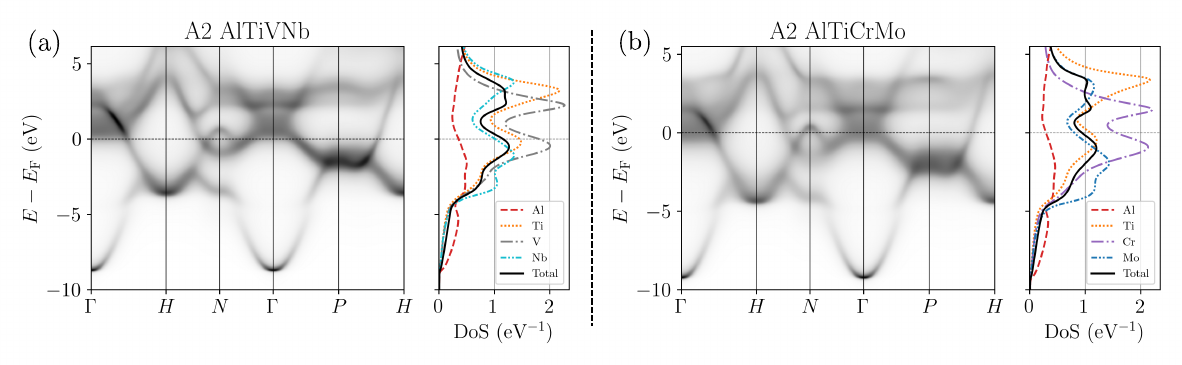}
    \caption{Plots of the Bloch spectral function (BSF) and electronic density of states (DoS) for (a) AlTiVNb and (b) AlTiCrMo, modelled assuming a chemically disordered bcc (A2) crystal structure. For both alloys, there is heavy smearing of all electronic bands on account of the substitutional disorder, which is associated with the finite lifetime of electronic states. The narrow $d$ bands of AlTiCrMo are shifted down relative to the Fermi level, $E_F$, compared to AlTiVNb, on account of the increased valence of Cr and Mo as compared to V and Nb.}
\label{fig:A2_AlTiVNb_and_AlTiCrMo_DoS_and_BSF}
\end{figure*}

\section{Results and Discussion}
\label{sec:results}

\subsection{Electronic Structure}
\label{sec:electronic_structure}

We begin our analysis by performing self-consistent DFT calculations to model the electronic structure of the disordered solid solutions. We use the all-electron \textit{SPR-KKR} package~\cite{ebert_munich_nodate,ebert_calculating_2011} to construct the self-consistent potentials of the KKR~\cite{korringa_calculation_1947, kohn_solution_1954, ebert_calculating_2011} formulation of DFT~\cite{hohenberg_inhomogeneous_1964, kohn_self-consistent_1965, martin_electronic_2004}, using the CPA~\cite{soven_coherent-potential_1967, gyorffy_coherent-potential_1972, stocks_complete_1978} to average over chemical disorder. We perform scalar-relativistic calculations within the atomic sphere approximation (ASA)~\cite{andersen_linear_1975}, employing an angular momentum cutoff of $l_\text{max} = 3$ for basis set expansions, and a 32-point semi-circular contour in the complex plane to integrate over valence energies. Sampling for integrations over the Brillouin zone during the self-consistency cycle used a parameter of \texttt{NKTAB=5000}, resulting in a dense $62\times62\times62$ $\mathbf{k}$-point mesh over the first Brillouin zone. We use the local density approximation (LDA) and the exchange-correlation functional of Vosko, Wilk, and Nusair~\cite{vosko_accurate_1980}. Both systems are treated as non-magnetic; for AlTiCrMo a self-consistent, spin-polarised, disordered local moment (DLM) calculation~\cite{staunton_disordered_1984} (to represent the paramagnetic state) was tested but the local moments collapsed during the self-consistency cycle, which is consistent with calculations for elemental Cr~\cite{pindor_disordered_1983}. We therefore believe that non-magnetic calculations are representative of the electronic structure of these alloys at typical processing temperatures. Further details of the self-consistent calculations and the relevant data can be found in the open-access dataset associated with this work~\cite{dataset}.

\begin{table}[b]
\begin{ruledtabular}
\begin{tabular}{lrr}
Alloy    & $a_0$ (\AA) & $B_0$ (GPa) \\ \hline
A2 AlTiVNb  & 3.113       & 151.2       \\
A2 AlTiCrMo & 3.025       & 190.1    
\end{tabular}
\end{ruledtabular}
\caption{DFT-optimised lattice parameters and bulk moduli for the two considered alloys when modelled with a chemically disordered bcc (A2) crystal structure. That AlTiCrMo has a smaller optimised lattice parameter than AlTiVNb is consistent with the decrease in atomic radii in the $3d$ transition metals from left to right across the periodic table.}
\label{table:a2_lattice_constants}
\end{table}

We optimise the lattice parameter and associated unit cell volume for the disordered bcc (A2) phase of both alloys by calculating the total DFT energy across a range of lattice parameters and fitting the results to a stabilised jellium equation of state (SJEOS)~\cite{alchagirov_energy_2001} as implemented in the Atomic Simulation Environment (ASE)~\cite{hjorth_larsen_atomic_2017}. This procedure also allows us to estimate the bulk modulus for both alloys. Our optimised lattice parameters and calculated bulk modulii are shown in Table~\ref{table:a2_lattice_constants}. Our optimised bcc cubic lattice parameters for the AlTiVNb and AlTiCrMo alloys of 3.113~\AA~and 3.025~\AA, respectively, compare reasonably with the experimentally determined values of 3.186~\AA~\cite{stepanov_structure_2015} and 3.100~\AA~\cite{chen_crystallographic_2019}, though are slight underestimates. This underestimation is typically expected when using LDA exchange correlation functionals on materials containing transition metals~\cite{grabowski_ab_2007}. In addition, our DFT calculations do not account for the modest thermal expansion of the lattice which would be present at room temperature.

Proceeding, in Fig.~\ref{fig:A2_AlTiVNb_and_AlTiCrMo_DoS_and_BSF}, we plot the electronic density of states (DoS) and Bloch spectral function (BSF) along high-symmetry lines of the Brillouin zone of the bcc lattice for both AlTiVNb and AlTiCrMo simulated in the chemically disordered A2 phase. The DoS and BSF are plotted for an energy range around $E_F$. The BSF can be thought of as a $\mathbf{k}$-resolved density of states~\cite{ebert_calculating_2011}. For a pristine crystal, it reduces simply to the conventional bandstructure. However, in a system with broken translational symmetry, such as the substitutional alloys of this work, pristine bands are smeared out by the disorder. The degree of smearing of bands can be related to the average lifetime (or mean free path) of electronic states in a material, which leads naturally to calculations of quantities such as the residual resistivity of a material.

Considering first the BSF data for the two alloys, it can be seen that the broken translational symmetry, associated with the substitutional disorder, heavily smears electronic states across the entire considered energy range. It can also be seen that the narrow $d$-bands associated with the transition metals are shifted down relative to the the Fermi level, $E_F$, for AlTiCrMo as compared to AlTiVNb, which is associated with the increased valence of Cr and Mo as compared to V and Nb.

\begin{figure*}[ht!]
\centering
\includegraphics[width=\textwidth]{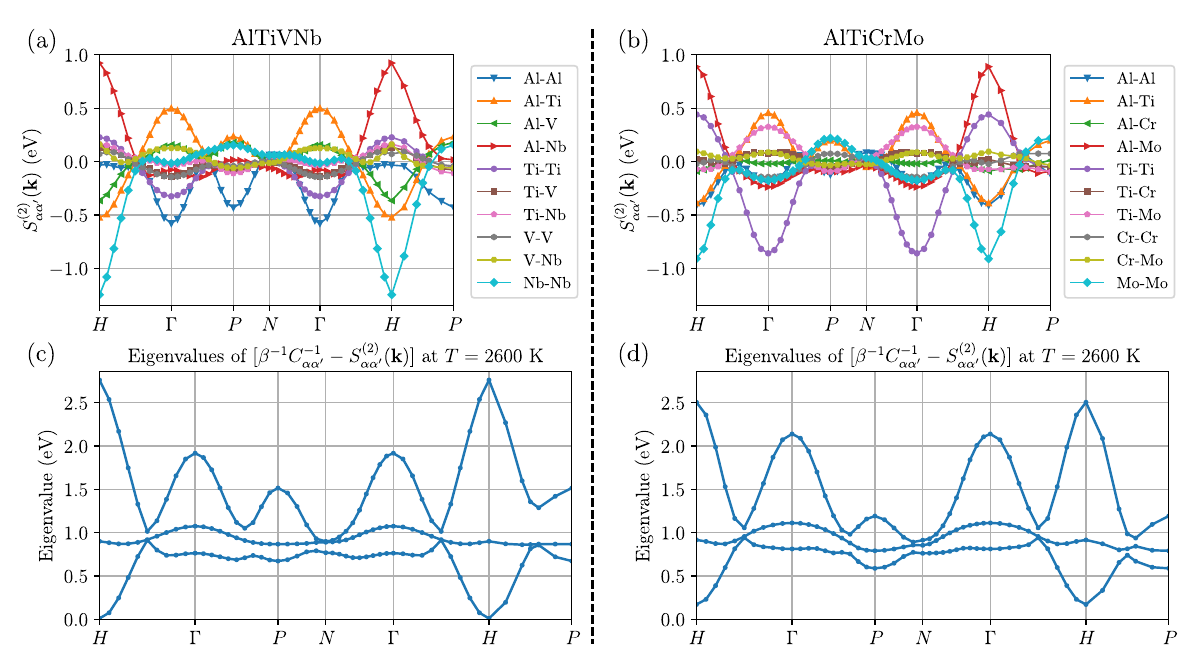}
    \caption{Plots of $S^{(2)}_{\alpha \alpha'}({\bf k})$ (top) and eigenvalues of the chemical stability matrix, $[\beta^{-1} C^{-1}_{\alpha \alpha'} -S^{(2)}_{\alpha \alpha'}({\bf k})]$, evaluated at a temperature of $T=2600$~K (bottom) along high-symmetry directions of the irreducible Brillouin zone of the bcc lattice for AlTiVNb and AlTiCrMo. The quantity $S^{(2)}_{\alpha \alpha'}({\bf k})$ can be thought of as the Fourier transform of an atom-atom effective pair interaction between chemical species, indicating the likely degree of atomic short-range order in the solid solution. The chemical stability matrix, and its eigenvalues, allow one to infer long-range chemical orderings using the language of concentration waves. For both alloys, the minimum eigenvalue at $H$, corresponding to $\mathbf{k}=(0,0,\frac{2\pi}{a})$, is associated with a B2 (CsCl) chemical ordering.}
    \label{fig:s2_comparison}
\end{figure*}

When considering the species-resolved DoS, contrasts should be made between the $sp$ states associated with Al (a `simple' metal); those associated with Ti, V, and Cr ($3d$ metals); and those associated with Nb and Mo ($4d$ metals). In elemental Al, electronic states are nearly free-electron like in character, and the DoS can be expected to be approximately proportional to $\sqrt{E}$. For both of the alloys considered here, that the species-resolved DoS for Al diverges from this behaviour and displays localised peaks around the $d$-states of the transition metals is indicative of the formation of hybridised $p$-$d$ bonding states between Al and the other elements present in the compositions. It should also be noted that the width of the $d$-bands associated with Ti, V, and Cr ($3d$ transition metals) are narrower than those associated with Nb and Mo ($4d$ transition metals). In previous studies, we have found that both bandwidth differences~\cite{woodgate_short-range_2023} and $p$-$d$ hybridisation~\cite{woodgate_structure_2024} can drive strong atomic ordering tendencies in multicomponent alloys such as the ones considered in this work.

\begin{table*}[ht!]
\begin{ruledtabular}
\begin{tabular}{lcccrrrr}
Alloy    & $T_\textrm{ord}$ (K) & $\mathbf{k}_\textrm{ord}$ ($2\pi/a$) & Structure & $\Delta c_\textrm{Al}$ & $\Delta c_\textrm{Ti}$ & $\Delta c_\textrm{V/Cr}$ & $\Delta c_\textrm{Nb/Mo}$ \\ \hline
AlTiVNb  & 2574             & (0,0,1)                   & B2        & 0.700        & $-$0.571     & $-$0.361     & 0.232        \\
AlTiCrMo & 2107             & (0,0,1)                   & B2        & 0.521        & $-$0.792     & $-$0.044        & 0.315       
\end{tabular}
\end{ruledtabular}
\caption{Predicted chemical ordering temperatures ($T_\textrm{ord}$), associated concentration wavevectors ($\mathbf{k}_\textrm{ord}$), and chemical polarisations ($\Delta c_\alpha$) for the two RSAs considered in this work. For both alloys, the wavevector describing the chemical ordering is $\mathbf{k}_\textrm{ord} = (0,0,1)$, which describes a B2 chemical ordering imposed on the bcc lattice. For AlTiVNb, the chemical polarisation of the concentration wave suggest that the B2 ordering will have one sublattice rich in Al and Nb, while the other will be rich in Ti and V, with Al and Ti having the strongest site preferences. For AlTiCrMo, Al and Mo are anticipated to move to one sublattice, Ti the other, with Cr remaining relatively disordered and spread across both sublattices.
}
\label{table:transition_temperatures}
\end{table*}

\subsection{Concentration Wave Analysis}
\label{sec:linear_response}

To study the nature of atomic short-range order in the solid solution, as well as to infer the temperature at which the solid solution becomes unstable to chemical fluctuations and a long-range crystallographically ordered structure emerges, we look to the $S(2)$ theory for multicomponent alloys, as outlined in Sec.~\ref{sec:concentration_wave_analysis}. We use a computational implementation of this theory of which the details have been discussed extensively in earlier works~\cite{khan_statistical_2016, woodgate_compositional_2022, woodgate_short-range_2023, woodgate_modelling_2024}. This methodology has previously been applied with success to the Cantor alloy and its derivatives~\cite{woodgate_compositional_2022, woodgate_interplay_2023}, the refractory high-entropy alloys without Al present in the composition~\cite{woodgate_short-range_2023, woodgate_competition_2024}, and to the Al$_x$CrFeCoNi system~\cite{woodgate_structure_2024}.

Shown in the top row of Fig.~\ref{fig:s2_comparison} are computed values of $S^{(2)}_{\alpha \alpha'}({\bf k})$ along high-symmetry lines of the irreducible Brillouin zone of the bcc lattice, while the bottom row shows the eigenvalues of the chemical stability matrix constructed from the \textit{ab initio} data for both alloys along the same high-symmetry directions and evaluated at a temperature of $T=2600$~K. The quantity $S^{(2)}_{\alpha \alpha'}({\bf k})$ can be thought of as the Fourier transform of an effective pair interaction between different chemical species in the alloy, indicating the strength and nature of various atom-atom correlations in the solid solution. Considering the data for (a) AlTiVNb and (b) AlTiCrMo, we see that there are strong interactions in both alloys for Al-Al, Al-Ti, Al-$4d$, Ti-Ti, and $4d$-$4d$ elemental pairs. (Note that Nb and Mo are the $4d$ elements present in the two compositions.) However, despite their presence in both compositions, the Al-Al, Al-Ti, and Ti-Ti data, although qualitatively similar, are substantially different in numerical value between the two alloys. This emphasises that extrapolation of atom-atom interactions from binary subsystems~\cite{troparevsky_criteria_2015, santodonato_predictive_2018} may not be a reliable approach for modelling the thermodynamics of high-entropy alloys.

Proceeding, we now consider the eigenvalues of the chemical stability matrix, $[\beta^{-1} C^{-1}_{\alpha \alpha'} -S^{(2)}_{\alpha \alpha'}({\bf k})]$ for (c) AlTiVNb and (d) AlTiCrMo evaluated at a temperature of $T=2600$~K, where both matrices are strictly positive definite, \textit{i.e.} all eigenvalues are greater than zero. For both alloys, the eigenvalue with lowest energy occurs at the $H$ point, corresponding to $\mathbf{k} = (0,0,\frac{2\pi}{a})$ and associated with a B2 chemical ordering as illustrated in Fig.~\ref{fig:A2_B2}. That, at a temperature of $T=2600$~K, the lowest-lying eigenvalue for AlTiVNb is lower than that for AlTiCrMo indicates that the B2 chemical ordering temperature for AlTiVNb will be higher than that for AlTiCrMo. The local minimum eigenvalue at the $P$ point (associated with B32 ordering tendencies~\cite{woodgate_short-range_2023}) for AlTiCrMo is perhaps suggestive of some weaker secondary ordering tendencies in this alloy.

\begin{table}[b]
\begin{ruledtabular}
\begin{tabular}{lrr}
Alloy    & $a_0$ (\AA) & $B_0$ (GPa) \\ \hline
B2 AlTiVNb  & 3.111       & 153.2       \\
B2 AlTiCrMo & 3.027       & 186.1    
\end{tabular}
\caption{DFT-optimised lattice parameters and bulk modulii for the two considered alloy when modelled with the predicted B2 (CsCl) crystal structures. The chemical orderings are found to have little impact on either the optimised lattice parameter or bulk modulus of the alloy within our modelling.}
\label{table:b2_lattice_constants}
\end{ruledtabular}
\end{table}

As outlined in Sec.~\ref{sec:concentration_wave_analysis}, we then search for the temperature at which an eigenvalue passess through zero for some wavevector in the irreducible Brillouin zone, as this provides a mean-field estimate of the chemical ordering temperature. Table~\ref{table:transition_temperatures} gives the predicted ordering temperature, wavevector, and associated eigenvector describing the chemical orderings for both AlTiVNb and AlTiCrMo RSAs. For both alloys, the wavevector describing the ordering is $\mathbf{k} = (0,0,\frac{2\pi}{a})$, associated with a B2 chemical ordering. For AlTiVNb this ordering is predicted to occur at 2574~K, while for AlTiCrMo it is predicted to occur at 2107~K. The eigenvectors or `chemical polarisations' associated with these orderings then provide insight as to which chemical species are preferentially sitting on which sublattice. For AlTiVNb, Al and Nb are predicted to move to one sublattice, while Ti and V move to the other, with Al and Ti having stronger site preferences than V and Nb. For AlTiCrMo by far the strongest site preference is for Ti, with Al and Mo moving to the other sublattice, and Cr remaining comparatively disordered, indicated by the small value of $\Delta c_\textrm{Cr}$.

\begin{figure}[b]
    \centering
\includegraphics[width=\linewidth]{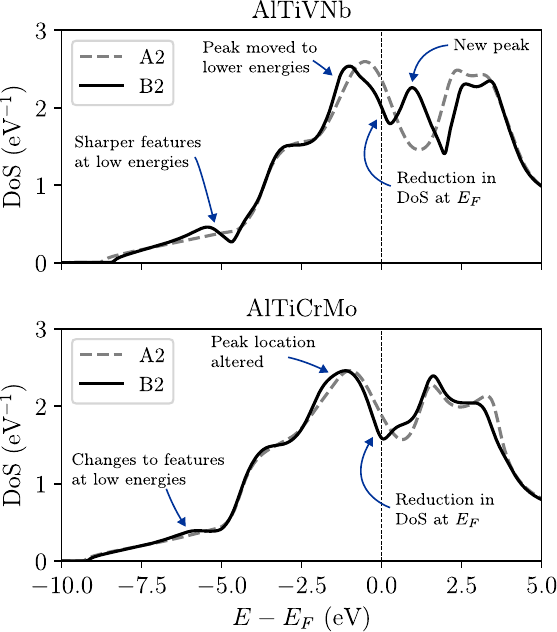}
    \caption{Comparison of the calculated total electronic density of states (DoS) for the AlTiVNb and AlTiCrMo RSAs in their chemically disordered (A2) structures compared to in their predicted chemically ordered (B2) structures. Key changes induced by the ordering include the appearance of new features at low energies, the movement of peaks, and a reduction in the DoS at the Fermi level, $E_F$. Note that here we use the cubic, 2-atom representation of the disordered A2 structure, for consistency with the B2 calculation.}
    \label{fig:a2_b2_dos_comparison}
\end{figure}

\begin{figure*}[ht]
    \centering
\includegraphics[width=\textwidth]{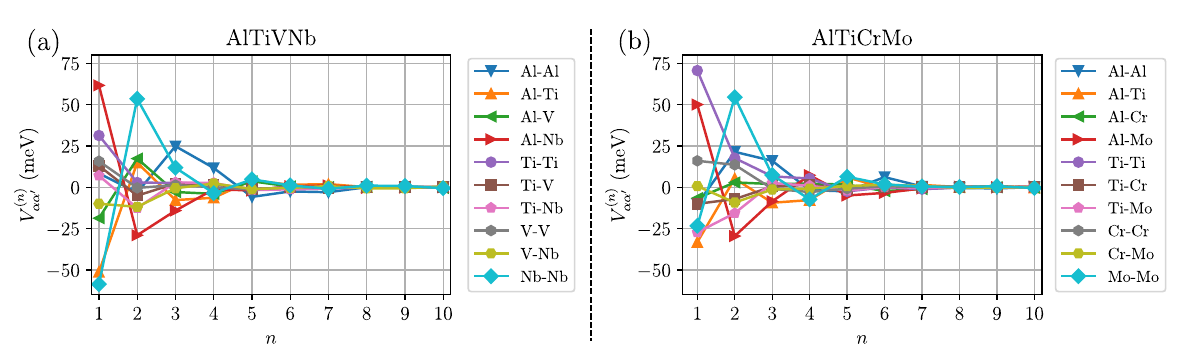}
    \caption{Plots of real-space effective pair interactions, $V_{\alpha \alpha'}^{(n)}$, as a function of coordination shell number, $n$, recovered from the reciprocal space $S^{(2)}_{\alpha \alpha'}({\bf k})$ data for (a) AlTiVNb and (b) AlTiCrMo. The notation $V_{\alpha \alpha'}^{(n)}$ indicates the interaction between chemical species $\alpha$ and $\alpha'$ on coordination shell $n$, \textit{i.e.} at $n$\textsuperscript{th} nearest neighbour distance. For both alloys, it can be seen that the comparative strength of interactions tails off quickly with increasing distance.}
    \label{fig:epi_comparison}
\end{figure*}

The temperature and nature of these B2 chemical orderings are in agreement with existing literature, both experimental and computational. (We note that our computed transition temperatures in Table~\ref{table:transition_temperatures} are likely to be overestimates as they are computed within a single-site or `mean-field' theory.) Experimentally, Ref.~\cite{yurchenko_effect_2018} reports that AlTiVNb samples annealed for 24~h at 1200~$^\circ$C were composed entirely of a B2 phase, while samples annealed at 1000~$^\circ$C and 800~$^\circ$C contained a majority B2 phase with some Nb$_2$Al-like $\sigma$-phase precipitates. Computationally, K\"ormann \textit{et al.}~\cite{kormann_b2_2021} simulated the phase stability of AlTiVNb across a wide temperature range using DFT calculations, a machine-learned interatomic potential, and Monte Carlo simulations. They report a B2 chemical ordering occurring at approximately 1700~K, with Al and Ti expressing strong site preferences, and Nb and V expressing weaker preferences, in good agreement with our own predictions. For AlTiCrMo, Ref.~\citenum{chen_crystallographic_2019} reports combined experimental results and thermodynamic calculations. The thermodynamic calculations estimate a B2 ordering temperature of just over 1100~$^\circ$C ($\approx1500$~K), in alignment with our result, though these authors also report that it is challenging to identify the B2 ordered structure in their experimental samples via X-ray diffraction due to counterbalancing of sublattice occupancies and atomic form factors in their predicted B2 ordered structure.

Though the eigenvectors of Table~\ref{table:transition_temperatures} describe initial (infinitesimal) change in partial lattice site occupancies, they can be used to infer partially ordered phases by allowing the size of the fluctuation to `grow' until one sublattice contains (at least) one atomic species whose concentration reaches zero. This condition identifies the largest permitted chemical fluctuation consistent with that concentration wave's chemical polarisation. We define an order parameter $\eta$ to quantify the degree of (partial) ordering, where $\eta=0$ corresponds to the disordered solid solution, and $\eta=1$ corresponds to the largest permitted chemical fluctuation consistent with the predicted chemical polarisation. For both alloys, the predicted partial occupancies of the two non-equivalent lattice sites as a function of $\eta$ are provided in the Supplementary Material~\cite{supplemental}.

To assess the impact of these chemical orderings on the mechanical properties of the alloys, we take the obtained partial lattice site occupancies for the case $\eta=1$ and compute revised lattice parameters and bulk moduli. These self-consistent calculations were again performed using \textit{SPR-KKR}~\cite{ebert_munich_nodate,ebert_calculating_2011}, with the same computational parameters as for calculations on the disordered phases, the results of which are provided in Table~\ref{table:b2_lattice_constants}. When these data are compared to the calculations performed on the disordered alloys (Table~\ref{table:a2_lattice_constants}), it can be seen that the predicted chemical orderings do not significantly impact interatomic spacing or the bulk modulus of either alloy. 

However, as a result of the B2 chemical orderings, the DFT internal energy of the alloys are found to be lowered by 49.6 meV/atom and 37.2 meV/atom for AlTiVNb and AlTiCrMo respectively. When we consider the electronic density of states for both alloys in their predicted B2 chemically ordered structures, shown in Fig.~\ref{fig:a2_b2_dos_comparison}, we see how this chemical ordering impacts the alloys' electronic structure. For both systems, there is a modest reduction in the DoS at the Fermi level, $E_F$, and the peak in the DoS below $E_F$ is also shifted to lower energies. Some sharper features at low energies are also seen to emerge for both alloys, suggestive of the formation of more localised bonding states. Finally, for AlTiVNb, a new peak in the DoS above $E_F$ can be seen to emerge.

\subsection{Effective pair interactions}
\label{sec:effective_pair_interactions}

From the \textit{ab initio} reciprocal space $S^{(2)}_{\alpha \alpha'}({\bf k})$ data, we Fourier transform and fit atom-atom effective pair interactions (EPIs) for both of the considered RSAs. These effective pair interactions are for the Bragg-Williams Hamiltonian, Eq.~\ref{eq:b-w1}. We assume that the interactions are homogeneous (translationally invariant) and isotropic, and write $V_{\alpha \alpha'}^{(n)}$ to denote the EPI between chemical species $\alpha$ and chemical species $\alpha'$ at $n$\textsuperscript{th} nearest neighbour distance.

We plot $V_{\alpha \alpha'}^{(n)}$ as a function of coordination number, $n$, for a fit to the first ten coordination shells of the bcc lattice for AlTiVNb and AlTiCrMo in Fig.~\ref{fig:epi_comparison}. For both alloys, it can be seen that interactions are dominated by the first few coordination shells and tail off quickly with increasing distance. It should be noted that, despite the fact that Al and Ti are present in both of the considered compositions, Al-Al, Al-Ti, and Ti-Ti interactions are different between the two systems. This confirms our earlier assertion that extrapolating atom-atom interactions from data for binary alloys~\cite{troparevsky_criteria_2015, santodonato_predictive_2018} (the so-called `pseudobinary' approximation) may be unreliable in the multicomponent setting.

For the Monte Carlo simulations which are to follow, we choose to truncate our fitted interactions and perform a fit limited to the first six coordination shells of the bcc lattice, corresponding to real-space radial `cutoffs' of 6.226~\AA~and 6.050~\AA~for AlTiVNb and AlTiCrMo respectively. Such cutoffs are supported by the data shown in Fig.~\ref{fig:epi_comparison} and are also consistent with typical radial cutoffs chosen for machine-learned interatomic potentials~\cite{zhang_roadmap_2025}. These truncated effective pair interactions are explicitly tabulated in the Supplementary Material~\cite{supplemental} and are also provided in the open-access dataset associated with this study~\cite{dataset}.

\subsection{Monte Carlo simulations}
\label{sec:atomistic_results}

\begin{figure}[ht]
    \centering
\includegraphics[width=0.99\linewidth]{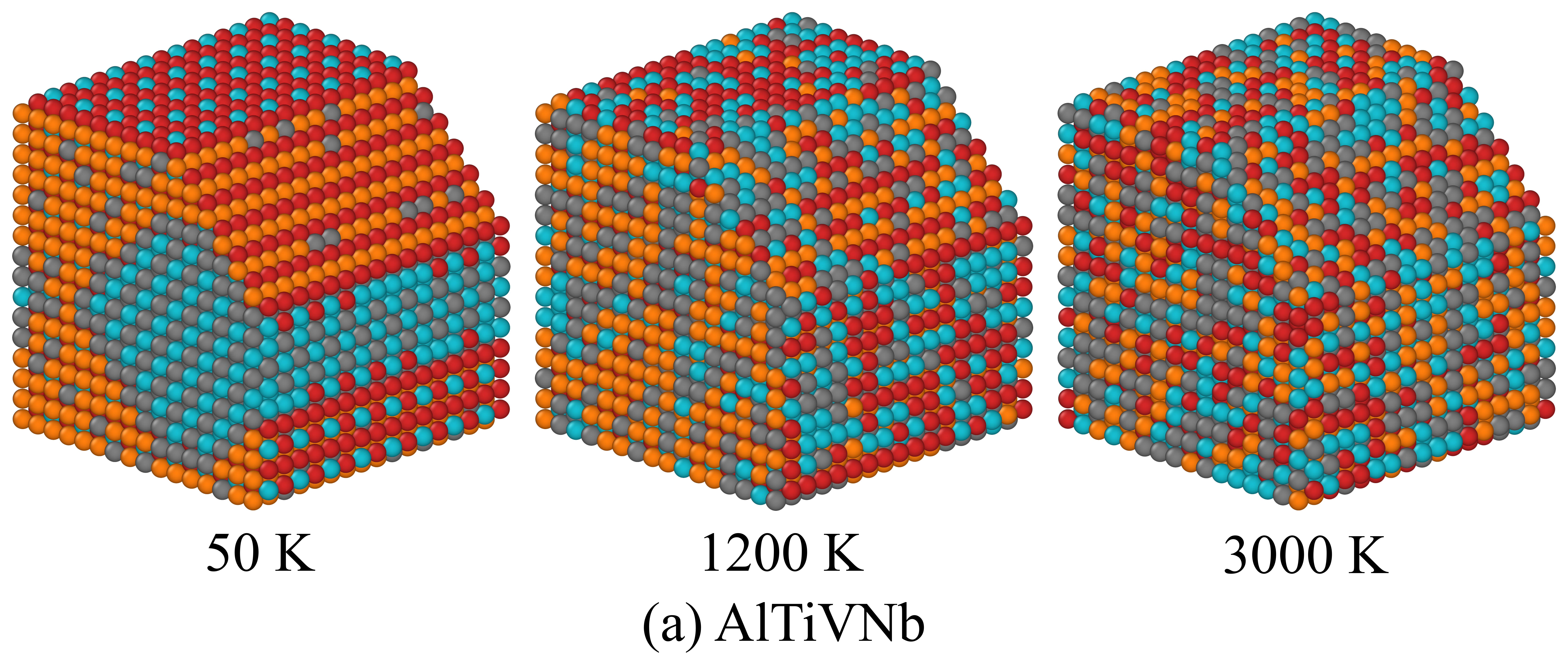}\\
\includegraphics[width=0.99\linewidth]{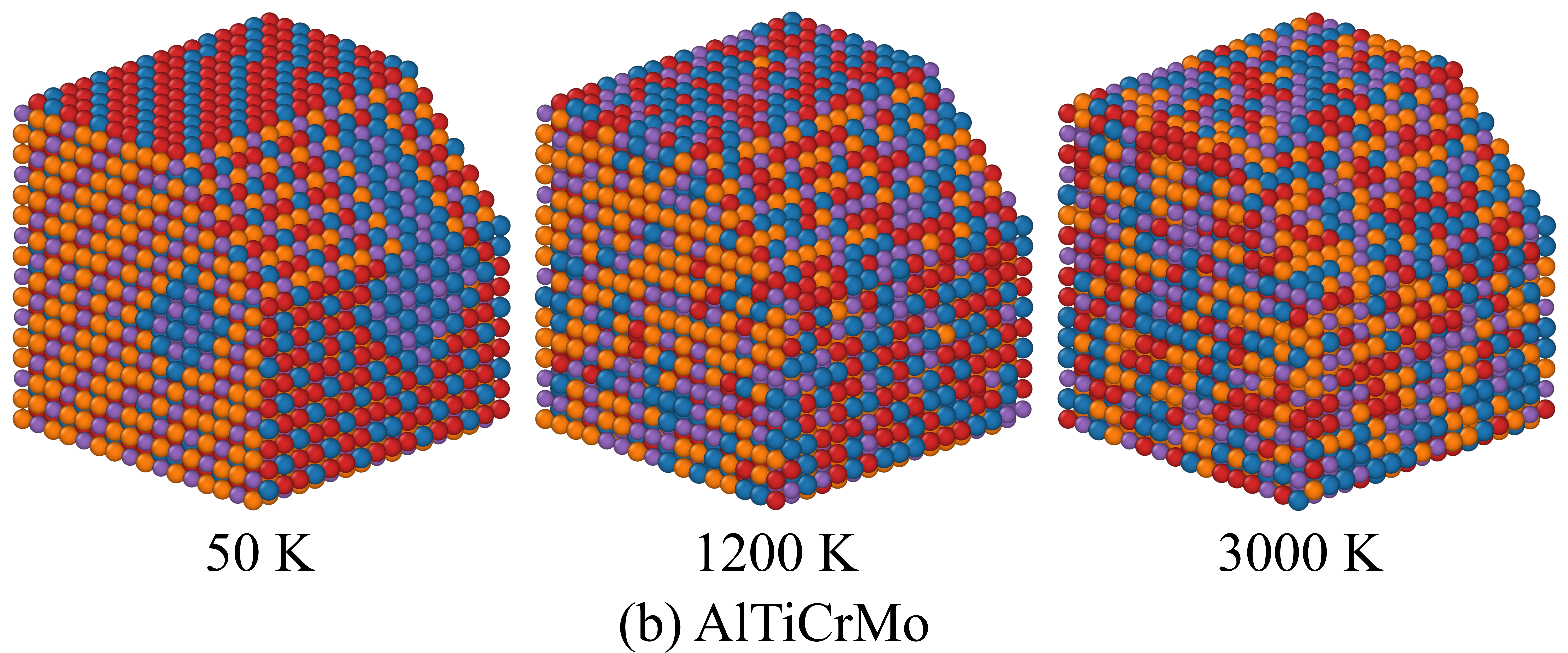}
    \caption{Representative configurations from Monte Carlo simulations for AlTiVNb and AlTiCrMo equilibrated at simulation temperatures of 3000, 1200, and 50~K. Al, Ti, V, Nb Cr, and Mo are coloured red, orange, grey, turquoise, purple, and blue respectively. A cut has been made through the simulation cell to make ordered structures more clearly visible. In both AlTiVNb and AlTiCrMo, the emergence of a layered, B2-like structure can be seen in the respective T = 1200~K configuration. At low temperatures, additional atom-atom correlations emerge and the simulations eventually decompose into multiple competing phases. Images generated using OVITO \cite{stukowski_visualization_2010}.}
    \label{fig:configurations}
\end{figure}

\begin{figure*}[ht]
    \centering
\includegraphics[width=\linewidth]{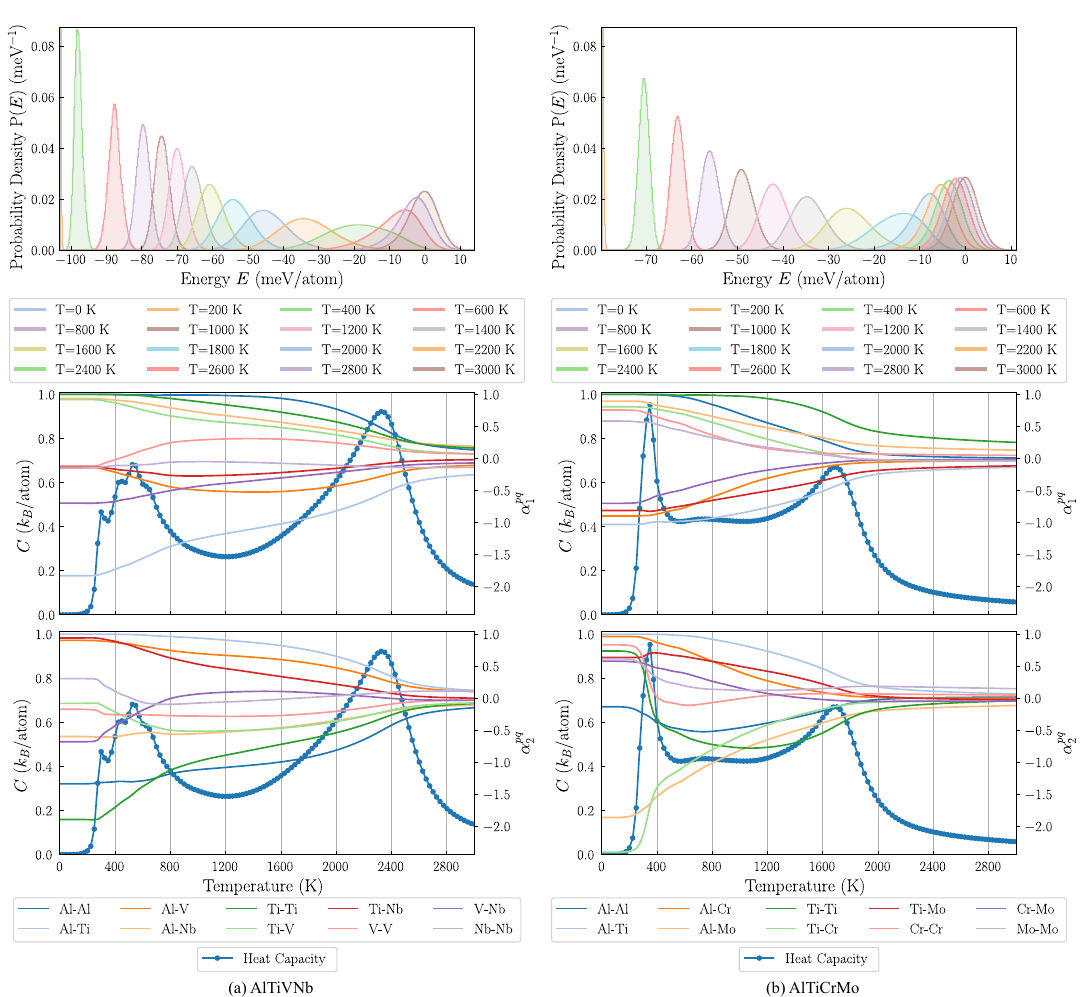}
    \caption{Plots of energy probability distributions, Warren-Cowley ASRO parameters ($\alpha^{pq}_n$) and simulation heat capacity ($C$) as a function of temperature for the two multicomponent systems obtained using lattice-based Monte Carlo simulations employing Wang-Landau sampling. We show $\alpha^{pq}_n$ for $n$ = 1, 2. The zero of the energy scale for the energy histograms for each alloy is set to be equal to the average internal energy of the alloy obtained at a simulation temperature of 3000~K. Both alloys exhibit a B2 chemical ordering, indicated by the peaks in SHC at approximately 2300~K and 1700~K for (a) AlTiVNb and (b) AlTiCrMo, respectively. Consistent with the higher ordering temperature, the phase transition in AlTiVNb results in a larger associated shift in the total energy-per-atom of the simulation cell compared to AlTiCrMo.}
    \label{fig:phase_diagrams}
\end{figure*}

To validate the results of the above concentration wave analysis, and to further explore the configuration space below the initial chemical ordering temperatures, we perform lattice-based Monte Carlo simulations using the BraWl package~\cite{naguszewski_brawl_nodate} as outlined in Sec.~\ref{sec:atomistic}. Using the atom-atom effective pair interactions obtained from the \textit{ab initio} data and plotted in Fig.~\ref{fig:epi_comparison}, we perform Monte Carlo simulations for both alloys using the Wang-Landau sampling algorithm \cite{wang_efficient_2001}. The AlTiCrMo and AlTiVNb density of state histograms were obtained using a system consisting of $6\times 6 \times 6$ bcc cubic unit cells for a total of 432 atoms in an initially random configuration with periodic boundaries. The applied Wang-Landau $\log(f)$ tolerance was $3\cdot10^{-6}$ with a desired flatness of 80\%. The AlTiCrMo simulation had 476 uniform energy bins across total energy range of width 96~meV/atom. The AlTiVNb simulation had 528 uniform energy bins across a total energy range of width 126 meV/atom. We do not consider total energies because the alloy $S^{(2)}$ theory evaluates \textit{derivatives} of the alloy internal energy, and it is therefore not meaningful to consider total energies, only relative differences in energies between structures. The width (or, equivalently, number) of energy bins for the Wang-Landau simulations was chosen such that the algorithm was just over the verge of being able to easily bias the system out of energy bins. The energy range was chosen based on the minimum and maximum energies obtained through equillibrium Metropolis--Hastings Monte Carlo simulations at the highest and lowest temperatures of interest. Fig.~\ref{fig:phase_diagrams} shows plots of energy probability distribution histograms, Warren-Cowley ASRO parameters and specific heat capacity (SHC) curves from our simulations, while Fig.~\ref{fig:configurations} shows visualised configuration for AlTiCrMo and AlTiVNb obtained using Metropolis--Hastings Monte Carlo equilibration at selected temperatures.

In AlTiVNb, an initial peak in the SHC can be seen at approximately 2300~K, indicating a phase transition, with a further transition appearing to occur at 500~K. The initial ordering temperature of 2300~K is reduced compared to the value of 2458~K predicted by the earlier concentration wave analysis, which is consistent with the fact that the concentration wave analysis provides a mean-field estimate of ordering temperatures. For the high temperature phase transition, the strongest ASRO is for Al-Al, Al-Ti, and Ti-Ti pairs. Al-Ti pairs are favoured on the first coordination shell and disfavoured on the second coordination shell, while Al-Al and Ti-Ti pairs are \textit{disfavoured} on the first coordination shell and \textit{favoured} on the second. This is indicative of a B2 ordering driven by Al and Ti, with Al moving to one simple cubic sublattice, Ti the other, and V and Nb expressing weaker site preferences. This analysis of the transition is supported by the visualised equilibrated configurations at 3000~K and 1200~K shown in Fig.~\ref{fig:configurations}. The peak in the heat capacity observed at around 500~K is associated with additional atom-atom correlations, clustering tendencies, and eventual phase decomposition, as can be clearly seen in the configuration equilibrated at low temperature in Fig.~\ref{fig:configurations}.

These findings for AlTiVNb are consistent with Ref.~\cite{kormann_b2_2021}, which reports a B2 chemical ordering occurring at approximately 1700~K, with Al and Ti expressing strong site preferences, and Nb and V expressing weaker preferences, in alignment with our own predictions. In addition, these authors report a secondary peak in their heat capacity data emerging around 600~K, consistent with our own findings. That our predicted B2 ordering temperature of 2300~K is higher than their prediction 1700~K we understand as originating in differences in lattice parameters and choices of exchange-correlation (XC) functional. Ref.~\cite{kormann_b2_2021} accounts for thermal expansion and uses a lattice parameter of $a=3.23$~\AA, in combination with a GGA XC  functional, whereas we use a DFT-optimised lattice parameter of $a=3.113$~\AA and an LDA XC functional. In previous work, we have found that a marginally contracted (expanded) lattice parameter results in an increased (decreased) chemical ordering temperature in our concentration wave analysis~\cite{woodgate_competition_2024, woodgate_modelling_2024}, and it is also known that LDA XC functionals typically over-bind transition metals~\cite{grabowski_ab_2007}. We understand these factors as the origin of the modest discrepancy between the two predicted ordering temperatures.

In AlTiCrMo, a peak in the SHC can be seen at approximately 1700~K, indicating a phase transition, with a further transition occurring at 350~K. The initial transition temperature of approximately 1700~K is reduced compared to the value of 1974~K predicted by the concentration wave analysis, which is again consistent with fact that the concentration wave analysis provides a mean-field estimate of chemical ordering temperatures. For the higher of the two ordering temperatures, the atom-atom correlations indicate that Al-Ti, Al-Cr, and Ti-Mo pairs are favoured on the first coordination shell, while Ti-Ti pairs are very heavily disfavoured. These pair preferences are consistent with the B2 ordering predicted by the earlier concentration wave analysis. Similarly to AlTiVNb, the transition occurring at lower simulation temperatures is associated with the emergence of additional sublattice atom-atom correlations and eventual phase decomposition. This is evidenced by the ASRO parameters on the second coordination shell showing that Al-Mo and Ti-Cr pairs are favoured, which is supported by the equilibrated low-temperature configuration shown in Fig.~\ref{fig:configurations}. That both alloys eventually segregate into multiple competing phases at low temperatures emphasises that entropy plays an important role in stabilising these single-phase systems.

\subsection{Effect of Chemical Ordering on Physical Properties: Residual Resistivity}
\label{sec:resistivity}

\begin{figure}[t]
    \centering
    \includegraphics[width=\linewidth]{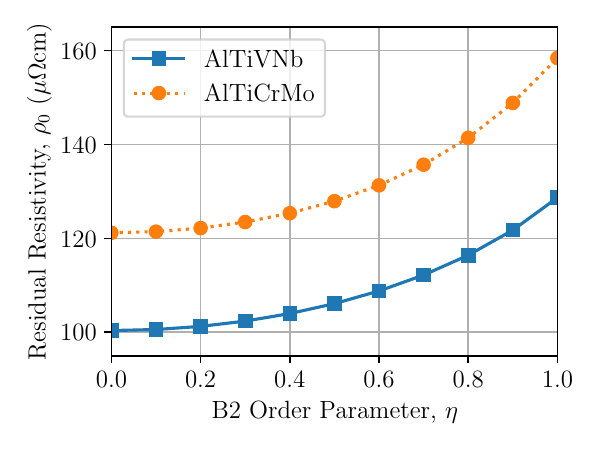}
    \caption{Calculated residual resistivity, $\rho_0$, as a function of atomic long-range order parameter, $\eta$, for the B2 chemical orderings predicted by our concentration wave analysis for AlTiVNb and AlTiCrMo. The case $\eta=0$ corresponds to the A2 (disordered bcc) structure, while $\eta=1$ corresponds to the B2 structure described by the maximal chemical fluctuation consistent with the chemical polarisations of Table~\ref{table:transition_temperatures}. Counter-intuitively, both alloys have an \textit{increased} residual resistivity once a B2 chemical ordering is established.}
    \label{fig:resistivity_comparison}
\end{figure}

As an example of how chemical orderings, such as the B2 orderings predicted in this work for the AlTiVNb and AlTiCrMo RSAs, can affect materials' properties, we now consider calculations of the residual resistivity for both alloys and examine how this intrinsic physical quantity is affected by the ordering. For the B2 orderings predicted for AlTiVNb and AlTiCrMo by the concentration wave analysis in Sec.~\ref{sec:linear_response}, we calculate the residual resistivity, $\rho_0$ as a function of atomic long-range order parameter, $\eta$, where $\eta=0$ corresponds to the disordered solid solution, and $\eta=1$ corresponds to the largest permitted chemical fluctuation consistent with the predicted chemical polarisation, \textit{i.e.} the maximal B2 ordering. (Explicit partial lattice site occupancies as a function of $\eta$ are provided in the Supplementary Material~\cite{supplemental}.) For these calculations, we again use the \textit{SPR-KKR} package~\cite{ebert_munich_nodate,ebert_calculating_2011}, with broadly the same settings as for our self-consistent calculations in Sec.~\ref{sec:electronic_structure}, but now with increased $\mathbf{k}$-point sampling (\texttt{NKTAB=150,000}) and angular momentum cutoff ($l_\textrm{max}=4$), as is necessary to capture the subtle details of the electronic scattering states around the Fermi level, $E_F$. For both alloys, these resistivity data are plotted in Fig.~\ref{fig:resistivity_comparison}.

For the case $\eta=0$, \textit{i.e.} the A2 (disordered bcc) solid solution, AlTiVNb has a calculated residual resistivity of 100.4~$\mu\Omega$cm, smaller than the value of 121.2~$\mu\Omega$cm calculated for AlTiCrMo. (Note that the conductivity tensor is diagonal due to isotropy of the alloys with $\sigma = \sigma_{xx} = \sigma_{yy} = \sigma_{zz}$, resulting in an isotropic resistivity.) We attribute this difference in calculated resistivity to A2 AlTiCrMo having a reduced electronic density of states and increased presence of heavily smeared states around the $E_F$ when compared to A2 AlTiVNb, as evidenced in Fig.~\ref{fig:A2_AlTiVNb_and_AlTiCrMo_DoS_and_BSF}. For the case $\eta=1$, \textit{i.e.} the predicted partially ordered B2 structures, both alloys have an increased residual resistivity compared to the disordered A2 phase. For AlTiVNb the predicted residual resistivity increases to 128.7~$\mu\Omega$cm, while for AlTiCrMo it increases to 158.4~$\mu\Omega$cm. Between $\eta=0$ and $\eta=1$, the resistivity varies smoothly, monotonically increasing with increasing atomic order parameter. Although counter-intuitive---chemical orderings usually represent a restoration of a degree of translational symmetry and a consequent reduction in resistivity---such behaviour is consistent with earlier calculations demonstrating that a degree of atomic short-range order can increase the residual resistivity of some transition metal alloys~\cite{lowitzer_ab_2010}. This is known as the ``komplex'' or $K$-state phenomenon, and was first observed in Ni-Cr alloys~\cite{thomas_uber_1951}. In part, we believe that the increase in residual resistivity for both alloys can be attributed to a reduction in the electronic DoS at $E_F$ for the B2 ordered structures compared to the disordered A2 phase, as seen in Fig.~\ref{fig:a2_b2_dos_comparison}. However, this does not represent the complete picture, and further analysis of the electronic structure of the alloys is required.

Understanding the underlying physical mechanisms governing the resistivity increase with increasing $\eta$ within the Kubo--Greenwood formalism remains challenging, as this calculation yields only the components of the conductivity tensor. To gain a more intuitive understanding, one may employ the semi-classical Boltzmann transport equation within the relaxation-time approximation \cite{mahan1996,scheidemantel2003}. In this framework, discussed in detail in the Supplementary Material~\cite{supplemental}, the zero temperature electrical conductivity is expressed in terms of the density of states (DoS) at the Fermi level, $g(E_F)$, as
\begin{equation} 
\sigma = e^2 \langle v_{\mathbf{k},x} v_{\mathbf{k},x} \tau_{\mathbf{k}} \rangle_{E_F} g(E_F), 
\end{equation}
where $e$ is the charge on the electron, $v_{\mathbf{k},i}$ represents the electronic group velocity component in the $i$-th direction, $\tau_{\mathbf{k}}$ is the $\mathbf{k}$-dependent electronic relaxation time, and $\langle \cdot \rangle_{E_F}$ denotes an average taken over $\mathbf{k}$ lying in the first Brillouin zone with the energy fixed at $E_F$. In addition, conductivity is directly influenced by $\mathbf{k}$-space smearing of the spectral function and the corresponding reduction in the electronic mean free path \cite{robarts_extreme_2020}.

A site- and orbital-resolved analysis of the electronic DoS---presented in the Supplementary Material~\cite{supplemental}---reveals that, with the increase in residual resistivity, there is a concurrent reduction in the number of delocalised $sp$ states at $E_F$. This is in alignment with a simple picture of the residual resistivity being proportional to the DoS at the Fermi level, $\rho_0 \propto g(E_F)^{-1}$~\cite{samolyuk_temperature_2018}. In particular, a decomposition of the $sp$ DoS by atomic sublattices shows that, in AlTiVNb, the relative decrease of the $sp$ states at $E_F$ on the 1b sublattice is in perfect agreement with the increase in relative resistivity. For AlTiCrMo, a similar trend is observed, albeit sublattice-inverted. We understand this decrease in available $sp$ states at the 1b (1a) sites for AlTiVNb (AlTiCrMo) as originating from Al moving to the 1a (1b) lattice site in the predicted B2 ordering. Since, for both alloys, both sublattices contribute to the total residual resistivity, the asymmetric decrease in $sp$ DoS at $E_F$ between sublattices 1a and 1b in AlTiVNb and AlTiCrMo, respectively, suggests associated changes to the electronic mean free path. An analysis of the Bloch spectral function (BSF) along high-symmetry directions and an $\mathbf{k}$-space cut through the Fermi surface resolved into its site-contributions supports this interpretation. As shown in the Supplementary Material~\cite{supplemental}, for AlTiVNb the 1a lattice exhibits a slight $\mathbf{k}$-space broadening (smaller change of intensity in BSF with $\Delta\mathbf{k}$) compared to the 1b lattice. For AlTiCrMo, however, the results are less clear without a direct calculation of the Fermi surface average of the electronic mean free path~\cite{robarts_extreme_2020}. 

Thus, we conclude that the increase in residual resistivity with increasing B2 chemical order parameter, $\eta$, in the alloys studied is primarily governed by two mechanisms: (i) a site-specific reduction in the electronic DoS associated with $sp$ bands at the Fermi level, which is the dominant effect, and (ii) a broadening or change of the spectral function in the $\mathbf{k}$ space, which leads to a shortening of the electronic mean free path, but which makes a less pronounced contribution and is more challenging to quantify.
 
\section{Conclusions}
\label{sec:conclusions}

In summary, we have studied the thermodynamics and phase stability of two refractory high-entropy superalloys (RSAs), AlTiVNb and AlTiCrMo, across a wide temperature range, using a combination of first-principles electronic structure calculations, a concentration wave analysis, and atomistic Monte Carlo simulations. In alignment with the experimental data, we predict a B2 chemical ordering in both systems emerging at high temperatures, with Al and Ti expressing strong site-preferences, and the other constituent elements (V, Nb, Cr, Mo) having weaker site preferences. The physical origins of these B2 chemical orderings have been discussed in terms of the alloys' electronic structure. Our Monte Carlo simulations then predict the emergence of additional atom-atom correlations in both systems with decreasing temperature. Finally, in a demonstration of the impact such chemical orderings can have on alloys' physical properties, we have examined the differences in bulk modulus and residual resistivity for both systems when simulated as chemically disordered, compared to when simulated in the B2 ordered structures as predicted by our modelling. Counter-intuitively, the residual resistivity is found to \textit{increase} for both alloys as a result of the chemical ordering, an outcome which is associated with a reduction in the electronic density of states at the Fermi level, as well as qualitative changes to the nature of the alloys' smeared-out Fermi surfaces.

These results provide detailed insight into the nature of the experimentally observed B2 chemical orderings in these complex materials, in particular providing information about which chemical species preferentially occupy which sublattice, information which can be challenging to determine experimentally. Further, by considering how the alloys' residual resistivities are affected by the chemical orderings, they serve to emphasise the close connections between the atomic-scale structure of high-entropy materials and their resultant physical properties. Finally, given the good agreement between these results and existing the existing computational and experimental data, these results reinforce that methodologies based on concentration waves provide powerful, computational efficient, and physically insightful tools for probing short- and long-range atomic ordering tendencies in multicomponent alloys. Future work could seek to adapt this modelling approach and use it to perform high-throughput screening and search for new alloy compositions.

\begin{acknowledgments}
C.D.W. thanks Dr Aki Pulkkinen (University of West Bohemia) for provision of Python scripts suitable for visualisation of \textit{SPR-KKR} outputs, and Dr Francesco Turci (University of Bristol) for helpful discussions. C.D.W. acknowledges support from a UK Engineering and Physical Sciences Research Council (EPSRC) Doctoral Prize Fellowship at the University of Bristol, Grant EP/W524414/1. H.J.N. is supported by a studentship within the EPSRC Centre for Doctoral Training in Modelling of Heterogeneous Systems, Grant EP/S022848/1. J.M. thanks the project MEBIOSYS, funded as project No. CZ.02.01.01/00/22$\_$008/0004634 by Programme Johannes Amos Commenius, call Excellent Research. J.B.S. acknowledges support from EPSRC Grant EP/W021331/1. We acknowledge use of the Sulis Tier 2 HPC platform hosted by the Scientific Computing Research Technology Platform (SCRTP) at the University of Warwick. Sulis is funded by EPSRC Grant EP/T022108/1 and the HPC Midlands+ consortium. Additional computing facilities were provided by the Advanced Computing Research Centre (ACRC) of the University of Bristol.

C.D.W. and J.B.S. conceived of the approach, with input from all authors. C.D.W. performed self-consistent \textit{SPR-KKR} calculations, the concentration wave analysis, and recovered the real-space effective pair interactions for the atomistic modelling. H.J.N. wrote the code implementing Wang-Landau sampling and ran the Monte Carlo simulations, with support from D.Q. and C.D.W. D.R. ran and analysed the residual resistivity calculations, with support from and J.M. and C.D.W. The first draft of the manuscript was written by C.D.W. and subsequently received input from all authors.
\end{acknowledgments}


\begin{thebibliography}{108}%
\makeatletter
\providecommand \@ifxundefined [1]{%
 \@ifx{#1\undefined}
}%
\providecommand \@ifnum [1]{%
 \ifnum #1\expandafter \@firstoftwo
 \else \expandafter \@secondoftwo
 \fi
}%
\providecommand \@ifx [1]{%
 \ifx #1\expandafter \@firstoftwo
 \else \expandafter \@secondoftwo
 \fi
}%
\providecommand \natexlab [1]{#1}%
\providecommand \enquote  [1]{``#1''}%
\providecommand \bibnamefont  [1]{#1}%
\providecommand \bibfnamefont [1]{#1}%
\providecommand \citenamefont [1]{#1}%
\providecommand \href@noop [0]{\@secondoftwo}%
\providecommand \href [0]{\begingroup \@sanitize@url \@href}%
\providecommand \@href[1]{\@@startlink{#1}\@@href}%
\providecommand \@@href[1]{\endgroup#1\@@endlink}%
\providecommand \@sanitize@url [0]{\catcode `\\12\catcode `\$12\catcode `\&12\catcode `\#12\catcode `\^12\catcode `\_12\catcode `\%12\relax}%
\providecommand \@@startlink[1]{}%
\providecommand \@@endlink[0]{}%
\providecommand \url  [0]{\begingroup\@sanitize@url \@url }%
\providecommand \@url [1]{\endgroup\@href {#1}{\urlprefix }}%
\providecommand \urlprefix  [0]{URL }%
\providecommand \Eprint [0]{\href }%
\providecommand \doibase [0]{https://doi.org/}%
\providecommand \selectlanguage [0]{\@gobble}%
\providecommand \bibinfo  [0]{\@secondoftwo}%
\providecommand \bibfield  [0]{\@secondoftwo}%
\providecommand \translation [1]{[#1]}%
\providecommand \BibitemOpen [0]{}%
\providecommand \bibitemStop [0]{}%
\providecommand \bibitemNoStop [0]{.\EOS\space}%
\providecommand \EOS [0]{\spacefactor3000\relax}%
\providecommand \BibitemShut  [1]{\csname bibitem#1\endcsname}%
\let\auto@bib@innerbib\@empty
\bibitem [{\citenamefont {Yeh}\ \emph {et~al.}(2004)\citenamefont {Yeh}, \citenamefont {Chen}, \citenamefont {Lin}, \citenamefont {Gan}, \citenamefont {Chin}, \citenamefont {Shun}, \citenamefont {Tsau},\ and\ \citenamefont {Chang}}]{yeh_nanostructured_2004}%
  \BibitemOpen
  \bibfield  {author} {\bibinfo {author} {\bibfnamefont {J.-W.}\ \bibnamefont {Yeh}}, \bibinfo {author} {\bibfnamefont {S.-K.}\ \bibnamefont {Chen}}, \bibinfo {author} {\bibfnamefont {S.-J.}\ \bibnamefont {Lin}}, \bibinfo {author} {\bibfnamefont {J.-Y.}\ \bibnamefont {Gan}}, \bibinfo {author} {\bibfnamefont {T.-S.}\ \bibnamefont {Chin}}, \bibinfo {author} {\bibfnamefont {T.-T.}\ \bibnamefont {Shun}}, \bibinfo {author} {\bibfnamefont {C.-H.}\ \bibnamefont {Tsau}},\ and\ \bibinfo {author} {\bibfnamefont {S.-Y.}\ \bibnamefont {Chang}},\ }\href {https://doi.org/10.1002/adem.200300567} {\bibfield  {journal} {\bibinfo  {journal} {Advanced Engineering Materials}\ }\textbf {\bibinfo {volume} {6}},\ \bibinfo {pages} {299} (\bibinfo {year} {2004})}\BibitemShut {NoStop}%
\bibitem [{\citenamefont {Cantor}\ \emph {et~al.}(2004)\citenamefont {Cantor}, \citenamefont {Chang}, \citenamefont {Knight},\ and\ \citenamefont {Vincent}}]{cantor_microstructural_2004}%
  \BibitemOpen
  \bibfield  {author} {\bibinfo {author} {\bibfnamefont {B.}~\bibnamefont {Cantor}}, \bibinfo {author} {\bibfnamefont {I.~T.~H.}\ \bibnamefont {Chang}}, \bibinfo {author} {\bibfnamefont {P.}~\bibnamefont {Knight}},\ and\ \bibinfo {author} {\bibfnamefont {A.~J.~B.}\ \bibnamefont {Vincent}},\ }\href {https://doi.org/10.1016/j.msea.2003.10.257} {\bibfield  {journal} {\bibinfo  {journal} {Materials Science and Engineering: A}\ }\textbf {\bibinfo {volume} {375-377}},\ \bibinfo {pages} {213} (\bibinfo {year} {2004})}\BibitemShut {NoStop}%
\bibitem [{\citenamefont {Miracle}\ and\ \citenamefont {Senkov}(2017)}]{miracle_critical_2017}%
  \BibitemOpen
  \bibfield  {author} {\bibinfo {author} {\bibfnamefont {D.~B.}\ \bibnamefont {Miracle}}\ and\ \bibinfo {author} {\bibfnamefont {O.~N.}\ \bibnamefont {Senkov}},\ }\href {https://doi.org/10.1016/j.actamat.2016.08.081} {\bibfield  {journal} {\bibinfo  {journal} {Acta Materialia}\ }\textbf {\bibinfo {volume} {122}},\ \bibinfo {pages} {448} (\bibinfo {year} {2017})}\BibitemShut {NoStop}%
\bibitem [{\citenamefont {George}\ \emph {et~al.}(2019)\citenamefont {George}, \citenamefont {Raabe},\ and\ \citenamefont {Ritchie}}]{george_high-entropy_2019}%
  \BibitemOpen
  \bibfield  {author} {\bibinfo {author} {\bibfnamefont {E.~P.}\ \bibnamefont {George}}, \bibinfo {author} {\bibfnamefont {D.}~\bibnamefont {Raabe}},\ and\ \bibinfo {author} {\bibfnamefont {R.~O.}\ \bibnamefont {Ritchie}},\ }\href {https://doi.org/10.1038/s41578-019-0121-4} {\bibfield  {journal} {\bibinfo  {journal} {Nature Reviews Materials}\ }\textbf {\bibinfo {volume} {4}},\ \bibinfo {pages} {515} (\bibinfo {year} {2019})}\BibitemShut {NoStop}%
\bibitem [{\citenamefont {Ayyagari}\ \emph {et~al.}(2018)\citenamefont {Ayyagari}, \citenamefont {Salloom}, \citenamefont {Muskeri},\ and\ \citenamefont {Mukherjee}}]{ayyagari_low_2018}%
  \BibitemOpen
  \bibfield  {author} {\bibinfo {author} {\bibfnamefont {A.}~\bibnamefont {Ayyagari}}, \bibinfo {author} {\bibfnamefont {R.}~\bibnamefont {Salloom}}, \bibinfo {author} {\bibfnamefont {S.}~\bibnamefont {Muskeri}},\ and\ \bibinfo {author} {\bibfnamefont {S.}~\bibnamefont {Mukherjee}},\ }\href {https://doi.org/10.1016/j.mtla.2018.09.014} {\bibfield  {journal} {\bibinfo  {journal} {Materialia}\ }\textbf {\bibinfo {volume} {4}},\ \bibinfo {pages} {99} (\bibinfo {year} {2018})}\BibitemShut {NoStop}%
\bibitem [{\citenamefont {El-Atwani}\ \emph {et~al.}(2019)\citenamefont {El-Atwani}, \citenamefont {Li}, \citenamefont {Li}, \citenamefont {Devaraj}, \citenamefont {Baldwin}, \citenamefont {Schneider}, \citenamefont {Sobieraj}, \citenamefont {Wróbel}, \citenamefont {Nguyen-Manh}, \citenamefont {Maloy},\ and\ \citenamefont {Martinez}}]{el-atwani_outstanding_2019}%
  \BibitemOpen
  \bibfield  {author} {\bibinfo {author} {\bibfnamefont {O.}~\bibnamefont {El-Atwani}}, \bibinfo {author} {\bibfnamefont {N.}~\bibnamefont {Li}}, \bibinfo {author} {\bibfnamefont {M.}~\bibnamefont {Li}}, \bibinfo {author} {\bibfnamefont {A.}~\bibnamefont {Devaraj}}, \bibinfo {author} {\bibfnamefont {J.~K.~S.}\ \bibnamefont {Baldwin}}, \bibinfo {author} {\bibfnamefont {M.~M.}\ \bibnamefont {Schneider}}, \bibinfo {author} {\bibfnamefont {D.}~\bibnamefont {Sobieraj}}, \bibinfo {author} {\bibfnamefont {J.~S.}\ \bibnamefont {Wróbel}}, \bibinfo {author} {\bibfnamefont {D.}~\bibnamefont {Nguyen-Manh}}, \bibinfo {author} {\bibfnamefont {S.~A.}\ \bibnamefont {Maloy}},\ and\ \bibinfo {author} {\bibfnamefont {E.}~\bibnamefont {Martinez}},\ }\href {https://doi.org/10.1126/sciadv.aav2002} {\bibfield  {journal} {\bibinfo  {journal} {Science Advances}\ }\textbf {\bibinfo {volume} {5}},\ \bibinfo {pages} {eaav2002} (\bibinfo {year} {2019})}\BibitemShut {NoStop}%
\bibitem [{\citenamefont {El~Atwani}\ \emph {et~al.}(2023)\citenamefont {El~Atwani}, \citenamefont {Vo}, \citenamefont {Tunes}, \citenamefont {Lee}, \citenamefont {Alvarado}, \citenamefont {Krienke}, \citenamefont {Poplawsky}, \citenamefont {Kohnert}, \citenamefont {Gigax}, \citenamefont {Chen}, \citenamefont {Li}, \citenamefont {Wang}, \citenamefont {Wróbel}, \citenamefont {Nguyen-Manh}, \citenamefont {Baldwin}, \citenamefont {Tukac}, \citenamefont {Aydogan}, \citenamefont {Fensin},\ and\ \citenamefont {Martinez}}]{el_atwani_quinary_2023}%
  \BibitemOpen
  \bibfield  {author} {\bibinfo {author} {\bibfnamefont {O.}~\bibnamefont {El~Atwani}}, \bibinfo {author} {\bibfnamefont {H.~T.}\ \bibnamefont {Vo}}, \bibinfo {author} {\bibfnamefont {M.~A.}\ \bibnamefont {Tunes}}, \bibinfo {author} {\bibfnamefont {C.}~\bibnamefont {Lee}}, \bibinfo {author} {\bibfnamefont {A.}~\bibnamefont {Alvarado}}, \bibinfo {author} {\bibfnamefont {N.}~\bibnamefont {Krienke}}, \bibinfo {author} {\bibfnamefont {J.~D.}\ \bibnamefont {Poplawsky}}, \bibinfo {author} {\bibfnamefont {A.~A.}\ \bibnamefont {Kohnert}}, \bibinfo {author} {\bibfnamefont {J.}~\bibnamefont {Gigax}}, \bibinfo {author} {\bibfnamefont {W.-Y.}\ \bibnamefont {Chen}}, \bibinfo {author} {\bibfnamefont {M.}~\bibnamefont {Li}}, \bibinfo {author} {\bibfnamefont {Y.~Q.}\ \bibnamefont {Wang}}, \bibinfo {author} {\bibfnamefont {J.~S.}\ \bibnamefont {Wróbel}}, \bibinfo {author} {\bibfnamefont {D.}~\bibnamefont {Nguyen-Manh}}, \bibinfo {author} {\bibfnamefont {J.~K.~S.}\ \bibnamefont {Baldwin}}, \bibinfo {author} {\bibfnamefont
  {O.~U.}\ \bibnamefont {Tukac}}, \bibinfo {author} {\bibfnamefont {E.}~\bibnamefont {Aydogan}}, \bibinfo {author} {\bibfnamefont {S.}~\bibnamefont {Fensin}},\ and\ \bibinfo {author} {\bibfnamefont {E.}~\bibnamefont {Martinez}},\ }\href {https://doi.org/10.1038/s41467-023-38000-y} {\bibfield  {journal} {\bibinfo  {journal} {Nature Communications}\ }\textbf {\bibinfo {volume} {14}},\ \bibinfo {pages} {2516} (\bibinfo {year} {2023})}\BibitemShut {NoStop}%
\bibitem [{\citenamefont {Gludovatz}\ \emph {et~al.}(2014)\citenamefont {Gludovatz}, \citenamefont {Hohenwarter}, \citenamefont {Catoor}, \citenamefont {Chang}, \citenamefont {George},\ and\ \citenamefont {Ritchie}}]{gludovatz_fracture-resistant_2014}%
  \BibitemOpen
  \bibfield  {author} {\bibinfo {author} {\bibfnamefont {B.}~\bibnamefont {Gludovatz}}, \bibinfo {author} {\bibfnamefont {A.}~\bibnamefont {Hohenwarter}}, \bibinfo {author} {\bibfnamefont {D.}~\bibnamefont {Catoor}}, \bibinfo {author} {\bibfnamefont {E.~H.}\ \bibnamefont {Chang}}, \bibinfo {author} {\bibfnamefont {E.~P.}\ \bibnamefont {George}},\ and\ \bibinfo {author} {\bibfnamefont {R.~O.}\ \bibnamefont {Ritchie}},\ }\href {https://doi.org/10.1126/science.1254581} {\bibfield  {journal} {\bibinfo  {journal} {Science}\ }\textbf {\bibinfo {volume} {345}},\ \bibinfo {pages} {1153} (\bibinfo {year} {2014})}\BibitemShut {NoStop}%
\bibitem [{\citenamefont {Gludovatz}\ \emph {et~al.}(2016)\citenamefont {Gludovatz}, \citenamefont {Hohenwarter}, \citenamefont {Thurston}, \citenamefont {Bei}, \citenamefont {Wu}, \citenamefont {George},\ and\ \citenamefont {Ritchie}}]{gludovatz_exceptional_2016}%
  \BibitemOpen
  \bibfield  {author} {\bibinfo {author} {\bibfnamefont {B.}~\bibnamefont {Gludovatz}}, \bibinfo {author} {\bibfnamefont {A.}~\bibnamefont {Hohenwarter}}, \bibinfo {author} {\bibfnamefont {K.~V.~S.}\ \bibnamefont {Thurston}}, \bibinfo {author} {\bibfnamefont {H.}~\bibnamefont {Bei}}, \bibinfo {author} {\bibfnamefont {Z.}~\bibnamefont {Wu}}, \bibinfo {author} {\bibfnamefont {E.~P.}\ \bibnamefont {George}},\ and\ \bibinfo {author} {\bibfnamefont {R.~O.}\ \bibnamefont {Ritchie}},\ }\href {https://doi.org/10.1038/ncomms10602} {\bibfield  {journal} {\bibinfo  {journal} {Nature Communications}\ }\textbf {\bibinfo {volume} {7}},\ \bibinfo {pages} {10602} (\bibinfo {year} {2016})}\BibitemShut {NoStop}%
\bibitem [{\citenamefont {Liu}\ \emph {et~al.}(2022)\citenamefont {Liu}, \citenamefont {Yu}, \citenamefont {Kabra}, \citenamefont {Jiang}, \citenamefont {Forna-Kreutzer}, \citenamefont {Zhang}, \citenamefont {Payne}, \citenamefont {Walsh}, \citenamefont {Gludovatz}, \citenamefont {Asta}, \citenamefont {Minor}, \citenamefont {George},\ and\ \citenamefont {Ritchie}}]{liu_exceptional_2022}%
  \BibitemOpen
  \bibfield  {author} {\bibinfo {author} {\bibfnamefont {D.}~\bibnamefont {Liu}}, \bibinfo {author} {\bibfnamefont {Q.}~\bibnamefont {Yu}}, \bibinfo {author} {\bibfnamefont {S.}~\bibnamefont {Kabra}}, \bibinfo {author} {\bibfnamefont {M.}~\bibnamefont {Jiang}}, \bibinfo {author} {\bibfnamefont {P.}~\bibnamefont {Forna-Kreutzer}}, \bibinfo {author} {\bibfnamefont {R.}~\bibnamefont {Zhang}}, \bibinfo {author} {\bibfnamefont {M.}~\bibnamefont {Payne}}, \bibinfo {author} {\bibfnamefont {F.}~\bibnamefont {Walsh}}, \bibinfo {author} {\bibfnamefont {B.}~\bibnamefont {Gludovatz}}, \bibinfo {author} {\bibfnamefont {M.}~\bibnamefont {Asta}}, \bibinfo {author} {\bibfnamefont {A.~M.}\ \bibnamefont {Minor}}, \bibinfo {author} {\bibfnamefont {E.~P.}\ \bibnamefont {George}},\ and\ \bibinfo {author} {\bibfnamefont {R.~O.}\ \bibnamefont {Ritchie}},\ }\href {https://doi.org/10.1126/science.abp8070} {\bibfield  {journal} {\bibinfo  {journal} {Science}\ }\textbf {\bibinfo {volume} {378}},\ \bibinfo {pages} {978} (\bibinfo {year}
  {2022})}\BibitemShut {NoStop}%
\bibitem [{\citenamefont {Praveen}\ and\ \citenamefont {Kim}(2018)}]{praveen_highentropy_2018}%
  \BibitemOpen
  \bibfield  {author} {\bibinfo {author} {\bibfnamefont {S.}~\bibnamefont {Praveen}}\ and\ \bibinfo {author} {\bibfnamefont {H.~S.}\ \bibnamefont {Kim}},\ }\href {https://doi.org/10.1002/adem.201700645} {\bibfield  {journal} {\bibinfo  {journal} {Advanced Engineering Materials}\ }\textbf {\bibinfo {volume} {20}},\ \bibinfo {pages} {1700645} (\bibinfo {year} {2018})}\BibitemShut {NoStop}%
\bibitem [{\citenamefont {Koželj}\ \emph {et~al.}(2014)\citenamefont {Koželj}, \citenamefont {Vrtnik}, \citenamefont {Jelen}, \citenamefont {Jazbec}, \citenamefont {Jagličić}, \citenamefont {Maiti}, \citenamefont {Feuerbacher}, \citenamefont {Steurer},\ and\ \citenamefont {Dolinšek}}]{kozelj_discovery_2014}%
  \BibitemOpen
  \bibfield  {author} {\bibinfo {author} {\bibfnamefont {P.}~\bibnamefont {Koželj}}, \bibinfo {author} {\bibfnamefont {S.}~\bibnamefont {Vrtnik}}, \bibinfo {author} {\bibfnamefont {A.}~\bibnamefont {Jelen}}, \bibinfo {author} {\bibfnamefont {S.}~\bibnamefont {Jazbec}}, \bibinfo {author} {\bibfnamefont {Z.}~\bibnamefont {Jagličić}}, \bibinfo {author} {\bibfnamefont {S.}~\bibnamefont {Maiti}}, \bibinfo {author} {\bibfnamefont {M.}~\bibnamefont {Feuerbacher}}, \bibinfo {author} {\bibfnamefont {W.}~\bibnamefont {Steurer}},\ and\ \bibinfo {author} {\bibfnamefont {J.}~\bibnamefont {Dolinšek}},\ }\href {https://doi.org/10.1103/PhysRevLett.113.107001} {\bibfield  {journal} {\bibinfo  {journal} {Physical Review Letters}\ }\textbf {\bibinfo {volume} {113}},\ \bibinfo {pages} {107001} (\bibinfo {year} {2014})}\BibitemShut {NoStop}%
\bibitem [{\citenamefont {Sales}\ \emph {et~al.}(2016)\citenamefont {Sales}, \citenamefont {Jin}, \citenamefont {Bei}, \citenamefont {Stocks}, \citenamefont {Samolyuk}, \citenamefont {May},\ and\ \citenamefont {McGuire}}]{sales_quantum_2016}%
  \BibitemOpen
  \bibfield  {author} {\bibinfo {author} {\bibfnamefont {B.~C.}\ \bibnamefont {Sales}}, \bibinfo {author} {\bibfnamefont {K.}~\bibnamefont {Jin}}, \bibinfo {author} {\bibfnamefont {H.}~\bibnamefont {Bei}}, \bibinfo {author} {\bibfnamefont {G.~M.}\ \bibnamefont {Stocks}}, \bibinfo {author} {\bibfnamefont {G.~D.}\ \bibnamefont {Samolyuk}}, \bibinfo {author} {\bibfnamefont {A.~F.}\ \bibnamefont {May}},\ and\ \bibinfo {author} {\bibfnamefont {M.~A.}\ \bibnamefont {McGuire}},\ }\href {https://doi.org/10.1038/srep26179} {\bibfield  {journal} {\bibinfo  {journal} {Scientific Reports}\ }\textbf {\bibinfo {volume} {6}},\ \bibinfo {pages} {26179} (\bibinfo {year} {2016})}\BibitemShut {NoStop}%
\bibitem [{\citenamefont {Robarts}\ \emph {et~al.}(2020)\citenamefont {Robarts}, \citenamefont {Millichamp}, \citenamefont {Lagos}, \citenamefont {Laverock}, \citenamefont {Billington}, \citenamefont {Duffy}, \citenamefont {O’Neill}, \citenamefont {Giblin}, \citenamefont {Taylor}, \citenamefont {Kontrym-Sznajd}, \citenamefont {Samsel-Czekała}, \citenamefont {Bei}, \citenamefont {Mu}, \citenamefont {Samolyuk}, \citenamefont {Stocks},\ and\ \citenamefont {Dugdale}}]{robarts_extreme_2020}%
  \BibitemOpen
  \bibfield  {author} {\bibinfo {author} {\bibfnamefont {H.~C.}\ \bibnamefont {Robarts}}, \bibinfo {author} {\bibfnamefont {T.~E.}\ \bibnamefont {Millichamp}}, \bibinfo {author} {\bibfnamefont {D.~A.}\ \bibnamefont {Lagos}}, \bibinfo {author} {\bibfnamefont {J.}~\bibnamefont {Laverock}}, \bibinfo {author} {\bibfnamefont {D.}~\bibnamefont {Billington}}, \bibinfo {author} {\bibfnamefont {J.~A.}\ \bibnamefont {Duffy}}, \bibinfo {author} {\bibfnamefont {D.}~\bibnamefont {O’Neill}}, \bibinfo {author} {\bibfnamefont {S.~R.}\ \bibnamefont {Giblin}}, \bibinfo {author} {\bibfnamefont {J.~W.}\ \bibnamefont {Taylor}}, \bibinfo {author} {\bibfnamefont {G.}~\bibnamefont {Kontrym-Sznajd}}, \bibinfo {author} {\bibfnamefont {M.}~\bibnamefont {Samsel-Czekała}}, \bibinfo {author} {\bibfnamefont {H.}~\bibnamefont {Bei}}, \bibinfo {author} {\bibfnamefont {S.}~\bibnamefont {Mu}}, \bibinfo {author} {\bibfnamefont {G.~D.}\ \bibnamefont {Samolyuk}}, \bibinfo {author} {\bibfnamefont {G.~M.}\ \bibnamefont {Stocks}},\ and\ \bibinfo
  {author} {\bibfnamefont {S.~B.}\ \bibnamefont {Dugdale}},\ }\href {https://doi.org/10.1103/PhysRevLett.124.046402} {\bibfield  {journal} {\bibinfo  {journal} {Physical Review Letters}\ }\textbf {\bibinfo {volume} {124}},\ \bibinfo {pages} {046402} (\bibinfo {year} {2020})}\BibitemShut {NoStop}%
\bibitem [{\citenamefont {Senkov}\ \emph {et~al.}(2018)\citenamefont {Senkov}, \citenamefont {Miracle}, \citenamefont {Chaput},\ and\ \citenamefont {Couzinie}}]{senkov_development_2018}%
  \BibitemOpen
  \bibfield  {author} {\bibinfo {author} {\bibfnamefont {O.~N.}\ \bibnamefont {Senkov}}, \bibinfo {author} {\bibfnamefont {D.~B.}\ \bibnamefont {Miracle}}, \bibinfo {author} {\bibfnamefont {K.~J.}\ \bibnamefont {Chaput}},\ and\ \bibinfo {author} {\bibfnamefont {J.-P.}\ \bibnamefont {Couzinie}},\ }\href {https://doi.org/10.1557/jmr.2018.153} {\bibfield  {journal} {\bibinfo  {journal} {Journal of Materials Research}\ }\textbf {\bibinfo {volume} {33}},\ \bibinfo {pages} {3092} (\bibinfo {year} {2018})}\BibitemShut {NoStop}%
\bibitem [{\citenamefont {Senkov}\ \emph {et~al.}(2010)\citenamefont {Senkov}, \citenamefont {Wilks}, \citenamefont {Miracle}, \citenamefont {Chuang},\ and\ \citenamefont {Liaw}}]{senkov_refractory_2010}%
  \BibitemOpen
  \bibfield  {author} {\bibinfo {author} {\bibfnamefont {O.}~\bibnamefont {Senkov}}, \bibinfo {author} {\bibfnamefont {G.}~\bibnamefont {Wilks}}, \bibinfo {author} {\bibfnamefont {D.}~\bibnamefont {Miracle}}, \bibinfo {author} {\bibfnamefont {C.}~\bibnamefont {Chuang}},\ and\ \bibinfo {author} {\bibfnamefont {P.}~\bibnamefont {Liaw}},\ }\href {https://doi.org/10.1016/j.intermet.2010.05.014} {\bibfield  {journal} {\bibinfo  {journal} {Intermetallics}\ }\textbf {\bibinfo {volume} {18}},\ \bibinfo {pages} {1758} (\bibinfo {year} {2010})}\BibitemShut {NoStop}%
\bibitem [{\citenamefont {Senkov}\ \emph {et~al.}(2011)\citenamefont {Senkov}, \citenamefont {Wilks}, \citenamefont {Scott},\ and\ \citenamefont {Miracle}}]{senkov_mechanical_2011}%
  \BibitemOpen
  \bibfield  {author} {\bibinfo {author} {\bibfnamefont {O.}~\bibnamefont {Senkov}}, \bibinfo {author} {\bibfnamefont {G.}~\bibnamefont {Wilks}}, \bibinfo {author} {\bibfnamefont {J.}~\bibnamefont {Scott}},\ and\ \bibinfo {author} {\bibfnamefont {D.}~\bibnamefont {Miracle}},\ }\href {https://doi.org/10.1016/j.intermet.2011.01.004} {\bibfield  {journal} {\bibinfo  {journal} {Intermetallics}\ }\textbf {\bibinfo {volume} {19}},\ \bibinfo {pages} {698} (\bibinfo {year} {2011})}\BibitemShut {NoStop}%
\bibitem [{\citenamefont {Lee}\ \emph {et~al.}(2018)\citenamefont {Lee}, \citenamefont {Song}, \citenamefont {Gao}, \citenamefont {Feng}, \citenamefont {Chen}, \citenamefont {Brechtl}, \citenamefont {Chen}, \citenamefont {An}, \citenamefont {Guo}, \citenamefont {Poplawsky}, \citenamefont {Li}, \citenamefont {Samaei}, \citenamefont {Chen}, \citenamefont {Hu}, \citenamefont {Choo},\ and\ \citenamefont {Liaw}}]{lee_lattice_2018}%
  \BibitemOpen
  \bibfield  {author} {\bibinfo {author} {\bibfnamefont {C.}~\bibnamefont {Lee}}, \bibinfo {author} {\bibfnamefont {G.}~\bibnamefont {Song}}, \bibinfo {author} {\bibfnamefont {M.~C.}\ \bibnamefont {Gao}}, \bibinfo {author} {\bibfnamefont {R.}~\bibnamefont {Feng}}, \bibinfo {author} {\bibfnamefont {P.}~\bibnamefont {Chen}}, \bibinfo {author} {\bibfnamefont {J.}~\bibnamefont {Brechtl}}, \bibinfo {author} {\bibfnamefont {Y.}~\bibnamefont {Chen}}, \bibinfo {author} {\bibfnamefont {K.}~\bibnamefont {An}}, \bibinfo {author} {\bibfnamefont {W.}~\bibnamefont {Guo}}, \bibinfo {author} {\bibfnamefont {J.~D.}\ \bibnamefont {Poplawsky}}, \bibinfo {author} {\bibfnamefont {S.}~\bibnamefont {Li}}, \bibinfo {author} {\bibfnamefont {A.}~\bibnamefont {Samaei}}, \bibinfo {author} {\bibfnamefont {W.}~\bibnamefont {Chen}}, \bibinfo {author} {\bibfnamefont {A.}~\bibnamefont {Hu}}, \bibinfo {author} {\bibfnamefont {H.}~\bibnamefont {Choo}},\ and\ \bibinfo {author} {\bibfnamefont {P.~K.}\ \bibnamefont {Liaw}},\ }\href
  {https://doi.org/10.1016/j.actamat.2018.08.053} {\bibfield  {journal} {\bibinfo  {journal} {Acta Materialia}\ }\textbf {\bibinfo {volume} {160}},\ \bibinfo {pages} {158} (\bibinfo {year} {2018})}\BibitemShut {NoStop}%
\bibitem [{\citenamefont {Maresca}\ and\ \citenamefont {Curtin}(2020)}]{maresca_mechanistic_2020}%
  \BibitemOpen
  \bibfield  {author} {\bibinfo {author} {\bibfnamefont {F.}~\bibnamefont {Maresca}}\ and\ \bibinfo {author} {\bibfnamefont {W.~A.}\ \bibnamefont {Curtin}},\ }\href {https://doi.org/10.1016/j.actamat.2019.10.015} {\bibfield  {journal} {\bibinfo  {journal} {Acta Materialia}\ }\textbf {\bibinfo {volume} {182}},\ \bibinfo {pages} {235} (\bibinfo {year} {2020})}\BibitemShut {NoStop}%
\bibitem [{\citenamefont {Senkov}\ \emph {et~al.}(2014{\natexlab{a}})\citenamefont {Senkov}, \citenamefont {Senkova},\ and\ \citenamefont {Woodward}}]{senkov_effect_2014}%
  \BibitemOpen
  \bibfield  {author} {\bibinfo {author} {\bibfnamefont {O.}~\bibnamefont {Senkov}}, \bibinfo {author} {\bibfnamefont {S.}~\bibnamefont {Senkova}},\ and\ \bibinfo {author} {\bibfnamefont {C.}~\bibnamefont {Woodward}},\ }\href {https://doi.org/10.1016/j.actamat.2014.01.029} {\bibfield  {journal} {\bibinfo  {journal} {Acta Materialia}\ }\textbf {\bibinfo {volume} {68}},\ \bibinfo {pages} {214} (\bibinfo {year} {2014}{\natexlab{a}})}\BibitemShut {NoStop}%
\bibitem [{\citenamefont {Senkov}\ \emph {et~al.}(2014{\natexlab{b}})\citenamefont {Senkov}, \citenamefont {Woodward},\ and\ \citenamefont {Miracle}}]{senkov_microstructure_2014}%
  \BibitemOpen
  \bibfield  {author} {\bibinfo {author} {\bibfnamefont {O.~N.}\ \bibnamefont {Senkov}}, \bibinfo {author} {\bibfnamefont {C.}~\bibnamefont {Woodward}},\ and\ \bibinfo {author} {\bibfnamefont {D.~B.}\ \bibnamefont {Miracle}},\ }\href {https://doi.org/10.1007/s11837-014-1066-0} {\bibfield  {journal} {\bibinfo  {journal} {JOM}\ }\textbf {\bibinfo {volume} {66}},\ \bibinfo {pages} {2030} (\bibinfo {year} {2014}{\natexlab{b}})}\BibitemShut {NoStop}%
\bibitem [{\citenamefont {Jensen}\ \emph {et~al.}(2016)\citenamefont {Jensen}, \citenamefont {Welk}, \citenamefont {Williams}, \citenamefont {Sosa}, \citenamefont {Huber}, \citenamefont {Senkov}, \citenamefont {Viswanathan},\ and\ \citenamefont {Fraser}}]{jensen_characterization_2016}%
  \BibitemOpen
  \bibfield  {author} {\bibinfo {author} {\bibfnamefont {J.}~\bibnamefont {Jensen}}, \bibinfo {author} {\bibfnamefont {B.}~\bibnamefont {Welk}}, \bibinfo {author} {\bibfnamefont {R.}~\bibnamefont {Williams}}, \bibinfo {author} {\bibfnamefont {J.}~\bibnamefont {Sosa}}, \bibinfo {author} {\bibfnamefont {D.}~\bibnamefont {Huber}}, \bibinfo {author} {\bibfnamefont {O.}~\bibnamefont {Senkov}}, \bibinfo {author} {\bibfnamefont {G.}~\bibnamefont {Viswanathan}},\ and\ \bibinfo {author} {\bibfnamefont {H.}~\bibnamefont {Fraser}},\ }\href {https://doi.org/10.1016/j.scriptamat.2016.04.017} {\bibfield  {journal} {\bibinfo  {journal} {Scripta Materialia}\ }\textbf {\bibinfo {volume} {121}},\ \bibinfo {pages} {1} (\bibinfo {year} {2016})}\BibitemShut {NoStop}%
\bibitem [{\citenamefont {Miracle}\ \emph {et~al.}(2020)\citenamefont {Miracle}, \citenamefont {Tsai}, \citenamefont {Senkov}, \citenamefont {Soni},\ and\ \citenamefont {Banerjee}}]{miracle_refractory_2020}%
  \BibitemOpen
  \bibfield  {author} {\bibinfo {author} {\bibfnamefont {D.~B.}\ \bibnamefont {Miracle}}, \bibinfo {author} {\bibfnamefont {M.-H.}\ \bibnamefont {Tsai}}, \bibinfo {author} {\bibfnamefont {O.~N.}\ \bibnamefont {Senkov}}, \bibinfo {author} {\bibfnamefont {V.}~\bibnamefont {Soni}},\ and\ \bibinfo {author} {\bibfnamefont {R.}~\bibnamefont {Banerjee}},\ }\href {https://doi.org/10.1016/j.scriptamat.2020.06.048} {\bibfield  {journal} {\bibinfo  {journal} {Scripta Materialia}\ }\textbf {\bibinfo {volume} {187}},\ \bibinfo {pages} {445} (\bibinfo {year} {2020})}\BibitemShut {NoStop}%
\bibitem [{\citenamefont {Stepanov}\ \emph {et~al.}(2015)\citenamefont {Stepanov}, \citenamefont {Yurchenko}, \citenamefont {Skibin}, \citenamefont {Tikhonovsky},\ and\ \citenamefont {Salishchev}}]{stepanov_structure_2015}%
  \BibitemOpen
  \bibfield  {author} {\bibinfo {author} {\bibfnamefont {N.}~\bibnamefont {Stepanov}}, \bibinfo {author} {\bibfnamefont {N.~Y.}\ \bibnamefont {Yurchenko}}, \bibinfo {author} {\bibfnamefont {D.}~\bibnamefont {Skibin}}, \bibinfo {author} {\bibfnamefont {M.}~\bibnamefont {Tikhonovsky}},\ and\ \bibinfo {author} {\bibfnamefont {G.}~\bibnamefont {Salishchev}},\ }\href {https://doi.org/10.1016/j.jallcom.2015.08.224} {\bibfield  {journal} {\bibinfo  {journal} {Journal of Alloys and Compounds}\ }\textbf {\bibinfo {volume} {652}},\ \bibinfo {pages} {266} (\bibinfo {year} {2015})}\BibitemShut {NoStop}%
\bibitem [{\citenamefont {Yurchenko}\ \emph {et~al.}(2018)\citenamefont {Yurchenko}, \citenamefont {Stepanov}, \citenamefont {Gridneva}, \citenamefont {Mishunin}, \citenamefont {Salishchev},\ and\ \citenamefont {Zherebtsov}}]{yurchenko_effect_2018}%
  \BibitemOpen
  \bibfield  {author} {\bibinfo {author} {\bibfnamefont {N.}~\bibnamefont {Yurchenko}}, \bibinfo {author} {\bibfnamefont {N.}~\bibnamefont {Stepanov}}, \bibinfo {author} {\bibfnamefont {A.}~\bibnamefont {Gridneva}}, \bibinfo {author} {\bibfnamefont {M.}~\bibnamefont {Mishunin}}, \bibinfo {author} {\bibfnamefont {G.}~\bibnamefont {Salishchev}},\ and\ \bibinfo {author} {\bibfnamefont {S.}~\bibnamefont {Zherebtsov}},\ }\href {https://doi.org/10.1016/j.jallcom.2018.05.099} {\bibfield  {journal} {\bibinfo  {journal} {Journal of Alloys and Compounds}\ }\textbf {\bibinfo {volume} {757}},\ \bibinfo {pages} {403} (\bibinfo {year} {2018})}\BibitemShut {NoStop}%
\bibitem [{\citenamefont {Chen}\ \emph {et~al.}(2019)\citenamefont {Chen}, \citenamefont {Kauffmann}, \citenamefont {Seils}, \citenamefont {Boll}, \citenamefont {Liebscher}, \citenamefont {Harding}, \citenamefont {Kumar}, \citenamefont {Szabó}, \citenamefont {Schlabach}, \citenamefont {Kauffmann-Weiss}, \citenamefont {Müller}, \citenamefont {Gorr}, \citenamefont {Christ},\ and\ \citenamefont {Heilmaier}}]{chen_crystallographic_2019}%
  \BibitemOpen
  \bibfield  {author} {\bibinfo {author} {\bibfnamefont {H.}~\bibnamefont {Chen}}, \bibinfo {author} {\bibfnamefont {A.}~\bibnamefont {Kauffmann}}, \bibinfo {author} {\bibfnamefont {S.}~\bibnamefont {Seils}}, \bibinfo {author} {\bibfnamefont {T.}~\bibnamefont {Boll}}, \bibinfo {author} {\bibfnamefont {C.}~\bibnamefont {Liebscher}}, \bibinfo {author} {\bibfnamefont {I.}~\bibnamefont {Harding}}, \bibinfo {author} {\bibfnamefont {K.}~\bibnamefont {Kumar}}, \bibinfo {author} {\bibfnamefont {D.}~\bibnamefont {Szabó}}, \bibinfo {author} {\bibfnamefont {S.}~\bibnamefont {Schlabach}}, \bibinfo {author} {\bibfnamefont {S.}~\bibnamefont {Kauffmann-Weiss}}, \bibinfo {author} {\bibfnamefont {F.}~\bibnamefont {Müller}}, \bibinfo {author} {\bibfnamefont {B.}~\bibnamefont {Gorr}}, \bibinfo {author} {\bibfnamefont {H.-J.}\ \bibnamefont {Christ}},\ and\ \bibinfo {author} {\bibfnamefont {M.}~\bibnamefont {Heilmaier}},\ }\href {https://doi.org/10.1016/j.actamat.2019.07.001} {\bibfield  {journal} {\bibinfo  {journal} {Acta
  Materialia}\ }\textbf {\bibinfo {volume} {176}},\ \bibinfo {pages} {123} (\bibinfo {year} {2019})}\BibitemShut {NoStop}%
\bibitem [{\citenamefont {Momma}\ and\ \citenamefont {Izumi}(2011)}]{momma_vesta_2011}%
  \BibitemOpen
  \bibfield  {author} {\bibinfo {author} {\bibfnamefont {K.}~\bibnamefont {Momma}}\ and\ \bibinfo {author} {\bibfnamefont {F.}~\bibnamefont {Izumi}},\ }\href {https://doi.org/10.1107/S0021889811038970} {\bibfield  {journal} {\bibinfo  {journal} {Journal of Applied Crystallography}\ }\textbf {\bibinfo {volume} {44}},\ \bibinfo {pages} {1272} (\bibinfo {year} {2011})}\BibitemShut {NoStop}%
\bibitem [{\citenamefont {Widom}(2018)}]{widom_modeling_2018}%
  \BibitemOpen
  \bibfield  {author} {\bibinfo {author} {\bibfnamefont {M.}~\bibnamefont {Widom}},\ }\href {https://doi.org/10.1557/jmr.2018.222} {\bibfield  {journal} {\bibinfo  {journal} {Journal of Materials Research}\ }\textbf {\bibinfo {volume} {33}},\ \bibinfo {pages} {2881} (\bibinfo {year} {2018})}\BibitemShut {NoStop}%
\bibitem [{\citenamefont {Ferrari}\ \emph {et~al.}(2020)\citenamefont {Ferrari}, \citenamefont {Dutta}, \citenamefont {Gubaev}, \citenamefont {Ikeda}, \citenamefont {Srinivasan}, \citenamefont {Grabowski},\ and\ \citenamefont {Körmann}}]{ferrari_frontiers_2020}%
  \BibitemOpen
  \bibfield  {author} {\bibinfo {author} {\bibfnamefont {A.}~\bibnamefont {Ferrari}}, \bibinfo {author} {\bibfnamefont {B.}~\bibnamefont {Dutta}}, \bibinfo {author} {\bibfnamefont {K.}~\bibnamefont {Gubaev}}, \bibinfo {author} {\bibfnamefont {Y.}~\bibnamefont {Ikeda}}, \bibinfo {author} {\bibfnamefont {P.}~\bibnamefont {Srinivasan}}, \bibinfo {author} {\bibfnamefont {B.}~\bibnamefont {Grabowski}},\ and\ \bibinfo {author} {\bibfnamefont {F.}~\bibnamefont {Körmann}},\ }\href {https://doi.org/10.1063/5.0025310} {\bibfield  {journal} {\bibinfo  {journal} {Journal of Applied Physics}\ }\textbf {\bibinfo {volume} {128}},\ \bibinfo {pages} {150901} (\bibinfo {year} {2020})}\BibitemShut {NoStop}%
\bibitem [{\citenamefont {Ferrari}\ \emph {et~al.}(2023)\citenamefont {Ferrari}, \citenamefont {Körmann}, \citenamefont {Asta},\ and\ \citenamefont {Neugebauer}}]{ferrari_simulating_2023}%
  \BibitemOpen
  \bibfield  {author} {\bibinfo {author} {\bibfnamefont {A.}~\bibnamefont {Ferrari}}, \bibinfo {author} {\bibfnamefont {F.}~\bibnamefont {Körmann}}, \bibinfo {author} {\bibfnamefont {M.}~\bibnamefont {Asta}},\ and\ \bibinfo {author} {\bibfnamefont {J.}~\bibnamefont {Neugebauer}},\ }\href {https://doi.org/10.1038/s43588-023-00407-4} {\bibfield  {journal} {\bibinfo  {journal} {Nature Computational Science}\ }\textbf {\bibinfo {volume} {3}},\ \bibinfo {pages} {221} (\bibinfo {year} {2023})}\BibitemShut {NoStop}%
\bibitem [{\citenamefont {Widom}\ \emph {et~al.}(2014)\citenamefont {Widom}, \citenamefont {Huhn}, \citenamefont {Maiti},\ and\ \citenamefont {Steurer}}]{widom_hybrid_2014}%
  \BibitemOpen
  \bibfield  {author} {\bibinfo {author} {\bibfnamefont {M.}~\bibnamefont {Widom}}, \bibinfo {author} {\bibfnamefont {W.~P.}\ \bibnamefont {Huhn}}, \bibinfo {author} {\bibfnamefont {S.}~\bibnamefont {Maiti}},\ and\ \bibinfo {author} {\bibfnamefont {W.}~\bibnamefont {Steurer}},\ }\href {https://doi.org/10.1007/s11661-013-2000-8} {\bibfield  {journal} {\bibinfo  {journal} {Metallurgical and Materials Transactions A}\ }\textbf {\bibinfo {volume} {45}},\ \bibinfo {pages} {196} (\bibinfo {year} {2014})}\BibitemShut {NoStop}%
\bibitem [{\citenamefont {Tamm}\ \emph {et~al.}(2015)\citenamefont {Tamm}, \citenamefont {Aabloo}, \citenamefont {Klintenberg}, \citenamefont {Stocks},\ and\ \citenamefont {Caro}}]{tamm_atomic-scale_2015}%
  \BibitemOpen
  \bibfield  {author} {\bibinfo {author} {\bibfnamefont {A.}~\bibnamefont {Tamm}}, \bibinfo {author} {\bibfnamefont {A.}~\bibnamefont {Aabloo}}, \bibinfo {author} {\bibfnamefont {M.}~\bibnamefont {Klintenberg}}, \bibinfo {author} {\bibfnamefont {M.}~\bibnamefont {Stocks}},\ and\ \bibinfo {author} {\bibfnamefont {A.}~\bibnamefont {Caro}},\ }\href {https://doi.org/10.1016/j.actamat.2015.08.015} {\bibfield  {journal} {\bibinfo  {journal} {Acta Materialia}\ }\textbf {\bibinfo {volume} {99}},\ \bibinfo {pages} {307} (\bibinfo {year} {2015})}\BibitemShut {NoStop}%
\bibitem [{\citenamefont {Widom}(2024)}]{widom_first-principles_2024}%
  \BibitemOpen
  \bibfield  {author} {\bibinfo {author} {\bibfnamefont {M.}~\bibnamefont {Widom}},\ }\href {https://doi.org/10.1103/PhysRevMaterials.8.093603} {\bibfield  {journal} {\bibinfo  {journal} {Physical Review Materials}\ }\textbf {\bibinfo {volume} {8}},\ \bibinfo {pages} {093603} (\bibinfo {year} {2024})}\BibitemShut {NoStop}%
\bibitem [{\citenamefont {Kostiuchenko}\ \emph {et~al.}(2019)\citenamefont {Kostiuchenko}, \citenamefont {Körmann}, \citenamefont {Neugebauer},\ and\ \citenamefont {Shapeev}}]{kostiuchenko_impact_2019}%
  \BibitemOpen
  \bibfield  {author} {\bibinfo {author} {\bibfnamefont {T.}~\bibnamefont {Kostiuchenko}}, \bibinfo {author} {\bibfnamefont {F.}~\bibnamefont {Körmann}}, \bibinfo {author} {\bibfnamefont {J.}~\bibnamefont {Neugebauer}},\ and\ \bibinfo {author} {\bibfnamefont {A.}~\bibnamefont {Shapeev}},\ }\href {https://doi.org/10.1038/s41524-019-0195-y} {\bibfield  {journal} {\bibinfo  {journal} {npj Computational Materials}\ }\textbf {\bibinfo {volume} {5}},\ \bibinfo {pages} {55} (\bibinfo {year} {2019})}\BibitemShut {NoStop}%
\bibitem [{\citenamefont {Rosenbrock}\ \emph {et~al.}(2021)\citenamefont {Rosenbrock}, \citenamefont {Gubaev}, \citenamefont {Shapeev}, \citenamefont {Pártay}, \citenamefont {Bernstein}, \citenamefont {Csányi},\ and\ \citenamefont {Hart}}]{rosenbrock_machine-learned_2021}%
  \BibitemOpen
  \bibfield  {author} {\bibinfo {author} {\bibfnamefont {C.~W.}\ \bibnamefont {Rosenbrock}}, \bibinfo {author} {\bibfnamefont {K.}~\bibnamefont {Gubaev}}, \bibinfo {author} {\bibfnamefont {A.~V.}\ \bibnamefont {Shapeev}}, \bibinfo {author} {\bibfnamefont {L.~B.}\ \bibnamefont {Pártay}}, \bibinfo {author} {\bibfnamefont {N.}~\bibnamefont {Bernstein}}, \bibinfo {author} {\bibfnamefont {G.}~\bibnamefont {Csányi}},\ and\ \bibinfo {author} {\bibfnamefont {G.~L.~W.}\ \bibnamefont {Hart}},\ }\href {https://doi.org/10.1038/s41524-020-00477-2} {\bibfield  {journal} {\bibinfo  {journal} {npj Computational Materials}\ }\textbf {\bibinfo {volume} {7}},\ \bibinfo {pages} {24} (\bibinfo {year} {2021})}\BibitemShut {NoStop}%
\bibitem [{\citenamefont {Körmann}\ \emph {et~al.}(2021)\citenamefont {Körmann}, \citenamefont {Kostiuchenko}, \citenamefont {Shapeev},\ and\ \citenamefont {Neugebauer}}]{kormann_b2_2021}%
  \BibitemOpen
  \bibfield  {author} {\bibinfo {author} {\bibfnamefont {F.}~\bibnamefont {Körmann}}, \bibinfo {author} {\bibfnamefont {T.}~\bibnamefont {Kostiuchenko}}, \bibinfo {author} {\bibfnamefont {A.}~\bibnamefont {Shapeev}},\ and\ \bibinfo {author} {\bibfnamefont {J.}~\bibnamefont {Neugebauer}},\ }\href {https://doi.org/10.1103/PhysRevMaterials.5.053803} {\bibfield  {journal} {\bibinfo  {journal} {Physical Review Materials}\ }\textbf {\bibinfo {volume} {5}},\ \bibinfo {pages} {053803} (\bibinfo {year} {2021})}\BibitemShut {NoStop}%
\bibitem [{\citenamefont {Zhou}\ \emph {et~al.}(2022)\citenamefont {Zhou}, \citenamefont {Srinivasan}, \citenamefont {Körmann}, \citenamefont {Grabowski}, \citenamefont {Smith}, \citenamefont {Goddard},\ and\ \citenamefont {Duff}}]{zhou_thermodynamics_2022}%
  \BibitemOpen
  \bibfield  {author} {\bibinfo {author} {\bibfnamefont {Y.}~\bibnamefont {Zhou}}, \bibinfo {author} {\bibfnamefont {P.}~\bibnamefont {Srinivasan}}, \bibinfo {author} {\bibfnamefont {F.}~\bibnamefont {Körmann}}, \bibinfo {author} {\bibfnamefont {B.}~\bibnamefont {Grabowski}}, \bibinfo {author} {\bibfnamefont {R.}~\bibnamefont {Smith}}, \bibinfo {author} {\bibfnamefont {P.}~\bibnamefont {Goddard}},\ and\ \bibinfo {author} {\bibfnamefont {A.~I.}\ \bibnamefont {Duff}},\ }\href {https://doi.org/10.1103/PhysRevB.105.214302} {\bibfield  {journal} {\bibinfo  {journal} {Physical Review B}\ }\textbf {\bibinfo {volume} {105}},\ \bibinfo {pages} {214302} (\bibinfo {year} {2022})}\BibitemShut {NoStop}%
\bibitem [{\citenamefont {Zhang}\ \emph {et~al.}(2025)\citenamefont {Zhang}, \citenamefont {Sorkin}, \citenamefont {Aitken}, \citenamefont {Politano}, \citenamefont {Behler}, \citenamefont {P~Thompson}, \citenamefont {Ko}, \citenamefont {Ong}, \citenamefont {Chalykh}, \citenamefont {Korogod}, \citenamefont {Podryabinkin}, \citenamefont {Shapeev}, \citenamefont {Li}, \citenamefont {Mishin}, \citenamefont {Pei}, \citenamefont {Liu}, \citenamefont {Kim}, \citenamefont {Park}, \citenamefont {Hwang}, \citenamefont {Han}, \citenamefont {Sheriff}, \citenamefont {Cao},\ and\ \citenamefont {Freitas}}]{zhang_roadmap_2025}%
  \BibitemOpen
  \bibfield  {author} {\bibinfo {author} {\bibfnamefont {Y.-W.}\ \bibnamefont {Zhang}}, \bibinfo {author} {\bibfnamefont {V.}~\bibnamefont {Sorkin}}, \bibinfo {author} {\bibfnamefont {Z.~H.}\ \bibnamefont {Aitken}}, \bibinfo {author} {\bibfnamefont {A.}~\bibnamefont {Politano}}, \bibinfo {author} {\bibfnamefont {J.}~\bibnamefont {Behler}}, \bibinfo {author} {\bibfnamefont {A.}~\bibnamefont {P~Thompson}}, \bibinfo {author} {\bibfnamefont {T.~W.}\ \bibnamefont {Ko}}, \bibinfo {author} {\bibfnamefont {S.~P.}\ \bibnamefont {Ong}}, \bibinfo {author} {\bibfnamefont {O.}~\bibnamefont {Chalykh}}, \bibinfo {author} {\bibfnamefont {D.}~\bibnamefont {Korogod}}, \bibinfo {author} {\bibfnamefont {E.}~\bibnamefont {Podryabinkin}}, \bibinfo {author} {\bibfnamefont {A.}~\bibnamefont {Shapeev}}, \bibinfo {author} {\bibfnamefont {J.}~\bibnamefont {Li}}, \bibinfo {author} {\bibfnamefont {Y.}~\bibnamefont {Mishin}}, \bibinfo {author} {\bibfnamefont {Z.}~\bibnamefont {Pei}}, \bibinfo {author} {\bibfnamefont {X.}~\bibnamefont
  {Liu}}, \bibinfo {author} {\bibfnamefont {J.}~\bibnamefont {Kim}}, \bibinfo {author} {\bibfnamefont {Y.}~\bibnamefont {Park}}, \bibinfo {author} {\bibfnamefont {S.}~\bibnamefont {Hwang}}, \bibinfo {author} {\bibfnamefont {S.}~\bibnamefont {Han}}, \bibinfo {author} {\bibfnamefont {K.}~\bibnamefont {Sheriff}}, \bibinfo {author} {\bibfnamefont {Y.}~\bibnamefont {Cao}},\ and\ \bibinfo {author} {\bibfnamefont {R.}~\bibnamefont {Freitas}},\ }\href {https://doi.org/10.1088/1361-651X/ad9d63} {\bibfield  {journal} {\bibinfo  {journal} {Modelling and Simulation in Materials Science and Engineering}\ }\textbf {\bibinfo {volume} {33}},\ \bibinfo {pages} {023301} (\bibinfo {year} {2025})}\BibitemShut {NoStop}%
\bibitem [{\citenamefont {Fernández-Caballero}\ \emph {et~al.}(2017)\citenamefont {Fernández-Caballero}, \citenamefont {Wróbel}, \citenamefont {Mummery},\ and\ \citenamefont {Nguyen-Manh}}]{fernandez-caballero_short-range_2017}%
  \BibitemOpen
  \bibfield  {author} {\bibinfo {author} {\bibfnamefont {A.}~\bibnamefont {Fernández-Caballero}}, \bibinfo {author} {\bibfnamefont {J.~S.}\ \bibnamefont {Wróbel}}, \bibinfo {author} {\bibfnamefont {P.~M.}\ \bibnamefont {Mummery}},\ and\ \bibinfo {author} {\bibfnamefont {D.}~\bibnamefont {Nguyen-Manh}},\ }\href {https://doi.org/10.1007/s11669-017-0582-3} {\bibfield  {journal} {\bibinfo  {journal} {Journal of Phase Equilibria and Diffusion}\ }\textbf {\bibinfo {volume} {38}},\ \bibinfo {pages} {391} (\bibinfo {year} {2017})}\BibitemShut {NoStop}%
\bibitem [{\citenamefont {Sobieraj}\ \emph {et~al.}(2020)\citenamefont {Sobieraj}, \citenamefont {Wróbel}, \citenamefont {Rygier}, \citenamefont {Kurzydłowski}, \citenamefont {El~Atwani}, \citenamefont {Devaraj}, \citenamefont {Martinez~Saez},\ and\ \citenamefont {Nguyen-Manh}}]{sobieraj_chemical_2020}%
  \BibitemOpen
  \bibfield  {author} {\bibinfo {author} {\bibfnamefont {D.}~\bibnamefont {Sobieraj}}, \bibinfo {author} {\bibfnamefont {J.~S.}\ \bibnamefont {Wróbel}}, \bibinfo {author} {\bibfnamefont {T.}~\bibnamefont {Rygier}}, \bibinfo {author} {\bibfnamefont {K.~J.}\ \bibnamefont {Kurzydłowski}}, \bibinfo {author} {\bibfnamefont {O.}~\bibnamefont {El~Atwani}}, \bibinfo {author} {\bibfnamefont {A.}~\bibnamefont {Devaraj}}, \bibinfo {author} {\bibfnamefont {E.}~\bibnamefont {Martinez~Saez}},\ and\ \bibinfo {author} {\bibfnamefont {D.}~\bibnamefont {Nguyen-Manh}},\ }\href {https://doi.org/10.1039/D0CP03764H} {\bibfield  {journal} {\bibinfo  {journal} {Physical Chemistry Chemical Physics}\ }\textbf {\bibinfo {volume} {22}},\ \bibinfo {pages} {23929} (\bibinfo {year} {2020})}\BibitemShut {NoStop}%
\bibitem [{\citenamefont {Kim}\ and\ \citenamefont {Widom}(2023)}]{kim_interaction_2023}%
  \BibitemOpen
  \bibfield  {author} {\bibinfo {author} {\bibfnamefont {A.~D.}\ \bibnamefont {Kim}}\ and\ \bibinfo {author} {\bibfnamefont {M.}~\bibnamefont {Widom}},\ }\href {https://doi.org/10.1103/PhysRevMaterials.7.063803} {\bibfield  {journal} {\bibinfo  {journal} {Physical Review Materials}\ }\textbf {\bibinfo {volume} {7}},\ \bibinfo {pages} {063803} (\bibinfo {year} {2023})}\BibitemShut {NoStop}%
\bibitem [{\citenamefont {Vazquez}\ \emph {et~al.}(2024)\citenamefont {Vazquez}, \citenamefont {Sauceda},\ and\ \citenamefont {Arróyave}}]{vazquez_deciphering_2024}%
  \BibitemOpen
  \bibfield  {author} {\bibinfo {author} {\bibfnamefont {G.}~\bibnamefont {Vazquez}}, \bibinfo {author} {\bibfnamefont {D.}~\bibnamefont {Sauceda}},\ and\ \bibinfo {author} {\bibfnamefont {R.}~\bibnamefont {Arróyave}},\ }\href {https://doi.org/10.1016/j.actamat.2024.120137} {\bibfield  {journal} {\bibinfo  {journal} {Acta Materialia}\ }\textbf {\bibinfo {volume} {276}},\ \bibinfo {pages} {120137} (\bibinfo {year} {2024})}\BibitemShut {NoStop}%
\bibitem [{\citenamefont {Singh}\ \emph {et~al.}(2015)\citenamefont {Singh}, \citenamefont {Smirnov},\ and\ \citenamefont {Johnson}}]{singh_atomic_2015}%
  \BibitemOpen
  \bibfield  {author} {\bibinfo {author} {\bibfnamefont {P.}~\bibnamefont {Singh}}, \bibinfo {author} {\bibfnamefont {A.~V.}\ \bibnamefont {Smirnov}},\ and\ \bibinfo {author} {\bibfnamefont {D.~D.}\ \bibnamefont {Johnson}},\ }\href {https://doi.org/10.1103/PhysRevB.91.224204} {\bibfield  {journal} {\bibinfo  {journal} {Physical Review B}\ }\textbf {\bibinfo {volume} {91}},\ \bibinfo {pages} {224204} (\bibinfo {year} {2015})}\BibitemShut {NoStop}%
\bibitem [{\citenamefont {Körmann}\ \emph {et~al.}(2017)\citenamefont {Körmann}, \citenamefont {Ruban},\ and\ \citenamefont {Sluiter}}]{kormann_long-ranged_2017}%
  \BibitemOpen
  \bibfield  {author} {\bibinfo {author} {\bibfnamefont {F.}~\bibnamefont {Körmann}}, \bibinfo {author} {\bibfnamefont {A.~V.}\ \bibnamefont {Ruban}},\ and\ \bibinfo {author} {\bibfnamefont {M.~H.}\ \bibnamefont {Sluiter}},\ }\href {https://doi.org/10.1080/21663831.2016.1198837} {\bibfield  {journal} {\bibinfo  {journal} {Materials Research Letters}\ }\textbf {\bibinfo {volume} {5}},\ \bibinfo {pages} {35} (\bibinfo {year} {2017})}\BibitemShut {NoStop}%
\bibitem [{\citenamefont {Singh}\ \emph {et~al.}(2018)\citenamefont {Singh}, \citenamefont {Smirnov},\ and\ \citenamefont {Johnson}}]{singh_ta-nb-mo-w_2018}%
  \BibitemOpen
  \bibfield  {author} {\bibinfo {author} {\bibfnamefont {P.}~\bibnamefont {Singh}}, \bibinfo {author} {\bibfnamefont {A.~V.}\ \bibnamefont {Smirnov}},\ and\ \bibinfo {author} {\bibfnamefont {D.~D.}\ \bibnamefont {Johnson}},\ }\href {https://doi.org/10.1103/PhysRevMaterials.2.055004} {\bibfield  {journal} {\bibinfo  {journal} {Physical Review Materials}\ }\textbf {\bibinfo {volume} {2}},\ \bibinfo {pages} {055004} (\bibinfo {year} {2018})}\BibitemShut {NoStop}%
\bibitem [{\citenamefont {Feng}\ \emph {et~al.}(2017)\citenamefont {Feng}, \citenamefont {Liaw}, \citenamefont {Gao},\ and\ \citenamefont {Widom}}]{feng_first-principles_2017}%
  \BibitemOpen
  \bibfield  {author} {\bibinfo {author} {\bibfnamefont {R.}~\bibnamefont {Feng}}, \bibinfo {author} {\bibfnamefont {P.~K.}\ \bibnamefont {Liaw}}, \bibinfo {author} {\bibfnamefont {M.~C.}\ \bibnamefont {Gao}},\ and\ \bibinfo {author} {\bibfnamefont {M.}~\bibnamefont {Widom}},\ }\href {https://doi.org/10.1038/s41524-017-0049-4} {\bibfield  {journal} {\bibinfo  {journal} {npj Computational Materials}\ }\textbf {\bibinfo {volume} {3}},\ \bibinfo {pages} {50} (\bibinfo {year} {2017})}\BibitemShut {NoStop}%
\bibitem [{\citenamefont {Li}\ \emph {et~al.}(2023)\citenamefont {Li}, \citenamefont {Wang}, \citenamefont {Fan}, \citenamefont {Lu}, \citenamefont {Wang}, \citenamefont {Li},\ and\ \citenamefont {Liaw}}]{li_calphad-aided_2023}%
  \BibitemOpen
  \bibfield  {author} {\bibinfo {author} {\bibfnamefont {T.}~\bibnamefont {Li}}, \bibinfo {author} {\bibfnamefont {S.}~\bibnamefont {Wang}}, \bibinfo {author} {\bibfnamefont {W.}~\bibnamefont {Fan}}, \bibinfo {author} {\bibfnamefont {Y.}~\bibnamefont {Lu}}, \bibinfo {author} {\bibfnamefont {T.}~\bibnamefont {Wang}}, \bibinfo {author} {\bibfnamefont {T.}~\bibnamefont {Li}},\ and\ \bibinfo {author} {\bibfnamefont {P.~K.}\ \bibnamefont {Liaw}},\ }\href {https://doi.org/10.1016/j.actamat.2023.118728} {\bibfield  {journal} {\bibinfo  {journal} {Acta Materialia}\ }\textbf {\bibinfo {volume} {246}},\ \bibinfo {pages} {118728} (\bibinfo {year} {2023})}\BibitemShut {NoStop}%
\bibitem [{\citenamefont {Khan}\ \emph {et~al.}(2016)\citenamefont {Khan}, \citenamefont {Staunton},\ and\ \citenamefont {Stocks}}]{khan_statistical_2016}%
  \BibitemOpen
  \bibfield  {author} {\bibinfo {author} {\bibfnamefont {S.~N.}\ \bibnamefont {Khan}}, \bibinfo {author} {\bibfnamefont {J.~B.}\ \bibnamefont {Staunton}},\ and\ \bibinfo {author} {\bibfnamefont {G.~M.}\ \bibnamefont {Stocks}},\ }\href {https://doi.org/10.1103/PhysRevB.93.054206} {\bibfield  {journal} {\bibinfo  {journal} {Physical Review B}\ }\textbf {\bibinfo {volume} {93}},\ \bibinfo {pages} {054206} (\bibinfo {year} {2016})}\BibitemShut {NoStop}%
\bibitem [{\citenamefont {Woodgate}\ and\ \citenamefont {Staunton}(2022)}]{woodgate_compositional_2022}%
  \BibitemOpen
  \bibfield  {author} {\bibinfo {author} {\bibfnamefont {C.~D.}\ \bibnamefont {Woodgate}}\ and\ \bibinfo {author} {\bibfnamefont {J.~B.}\ \bibnamefont {Staunton}},\ }\href {https://doi.org/10.1103/PhysRevB.105.115124} {\bibfield  {journal} {\bibinfo  {journal} {Physical Review B}\ }\textbf {\bibinfo {volume} {105}},\ \bibinfo {pages} {115124} (\bibinfo {year} {2022})}\BibitemShut {NoStop}%
\bibitem [{\citenamefont {Woodgate}\ and\ \citenamefont {Staunton}(2023)}]{woodgate_short-range_2023}%
  \BibitemOpen
  \bibfield  {author} {\bibinfo {author} {\bibfnamefont {C.~D.}\ \bibnamefont {Woodgate}}\ and\ \bibinfo {author} {\bibfnamefont {J.~B.}\ \bibnamefont {Staunton}},\ }\href {https://doi.org/10.1103/PhysRevMaterials.7.013801} {\bibfield  {journal} {\bibinfo  {journal} {Physical Review Materials}\ }\textbf {\bibinfo {volume} {7}},\ \bibinfo {pages} {013801} (\bibinfo {year} {2023})}\BibitemShut {NoStop}%
\bibitem [{\citenamefont {Woodgate}(2024)}]{woodgate_modelling_2024}%
  \BibitemOpen
  \bibfield  {author} {\bibinfo {author} {\bibfnamefont {C.~D.}\ \bibnamefont {Woodgate}},\ }\href {https://doi.org/10.1007/978-3-031-62021-8} {\emph {\bibinfo {title} {Modelling {Atomic} {Arrangements} in {Multicomponent} {Alloys}: {A} {Perturbative}, {First}-{Principles}-{Based} {Approach}}}},\ \bibinfo {series} {Springer {Series} in {Materials} {Science}}, Vol.\ \bibinfo {volume} {346}\ (\bibinfo  {publisher} {Springer Nature Switzerland},\ \bibinfo {address} {Cham},\ \bibinfo {year} {2024})\BibitemShut {NoStop}%
\bibitem [{\citenamefont {Soven}(1967)}]{soven_coherent-potential_1967}%
  \BibitemOpen
  \bibfield  {author} {\bibinfo {author} {\bibfnamefont {P.}~\bibnamefont {Soven}},\ }\href {https://doi.org/10.1103/PhysRev.156.809} {\bibfield  {journal} {\bibinfo  {journal} {Physical Review}\ }\textbf {\bibinfo {volume} {156}},\ \bibinfo {pages} {809} (\bibinfo {year} {1967})}\BibitemShut {NoStop}%
\bibitem [{\citenamefont {Gyorffy}(1972)}]{gyorffy_coherent-potential_1972}%
  \BibitemOpen
  \bibfield  {author} {\bibinfo {author} {\bibfnamefont {B.~L.}\ \bibnamefont {Gyorffy}},\ }\href {https://doi.org/10.1103/PhysRevB.5.2382} {\bibfield  {journal} {\bibinfo  {journal} {Physical Review B}\ }\textbf {\bibinfo {volume} {5}},\ \bibinfo {pages} {2382} (\bibinfo {year} {1972})}\BibitemShut {NoStop}%
\bibitem [{\citenamefont {Stocks}\ \emph {et~al.}(1978)\citenamefont {Stocks}, \citenamefont {Temmerman},\ and\ \citenamefont {Gyorffy}}]{stocks_complete_1978}%
  \BibitemOpen
  \bibfield  {author} {\bibinfo {author} {\bibfnamefont {G.~M.}\ \bibnamefont {Stocks}}, \bibinfo {author} {\bibfnamefont {W.~M.}\ \bibnamefont {Temmerman}},\ and\ \bibinfo {author} {\bibfnamefont {B.~L.}\ \bibnamefont {Gyorffy}},\ }\href {https://doi.org/10.1103/PhysRevLett.41.339} {\bibfield  {journal} {\bibinfo  {journal} {Physical Review Letters}\ }\textbf {\bibinfo {volume} {41}},\ \bibinfo {pages} {339} (\bibinfo {year} {1978})}\BibitemShut {NoStop}%
\bibitem [{\citenamefont {Korringa}(1947)}]{korringa_calculation_1947}%
  \BibitemOpen
  \bibfield  {author} {\bibinfo {author} {\bibfnamefont {J.}~\bibnamefont {Korringa}},\ }\href {https://doi.org/10.1016/0031-8914(47)90013-X} {\bibfield  {journal} {\bibinfo  {journal} {Physica}\ }\textbf {\bibinfo {volume} {13}},\ \bibinfo {pages} {392} (\bibinfo {year} {1947})}\BibitemShut {NoStop}%
\bibitem [{\citenamefont {Kohn}\ and\ \citenamefont {Rostoker}(1954)}]{kohn_solution_1954}%
  \BibitemOpen
  \bibfield  {author} {\bibinfo {author} {\bibfnamefont {W.}~\bibnamefont {Kohn}}\ and\ \bibinfo {author} {\bibfnamefont {N.}~\bibnamefont {Rostoker}},\ }\href {https://doi.org/10.1103/PhysRev.94.1111} {\bibfield  {journal} {\bibinfo  {journal} {Physical Review}\ }\textbf {\bibinfo {volume} {94}},\ \bibinfo {pages} {1111} (\bibinfo {year} {1954})}\BibitemShut {NoStop}%
\bibitem [{\citenamefont {Ebert}\ \emph {et~al.}(2011)\citenamefont {Ebert}, \citenamefont {Ködderitzsch},\ and\ \citenamefont {Minár}}]{ebert_calculating_2011}%
  \BibitemOpen
  \bibfield  {author} {\bibinfo {author} {\bibfnamefont {H.}~\bibnamefont {Ebert}}, \bibinfo {author} {\bibfnamefont {D.}~\bibnamefont {Ködderitzsch}},\ and\ \bibinfo {author} {\bibfnamefont {J.}~\bibnamefont {Minár}},\ }\href {https://doi.org/10.1088/0034-4885/74/9/096501} {\bibfield  {journal} {\bibinfo  {journal} {Reports on Progress in Physics}\ }\textbf {\bibinfo {volume} {74}},\ \bibinfo {pages} {096501} (\bibinfo {year} {2011})}\BibitemShut {NoStop}%
\bibitem [{\citenamefont {Hohenberg}\ and\ \citenamefont {Kohn}(1964)}]{hohenberg_inhomogeneous_1964}%
  \BibitemOpen
  \bibfield  {author} {\bibinfo {author} {\bibfnamefont {P.}~\bibnamefont {Hohenberg}}\ and\ \bibinfo {author} {\bibfnamefont {W.}~\bibnamefont {Kohn}},\ }\href {https://doi.org/10.1103/PhysRev.136.B864} {\bibfield  {journal} {\bibinfo  {journal} {Phys. Rev.}\ }\textbf {\bibinfo {volume} {136}},\ \bibinfo {pages} {B864} (\bibinfo {year} {1964})}\BibitemShut {NoStop}%
\bibitem [{\citenamefont {Kohn}\ and\ \citenamefont {Sham}(1965)}]{kohn_self-consistent_1965}%
  \BibitemOpen
  \bibfield  {author} {\bibinfo {author} {\bibfnamefont {W.}~\bibnamefont {Kohn}}\ and\ \bibinfo {author} {\bibfnamefont {L.~J.}\ \bibnamefont {Sham}},\ }\href {https://doi.org/10.1103/PhysRev.140.A1133} {\bibfield  {journal} {\bibinfo  {journal} {Physical Review}\ }\textbf {\bibinfo {volume} {140}},\ \bibinfo {pages} {A1133} (\bibinfo {year} {1965})}\BibitemShut {NoStop}%
\bibitem [{\citenamefont {Martin}(2004)}]{martin_electronic_2004}%
  \BibitemOpen
  \bibfield  {author} {\bibinfo {author} {\bibfnamefont {R.~M.}\ \bibnamefont {Martin}},\ }\href@noop {} {\emph {\bibinfo {title} {Electronic {Structure}: {Basic} {Theory} and {Practical} {Methods}}}}\ (\bibinfo  {publisher} {Cambridge University Press},\ \bibinfo {address} {Cambridge, UK},\ \bibinfo {year} {2004})\BibitemShut {NoStop}%
\bibitem [{\citenamefont {Faulkner}\ \emph {et~al.}(2018)\citenamefont {Faulkner}, \citenamefont {Stocks},\ and\ \citenamefont {Wang}}]{faulkner_multiple_2018}%
  \BibitemOpen
  \bibfield  {author} {\bibinfo {author} {\bibfnamefont {J.~S.}\ \bibnamefont {Faulkner}}, \bibinfo {author} {\bibfnamefont {G.~M.}\ \bibnamefont {Stocks}},\ and\ \bibinfo {author} {\bibfnamefont {Y.}~\bibnamefont {Wang}},\ }\href {https://doi.org/10.1088/2053-2563/aae7d8} {\emph {\bibinfo {title} {Multiple {Scattering} {Theory}: {Electronic} {Structure} of {Solids}}}},\ \bibinfo {edition} {1st}\ ed.\ (\bibinfo  {publisher} {IOP Publishing},\ \bibinfo {address} {Bristol, UK},\ \bibinfo {year} {2018})\BibitemShut {NoStop}%
\bibitem [{\citenamefont {Redka}\ \emph {et~al.}(2024)\citenamefont {Redka}, \citenamefont {Khan}, \citenamefont {Martino}, \citenamefont {Mettan}, \citenamefont {Ciric}, \citenamefont {Tolj}, \citenamefont {Ivšić}, \citenamefont {Held}, \citenamefont {Caputo}, \citenamefont {Guedes}, \citenamefont {Strocov}, \citenamefont {Di~Marco}, \citenamefont {Ebert}, \citenamefont {Huber}, \citenamefont {Dil}, \citenamefont {Forró},\ and\ \citenamefont {Minár}}]{redka_interplay_2024}%
  \BibitemOpen
  \bibfield  {author} {\bibinfo {author} {\bibfnamefont {D.}~\bibnamefont {Redka}}, \bibinfo {author} {\bibfnamefont {S.~A.}\ \bibnamefont {Khan}}, \bibinfo {author} {\bibfnamefont {E.}~\bibnamefont {Martino}}, \bibinfo {author} {\bibfnamefont {X.}~\bibnamefont {Mettan}}, \bibinfo {author} {\bibfnamefont {L.}~\bibnamefont {Ciric}}, \bibinfo {author} {\bibfnamefont {D.}~\bibnamefont {Tolj}}, \bibinfo {author} {\bibfnamefont {T.}~\bibnamefont {Ivšić}}, \bibinfo {author} {\bibfnamefont {A.}~\bibnamefont {Held}}, \bibinfo {author} {\bibfnamefont {M.}~\bibnamefont {Caputo}}, \bibinfo {author} {\bibfnamefont {E.~B.}\ \bibnamefont {Guedes}}, \bibinfo {author} {\bibfnamefont {V.~N.}\ \bibnamefont {Strocov}}, \bibinfo {author} {\bibfnamefont {I.}~\bibnamefont {Di~Marco}}, \bibinfo {author} {\bibfnamefont {H.}~\bibnamefont {Ebert}}, \bibinfo {author} {\bibfnamefont {H.~P.}\ \bibnamefont {Huber}}, \bibinfo {author} {\bibfnamefont {J.~H.}\ \bibnamefont {Dil}}, \bibinfo {author} {\bibfnamefont {L.}~\bibnamefont
  {Forró}},\ and\ \bibinfo {author} {\bibfnamefont {J.}~\bibnamefont {Minár}},\ }\href {https://doi.org/10.1038/s41467-024-52349-8} {\bibfield  {journal} {\bibinfo  {journal} {Nature Communications}\ }\textbf {\bibinfo {volume} {15}},\ \bibinfo {pages} {7983} (\bibinfo {year} {2024})}\BibitemShut {NoStop}%
\bibitem [{\citenamefont {Billington}\ \emph {et~al.}(2020)\citenamefont {Billington}, \citenamefont {James}, \citenamefont {Harris-Lee}, \citenamefont {Lagos}, \citenamefont {O'Neill}, \citenamefont {Tsuda}, \citenamefont {Toyoki}, \citenamefont {Kotani}, \citenamefont {Nakamura}, \citenamefont {Bei}, \citenamefont {Mu}, \citenamefont {Samolyuk}, \citenamefont {Stocks}, \citenamefont {Duffy}, \citenamefont {Taylor}, \citenamefont {Giblin},\ and\ \citenamefont {Dugdale}}]{billington_bulk_2020}%
  \BibitemOpen
  \bibfield  {author} {\bibinfo {author} {\bibfnamefont {D.}~\bibnamefont {Billington}}, \bibinfo {author} {\bibfnamefont {A.~D.~N.}\ \bibnamefont {James}}, \bibinfo {author} {\bibfnamefont {E.~I.}\ \bibnamefont {Harris-Lee}}, \bibinfo {author} {\bibfnamefont {D.~A.}\ \bibnamefont {Lagos}}, \bibinfo {author} {\bibfnamefont {D.}~\bibnamefont {O'Neill}}, \bibinfo {author} {\bibfnamefont {N.}~\bibnamefont {Tsuda}}, \bibinfo {author} {\bibfnamefont {K.}~\bibnamefont {Toyoki}}, \bibinfo {author} {\bibfnamefont {Y.}~\bibnamefont {Kotani}}, \bibinfo {author} {\bibfnamefont {T.}~\bibnamefont {Nakamura}}, \bibinfo {author} {\bibfnamefont {H.}~\bibnamefont {Bei}}, \bibinfo {author} {\bibfnamefont {S.}~\bibnamefont {Mu}}, \bibinfo {author} {\bibfnamefont {G.~D.}\ \bibnamefont {Samolyuk}}, \bibinfo {author} {\bibfnamefont {G.~M.}\ \bibnamefont {Stocks}}, \bibinfo {author} {\bibfnamefont {J.~A.}\ \bibnamefont {Duffy}}, \bibinfo {author} {\bibfnamefont {J.~W.}\ \bibnamefont {Taylor}}, \bibinfo {author} {\bibfnamefont
  {S.~R.}\ \bibnamefont {Giblin}},\ and\ \bibinfo {author} {\bibfnamefont {S.~B.}\ \bibnamefont {Dugdale}},\ }\href {https://doi.org/10.1103/PhysRevB.102.174405} {\bibfield  {journal} {\bibinfo  {journal} {Physical Review B}\ }\textbf {\bibinfo {volume} {102}},\ \bibinfo {pages} {174405} (\bibinfo {year} {2020})}\BibitemShut {NoStop}%
\bibitem [{\citenamefont {Bista}\ \emph {et~al.}(2025)\citenamefont {Bista}, \citenamefont {Beeson}, \citenamefont {Sengupta}, \citenamefont {Jackson}, \citenamefont {Khanna}, \citenamefont {Liu},\ and\ \citenamefont {Yin}}]{bista_fast_2025}%
  \BibitemOpen
  \bibfield  {author} {\bibinfo {author} {\bibfnamefont {D.}~\bibnamefont {Bista}}, \bibinfo {author} {\bibfnamefont {W.~B.}\ \bibnamefont {Beeson}}, \bibinfo {author} {\bibfnamefont {T.}~\bibnamefont {Sengupta}}, \bibinfo {author} {\bibfnamefont {J.}~\bibnamefont {Jackson}}, \bibinfo {author} {\bibfnamefont {S.~N.}\ \bibnamefont {Khanna}}, \bibinfo {author} {\bibfnamefont {K.}~\bibnamefont {Liu}},\ and\ \bibinfo {author} {\bibfnamefont {G.}~\bibnamefont {Yin}},\ }\href {https://doi.org/10.1103/PhysRevMaterials.9.L031401} {\bibfield  {journal} {\bibinfo  {journal} {Physical Review Materials}\ }\textbf {\bibinfo {volume} {9}},\ \bibinfo {pages} {L031401} (\bibinfo {year} {2025})}\BibitemShut {NoStop}%
\bibitem [{\citenamefont {Samolyuk}\ \emph {et~al.}(2018)\citenamefont {Samolyuk}, \citenamefont {Mu}, \citenamefont {May}, \citenamefont {Sales}, \citenamefont {Wimmer}, \citenamefont {Mankovsky}, \citenamefont {Ebert},\ and\ \citenamefont {Stocks}}]{samolyuk_temperature_2018}%
  \BibitemOpen
  \bibfield  {author} {\bibinfo {author} {\bibfnamefont {G.~D.}\ \bibnamefont {Samolyuk}}, \bibinfo {author} {\bibfnamefont {S.}~\bibnamefont {Mu}}, \bibinfo {author} {\bibfnamefont {A.~F.}\ \bibnamefont {May}}, \bibinfo {author} {\bibfnamefont {B.~C.}\ \bibnamefont {Sales}}, \bibinfo {author} {\bibfnamefont {S.}~\bibnamefont {Wimmer}}, \bibinfo {author} {\bibfnamefont {S.}~\bibnamefont {Mankovsky}}, \bibinfo {author} {\bibfnamefont {H.}~\bibnamefont {Ebert}},\ and\ \bibinfo {author} {\bibfnamefont {G.~M.}\ \bibnamefont {Stocks}},\ }\href {https://doi.org/10.1103/PhysRevB.98.165141} {\bibfield  {journal} {\bibinfo  {journal} {Physical Review B}\ }\textbf {\bibinfo {volume} {98}},\ \bibinfo {pages} {165141} (\bibinfo {year} {2018})}\BibitemShut {NoStop}%
\bibitem [{\citenamefont {Mu}\ \emph {et~al.}(2019)\citenamefont {Mu}, \citenamefont {Samolyuk}, \citenamefont {Wimmer}, \citenamefont {Troparevsky}, \citenamefont {Khan}, \citenamefont {Mankovsky}, \citenamefont {Ebert},\ and\ \citenamefont {Stocks}}]{mu_uncovering_2019}%
  \BibitemOpen
  \bibfield  {author} {\bibinfo {author} {\bibfnamefont {S.}~\bibnamefont {Mu}}, \bibinfo {author} {\bibfnamefont {G.~D.}\ \bibnamefont {Samolyuk}}, \bibinfo {author} {\bibfnamefont {S.}~\bibnamefont {Wimmer}}, \bibinfo {author} {\bibfnamefont {M.~C.}\ \bibnamefont {Troparevsky}}, \bibinfo {author} {\bibfnamefont {S.~N.}\ \bibnamefont {Khan}}, \bibinfo {author} {\bibfnamefont {S.}~\bibnamefont {Mankovsky}}, \bibinfo {author} {\bibfnamefont {H.}~\bibnamefont {Ebert}},\ and\ \bibinfo {author} {\bibfnamefont {G.~M.}\ \bibnamefont {Stocks}},\ }\href {https://doi.org/10.1038/s41524-018-0138-z} {\bibfield  {journal} {\bibinfo  {journal} {npj Computational Materials}\ }\textbf {\bibinfo {volume} {5}},\ \bibinfo {pages} {1} (\bibinfo {year} {2019})}\BibitemShut {NoStop}%
\bibitem [{\citenamefont {Tian}\ \emph {et~al.}(2013)\citenamefont {Tian}, \citenamefont {Delczeg}, \citenamefont {Chen}, \citenamefont {Varga}, \citenamefont {Shen},\ and\ \citenamefont {Vitos}}]{tian_structural_2013}%
  \BibitemOpen
  \bibfield  {author} {\bibinfo {author} {\bibfnamefont {F.}~\bibnamefont {Tian}}, \bibinfo {author} {\bibfnamefont {L.}~\bibnamefont {Delczeg}}, \bibinfo {author} {\bibfnamefont {N.}~\bibnamefont {Chen}}, \bibinfo {author} {\bibfnamefont {L.~K.}\ \bibnamefont {Varga}}, \bibinfo {author} {\bibfnamefont {J.}~\bibnamefont {Shen}},\ and\ \bibinfo {author} {\bibfnamefont {L.}~\bibnamefont {Vitos}},\ }\href {https://doi.org/10.1103/PhysRevB.88.085128} {\bibfield  {journal} {\bibinfo  {journal} {Physical Review B}\ }\textbf {\bibinfo {volume} {88}},\ \bibinfo {pages} {085128} (\bibinfo {year} {2013})}\BibitemShut {NoStop}%
\bibitem [{\citenamefont {Tian}\ \emph {et~al.}(2017)\citenamefont {Tian}, \citenamefont {Wang},\ and\ \citenamefont {Vitos}}]{tian_impact_2017}%
  \BibitemOpen
  \bibfield  {author} {\bibinfo {author} {\bibfnamefont {F.}~\bibnamefont {Tian}}, \bibinfo {author} {\bibfnamefont {Y.}~\bibnamefont {Wang}},\ and\ \bibinfo {author} {\bibfnamefont {L.}~\bibnamefont {Vitos}},\ }\href {https://doi.org/10.1063/1.4973489} {\bibfield  {journal} {\bibinfo  {journal} {Journal of Applied Physics}\ }\textbf {\bibinfo {volume} {121}},\ \bibinfo {pages} {015105} (\bibinfo {year} {2017})}\BibitemShut {NoStop}%
\bibitem [{\citenamefont {Huang}\ \emph {et~al.}(2018)\citenamefont {Huang}, \citenamefont {Tian},\ and\ \citenamefont {Vitos}}]{huang_elasticity_2018}%
  \BibitemOpen
  \bibfield  {author} {\bibinfo {author} {\bibfnamefont {S.}~\bibnamefont {Huang}}, \bibinfo {author} {\bibfnamefont {F.}~\bibnamefont {Tian}},\ and\ \bibinfo {author} {\bibfnamefont {L.}~\bibnamefont {Vitos}},\ }\href {https://doi.org/10.1557/jmr.2018.237} {\bibfield  {journal} {\bibinfo  {journal} {Journal of Materials Research}\ }\textbf {\bibinfo {volume} {33}},\ \bibinfo {pages} {2938} (\bibinfo {year} {2018})}\BibitemShut {NoStop}%
\bibitem [{\citenamefont {Khachaturyan}(1978)}]{khachaturyan_ordering_1978}%
  \BibitemOpen
  \bibfield  {author} {\bibinfo {author} {\bibfnamefont {A.~G.}\ \bibnamefont {Khachaturyan}},\ }\href {https://doi.org/10.1016/0079-6425(78)90003-8} {\bibfield  {journal} {\bibinfo  {journal} {Progress in Materials Science}\ }\textbf {\bibinfo {volume} {22}},\ \bibinfo {pages} {1} (\bibinfo {year} {1978})}\BibitemShut {NoStop}%
\bibitem [{\citenamefont {Gyorffy}\ and\ \citenamefont {Stocks}(1983)}]{gyorffy_concentration_1983}%
  \BibitemOpen
  \bibfield  {author} {\bibinfo {author} {\bibfnamefont {B.~L.}\ \bibnamefont {Gyorffy}}\ and\ \bibinfo {author} {\bibfnamefont {G.~M.}\ \bibnamefont {Stocks}},\ }\href {https://doi.org/10.1103/PhysRevLett.50.374} {\bibfield  {journal} {\bibinfo  {journal} {Physical Review Letters}\ }\textbf {\bibinfo {volume} {50}},\ \bibinfo {pages} {374} (\bibinfo {year} {1983})}\BibitemShut {NoStop}%
\bibitem [{\citenamefont {Staunton}\ \emph {et~al.}(1994)\citenamefont {Staunton}, \citenamefont {Johnson},\ and\ \citenamefont {Pinski}}]{staunton_compositional_1994}%
  \BibitemOpen
  \bibfield  {author} {\bibinfo {author} {\bibfnamefont {J.~B.}\ \bibnamefont {Staunton}}, \bibinfo {author} {\bibfnamefont {D.~D.}\ \bibnamefont {Johnson}},\ and\ \bibinfo {author} {\bibfnamefont {F.~J.}\ \bibnamefont {Pinski}},\ }\href {https://doi.org/10.1103/PhysRevB.50.1450} {\bibfield  {journal} {\bibinfo  {journal} {Physical Review B}\ }\textbf {\bibinfo {volume} {50}},\ \bibinfo {pages} {1450} (\bibinfo {year} {1994})}\BibitemShut {NoStop}%
\bibitem [{\citenamefont {Johnson}\ \emph {et~al.}(1994)\citenamefont {Johnson}, \citenamefont {Staunton},\ and\ \citenamefont {Pinski}}]{johnson_first-principles_1994}%
  \BibitemOpen
  \bibfield  {author} {\bibinfo {author} {\bibfnamefont {D.~D.}\ \bibnamefont {Johnson}}, \bibinfo {author} {\bibfnamefont {J.~B.}\ \bibnamefont {Staunton}},\ and\ \bibinfo {author} {\bibfnamefont {F.~J.}\ \bibnamefont {Pinski}},\ }\href {https://doi.org/10.1103/PhysRevB.50.1473} {\bibfield  {journal} {\bibinfo  {journal} {Physical Review B}\ }\textbf {\bibinfo {volume} {50}},\ \bibinfo {pages} {1473} (\bibinfo {year} {1994})}\BibitemShut {NoStop}%
\bibitem [{\citenamefont {Clark}\ \emph {et~al.}(1995)\citenamefont {Clark}, \citenamefont {Pinski}, \citenamefont {Johnson}, \citenamefont {Sterne}, \citenamefont {Staunton},\ and\ \citenamefont {Ginatempo}}]{clark_van_1995}%
  \BibitemOpen
  \bibfield  {author} {\bibinfo {author} {\bibfnamefont {J.~F.}\ \bibnamefont {Clark}}, \bibinfo {author} {\bibfnamefont {F.~J.}\ \bibnamefont {Pinski}}, \bibinfo {author} {\bibfnamefont {D.~D.}\ \bibnamefont {Johnson}}, \bibinfo {author} {\bibfnamefont {P.~A.}\ \bibnamefont {Sterne}}, \bibinfo {author} {\bibfnamefont {J.~B.}\ \bibnamefont {Staunton}},\ and\ \bibinfo {author} {\bibfnamefont {B.}~\bibnamefont {Ginatempo}},\ }\href {https://doi.org/10.1103/PhysRevLett.74.3225} {\bibfield  {journal} {\bibinfo  {journal} {Physical Review Letters}\ }\textbf {\bibinfo {volume} {74}},\ \bibinfo {pages} {3225} (\bibinfo {year} {1995})}\BibitemShut {NoStop}%
\bibitem [{\citenamefont {Bragg}\ and\ \citenamefont {Williams}(1934)}]{bragg_effect_1934}%
  \BibitemOpen
  \bibfield  {author} {\bibinfo {author} {\bibfnamefont {W.~L.}\ \bibnamefont {Bragg}}\ and\ \bibinfo {author} {\bibfnamefont {E.~J.}\ \bibnamefont {Williams}},\ }\href {https://doi.org/10.1098/rspa.1934.0132} {\bibfield  {journal} {\bibinfo  {journal} {Proceedings of the Royal Society of London. Series A, Containing Papers of a Mathematical and Physical Character}\ }\textbf {\bibinfo {volume} {145}},\ \bibinfo {pages} {699} (\bibinfo {year} {1934})}\BibitemShut {NoStop}%
\bibitem [{\citenamefont {Bragg}\ and\ \citenamefont {Williams}(1935)}]{bragg_effect_1935}%
  \BibitemOpen
  \bibfield  {author} {\bibinfo {author} {\bibfnamefont {W.~L.}\ \bibnamefont {Bragg}}\ and\ \bibinfo {author} {\bibfnamefont {E.~J.}\ \bibnamefont {Williams}},\ }\href {https://doi.org/10.1098/rspa.1935.0165} {\bibfield  {journal} {\bibinfo  {journal} {Proceedings of the Royal Society of London. Series A - Mathematical and Physical Sciences}\ }\textbf {\bibinfo {volume} {151}},\ \bibinfo {pages} {540} (\bibinfo {year} {1935})}\BibitemShut {NoStop}%
\bibitem [{\citenamefont {Brush}(1967)}]{brush_history_1967}%
  \BibitemOpen
  \bibfield  {author} {\bibinfo {author} {\bibfnamefont {S.~G.}\ \bibnamefont {Brush}},\ }\href {https://doi.org/10.1103/RevModPhys.39.883} {\bibfield  {journal} {\bibinfo  {journal} {Reviews of Modern Physics}\ }\textbf {\bibinfo {volume} {39}},\ \bibinfo {pages} {883} (\bibinfo {year} {1967})}\BibitemShut {NoStop}%
\bibitem [{\citenamefont {Kawasaki}(1966)}]{kawasaki_diffusion_1966}%
  \BibitemOpen
  \bibfield  {author} {\bibinfo {author} {\bibfnamefont {K.}~\bibnamefont {Kawasaki}},\ }\href {https://doi.org/10.1103/PhysRev.145.224} {\bibfield  {journal} {\bibinfo  {journal} {Physical Review}\ }\textbf {\bibinfo {volume} {145}},\ \bibinfo {pages} {224} (\bibinfo {year} {1966})}\BibitemShut {NoStop}%
\bibitem [{\citenamefont {Metropolis}\ \emph {et~al.}(1953)\citenamefont {Metropolis}, \citenamefont {Rosenbluth}, \citenamefont {Rosenbluth}, \citenamefont {Teller},\ and\ \citenamefont {Teller}}]{metropolis_equation_1953}%
  \BibitemOpen
  \bibfield  {author} {\bibinfo {author} {\bibfnamefont {N.}~\bibnamefont {Metropolis}}, \bibinfo {author} {\bibfnamefont {A.~W.}\ \bibnamefont {Rosenbluth}}, \bibinfo {author} {\bibfnamefont {M.~N.}\ \bibnamefont {Rosenbluth}}, \bibinfo {author} {\bibfnamefont {A.~H.}\ \bibnamefont {Teller}},\ and\ \bibinfo {author} {\bibfnamefont {E.}~\bibnamefont {Teller}},\ }\href {https://doi.org/10.1063/1.1699114} {\bibfield  {journal} {\bibinfo  {journal} {The Journal of Chemical Physics}\ }\textbf {\bibinfo {volume} {21}},\ \bibinfo {pages} {1087} (\bibinfo {year} {1953})}\BibitemShut {NoStop}%
\bibitem [{\citenamefont {Wang}\ and\ \citenamefont {Landau}(2001)}]{wang_efficient_2001}%
  \BibitemOpen
  \bibfield  {author} {\bibinfo {author} {\bibfnamefont {F.}~\bibnamefont {Wang}}\ and\ \bibinfo {author} {\bibfnamefont {D.~P.}\ \bibnamefont {Landau}},\ }\href {https://doi.org/10.1103/PhysRevLett.86.2050} {\bibfield  {journal} {\bibinfo  {journal} {Physical Review Letters}\ }\textbf {\bibinfo {volume} {86}},\ \bibinfo {pages} {2050} (\bibinfo {year} {2001})}\BibitemShut {NoStop}%
\bibitem [{\citenamefont {Landau}\ and\ \citenamefont {Binder}(2014)}]{landau_guide_2014}%
  \BibitemOpen
  \bibfield  {author} {\bibinfo {author} {\bibfnamefont {D.~P.}\ \bibnamefont {Landau}}\ and\ \bibinfo {author} {\bibfnamefont {K.}~\bibnamefont {Binder}},\ }\href {https://doi.org/10.1017/CBO9781139696463} {\emph {\bibinfo {title} {A {Guide} to {Monte} {Carlo} {Simulations} in {Statistical} {Physics}}}},\ \bibinfo {edition} {4th}\ ed.\ (\bibinfo  {publisher} {Cambridge University Press},\ \bibinfo {address} {Cambridge, UK},\ \bibinfo {year} {2014})\BibitemShut {NoStop}%
\bibitem [{\citenamefont {Cowley}(1950)}]{cowley_approximate_1950}%
  \BibitemOpen
  \bibfield  {author} {\bibinfo {author} {\bibfnamefont {J.~M.}\ \bibnamefont {Cowley}},\ }\href {https://doi.org/10.1103/PhysRev.77.669} {\bibfield  {journal} {\bibinfo  {journal} {Physical Review}\ }\textbf {\bibinfo {volume} {77}},\ \bibinfo {pages} {669} (\bibinfo {year} {1950})}\BibitemShut {NoStop}%
\bibitem [{\citenamefont {Norman}\ and\ \citenamefont {Warren}(1951)}]{norman_x-ray_1951}%
  \BibitemOpen
  \bibfield  {author} {\bibinfo {author} {\bibfnamefont {N.}~\bibnamefont {Norman}}\ and\ \bibinfo {author} {\bibfnamefont {B.~E.}\ \bibnamefont {Warren}},\ }\href {https://doi.org/10.1063/1.1699988} {\bibfield  {journal} {\bibinfo  {journal} {Journal of Applied Physics}\ }\textbf {\bibinfo {volume} {22}},\ \bibinfo {pages} {483} (\bibinfo {year} {1951})}\BibitemShut {NoStop}%
\bibitem [{\citenamefont {Cowley}(1965)}]{cowley_short-range_1965}%
  \BibitemOpen
  \bibfield  {author} {\bibinfo {author} {\bibfnamefont {J.~M.}\ \bibnamefont {Cowley}},\ }\href {https://doi.org/10.1103/PhysRev.138.A1384} {\bibfield  {journal} {\bibinfo  {journal} {Physical Review}\ }\textbf {\bibinfo {volume} {138}},\ \bibinfo {pages} {A1384} (\bibinfo {year} {1965})}\BibitemShut {NoStop}%
\bibitem [{\citenamefont {Kubo}(1957)}]{kubo_statistical-mechanical_1957}%
  \BibitemOpen
  \bibfield  {author} {\bibinfo {author} {\bibfnamefont {R.}~\bibnamefont {Kubo}},\ }\href {https://doi.org/10.1143/JPSJ.12.570} {\bibfield  {journal} {\bibinfo  {journal} {Journal of the Physical Society of Japan}\ }\textbf {\bibinfo {volume} {12}},\ \bibinfo {pages} {570} (\bibinfo {year} {1957})}\BibitemShut {NoStop}%
\bibitem [{\citenamefont {Greenwood}(1958)}]{greenwood_boltzmann_1958}%
  \BibitemOpen
  \bibfield  {author} {\bibinfo {author} {\bibfnamefont {D.~A.}\ \bibnamefont {Greenwood}},\ }\href {https://doi.org/10.1088/0370-1328/71/4/306} {\bibfield  {journal} {\bibinfo  {journal} {Proceedings of the Physical Society}\ }\textbf {\bibinfo {volume} {71}},\ \bibinfo {pages} {585} (\bibinfo {year} {1958})}\BibitemShut {NoStop}%
\bibitem [{\citenamefont {Butler}(1985)}]{butler_theory_1985}%
  \BibitemOpen
  \bibfield  {author} {\bibinfo {author} {\bibfnamefont {W.~H.}\ \bibnamefont {Butler}},\ }\href {https://doi.org/10.1103/PhysRevB.31.3260} {\bibfield  {journal} {\bibinfo  {journal} {Physical Review B}\ }\textbf {\bibinfo {volume} {31}},\ \bibinfo {pages} {3260} (\bibinfo {year} {1985})}\BibitemShut {NoStop}%
\bibitem [{\citenamefont {Ebert}\ \emph {et~al.}()\citenamefont {Ebert} \emph {et~al.}}]{ebert_munich_nodate}%
  \BibitemOpen
  \bibfield  {author} {\bibinfo {author} {\bibfnamefont {H.}~\bibnamefont {Ebert}} \emph {et~al.},\ }\href@noop {} {\bibinfo {title} {The {Munich} {SPR}-{KKR} package}},\ \bibinfo {note} {\href{https://www.ebert.cup.uni-muenchen.de/old/index.php}{https://www.ebert.cup.uni-muenchen.de/old/index.php}}\BibitemShut {NoStop}%
\bibitem [{\citenamefont {Andersen}(1975)}]{andersen_linear_1975}%
  \BibitemOpen
  \bibfield  {author} {\bibinfo {author} {\bibfnamefont {O.~K.}\ \bibnamefont {Andersen}},\ }\href {https://doi.org/10.1103/PhysRevB.12.3060} {\bibfield  {journal} {\bibinfo  {journal} {Physical Review B}\ }\textbf {\bibinfo {volume} {12}},\ \bibinfo {pages} {3060} (\bibinfo {year} {1975})}\BibitemShut {NoStop}%
\bibitem [{\citenamefont {Vosko}\ \emph {et~al.}(1980)\citenamefont {Vosko}, \citenamefont {Wilk},\ and\ \citenamefont {Nusair}}]{vosko_accurate_1980}%
  \BibitemOpen
  \bibfield  {author} {\bibinfo {author} {\bibfnamefont {S.~H.}\ \bibnamefont {Vosko}}, \bibinfo {author} {\bibfnamefont {L.}~\bibnamefont {Wilk}},\ and\ \bibinfo {author} {\bibfnamefont {M.}~\bibnamefont {Nusair}},\ }\href {https://doi.org/10.1139/p80-159} {\bibfield  {journal} {\bibinfo  {journal} {Canadian Journal of Physics}\ }\textbf {\bibinfo {volume} {58}},\ \bibinfo {pages} {1200} (\bibinfo {year} {1980})}\BibitemShut {NoStop}%
\bibitem [{\citenamefont {Staunton}\ \emph {et~al.}(1984)\citenamefont {Staunton}, \citenamefont {Gyorffy}, \citenamefont {Pindor}, \citenamefont {Stocks},\ and\ \citenamefont {Winter}}]{staunton_disordered_1984}%
  \BibitemOpen
  \bibfield  {author} {\bibinfo {author} {\bibfnamefont {J.}~\bibnamefont {Staunton}}, \bibinfo {author} {\bibfnamefont {B.}~\bibnamefont {Gyorffy}}, \bibinfo {author} {\bibfnamefont {A.}~\bibnamefont {Pindor}}, \bibinfo {author} {\bibfnamefont {G.}~\bibnamefont {Stocks}},\ and\ \bibinfo {author} {\bibfnamefont {H.}~\bibnamefont {Winter}},\ }\href {https://doi.org/10.1016/0304-8853(84)90367-6} {\bibfield  {journal} {\bibinfo  {journal} {Journal of Magnetism and Magnetic Materials}\ }\textbf {\bibinfo {volume} {45}},\ \bibinfo {pages} {15} (\bibinfo {year} {1984})}\BibitemShut {NoStop}%
\bibitem [{\citenamefont {Pindor}\ \emph {et~al.}(1983)\citenamefont {Pindor}, \citenamefont {Staunton}, \citenamefont {Stocks},\ and\ \citenamefont {Winter}}]{pindor_disordered_1983}%
  \BibitemOpen
  \bibfield  {author} {\bibinfo {author} {\bibfnamefont {A.~J.}\ \bibnamefont {Pindor}}, \bibinfo {author} {\bibfnamefont {J.}~\bibnamefont {Staunton}}, \bibinfo {author} {\bibfnamefont {G.~M.}\ \bibnamefont {Stocks}},\ and\ \bibinfo {author} {\bibfnamefont {H.}~\bibnamefont {Winter}},\ }\href {https://doi.org/10.1088/0305-4608/13/5/012} {\bibfield  {journal} {\bibinfo  {journal} {Journal of Physics F: Metal Physics}\ }\textbf {\bibinfo {volume} {13}},\ \bibinfo {pages} {979} (\bibinfo {year} {1983})}\BibitemShut {NoStop}%
\bibitem [{dat()}]{dataset}%
  \BibitemOpen
  \href@noop {} {}\bibinfo {note} {The data supporting the findings of this study are available at \href{https://doi.org/10.5281/zenodo.14849315}{https://doi.org/10.5281/zenodo.14849315}}\BibitemShut {NoStop}%
\bibitem [{\citenamefont {Alchagirov}\ \emph {et~al.}(2001)\citenamefont {Alchagirov}, \citenamefont {Perdew}, \citenamefont {Boettger}, \citenamefont {Albers},\ and\ \citenamefont {Fiolhais}}]{alchagirov_energy_2001}%
  \BibitemOpen
  \bibfield  {author} {\bibinfo {author} {\bibfnamefont {A.~B.}\ \bibnamefont {Alchagirov}}, \bibinfo {author} {\bibfnamefont {J.~P.}\ \bibnamefont {Perdew}}, \bibinfo {author} {\bibfnamefont {J.~C.}\ \bibnamefont {Boettger}}, \bibinfo {author} {\bibfnamefont {R.~C.}\ \bibnamefont {Albers}},\ and\ \bibinfo {author} {\bibfnamefont {C.}~\bibnamefont {Fiolhais}},\ }\href {https://doi.org/10.1103/PhysRevB.63.224115} {\bibfield  {journal} {\bibinfo  {journal} {Physical Review B}\ }\textbf {\bibinfo {volume} {63}},\ \bibinfo {pages} {224115} (\bibinfo {year} {2001})}\BibitemShut {NoStop}%
\bibitem [{\citenamefont {Hjorth~Larsen}\ \emph {et~al.}(2017)\citenamefont {Hjorth~Larsen}, \citenamefont {Jørgen~Mortensen}, \citenamefont {Blomqvist}, \citenamefont {Castelli}, \citenamefont {Christensen}, \citenamefont {Dułak}, \citenamefont {Friis}, \citenamefont {Groves}, \citenamefont {Hammer}, \citenamefont {Hargus}, \citenamefont {Hermes}, \citenamefont {Jennings}, \citenamefont {Bjerre~Jensen}, \citenamefont {Kermode}, \citenamefont {Kitchin}, \citenamefont {Leonhard~Kolsbjerg}, \citenamefont {Kubal}, \citenamefont {Kaasbjerg}, \citenamefont {Lysgaard}, \citenamefont {Bergmann~Maronsson}, \citenamefont {Maxson}, \citenamefont {Olsen}, \citenamefont {Pastewka}, \citenamefont {Peterson}, \citenamefont {Rostgaard}, \citenamefont {Schiøtz}, \citenamefont {Schütt}, \citenamefont {Strange}, \citenamefont {Thygesen}, \citenamefont {Vegge}, \citenamefont {Vilhelmsen}, \citenamefont {Walter}, \citenamefont {Zeng},\ and\ \citenamefont {Jacobsen}}]{hjorth_larsen_atomic_2017}%
  \BibitemOpen
  \bibfield  {author} {\bibinfo {author} {\bibfnamefont {A.}~\bibnamefont {Hjorth~Larsen}}, \bibinfo {author} {\bibfnamefont {J.}~\bibnamefont {Jørgen~Mortensen}}, \bibinfo {author} {\bibfnamefont {J.}~\bibnamefont {Blomqvist}}, \bibinfo {author} {\bibfnamefont {I.~E.}\ \bibnamefont {Castelli}}, \bibinfo {author} {\bibfnamefont {R.}~\bibnamefont {Christensen}}, \bibinfo {author} {\bibfnamefont {M.}~\bibnamefont {Dułak}}, \bibinfo {author} {\bibfnamefont {J.}~\bibnamefont {Friis}}, \bibinfo {author} {\bibfnamefont {M.~N.}\ \bibnamefont {Groves}}, \bibinfo {author} {\bibfnamefont {B.}~\bibnamefont {Hammer}}, \bibinfo {author} {\bibfnamefont {C.}~\bibnamefont {Hargus}}, \bibinfo {author} {\bibfnamefont {E.~D.}\ \bibnamefont {Hermes}}, \bibinfo {author} {\bibfnamefont {P.~C.}\ \bibnamefont {Jennings}}, \bibinfo {author} {\bibfnamefont {P.}~\bibnamefont {Bjerre~Jensen}}, \bibinfo {author} {\bibfnamefont {J.}~\bibnamefont {Kermode}}, \bibinfo {author} {\bibfnamefont {J.~R.}\ \bibnamefont {Kitchin}}, \bibinfo
  {author} {\bibfnamefont {E.}~\bibnamefont {Leonhard~Kolsbjerg}}, \bibinfo {author} {\bibfnamefont {J.}~\bibnamefont {Kubal}}, \bibinfo {author} {\bibfnamefont {K.}~\bibnamefont {Kaasbjerg}}, \bibinfo {author} {\bibfnamefont {S.}~\bibnamefont {Lysgaard}}, \bibinfo {author} {\bibfnamefont {J.}~\bibnamefont {Bergmann~Maronsson}}, \bibinfo {author} {\bibfnamefont {T.}~\bibnamefont {Maxson}}, \bibinfo {author} {\bibfnamefont {T.}~\bibnamefont {Olsen}}, \bibinfo {author} {\bibfnamefont {L.}~\bibnamefont {Pastewka}}, \bibinfo {author} {\bibfnamefont {A.}~\bibnamefont {Peterson}}, \bibinfo {author} {\bibfnamefont {C.}~\bibnamefont {Rostgaard}}, \bibinfo {author} {\bibfnamefont {J.}~\bibnamefont {Schiøtz}}, \bibinfo {author} {\bibfnamefont {O.}~\bibnamefont {Schütt}}, \bibinfo {author} {\bibfnamefont {M.}~\bibnamefont {Strange}}, \bibinfo {author} {\bibfnamefont {K.~S.}\ \bibnamefont {Thygesen}}, \bibinfo {author} {\bibfnamefont {T.}~\bibnamefont {Vegge}}, \bibinfo {author} {\bibfnamefont {L.}~\bibnamefont
  {Vilhelmsen}}, \bibinfo {author} {\bibfnamefont {M.}~\bibnamefont {Walter}}, \bibinfo {author} {\bibfnamefont {Z.}~\bibnamefont {Zeng}},\ and\ \bibinfo {author} {\bibfnamefont {K.~W.}\ \bibnamefont {Jacobsen}},\ }\href {https://doi.org/10.1088/1361-648X/aa680e} {\bibfield  {journal} {\bibinfo  {journal} {Journal of Physics: Condensed Matter}\ }\textbf {\bibinfo {volume} {29}},\ \bibinfo {pages} {273002} (\bibinfo {year} {2017})}\BibitemShut {NoStop}%
\bibitem [{\citenamefont {Grabowski}\ \emph {et~al.}(2007)\citenamefont {Grabowski}, \citenamefont {Hickel},\ and\ \citenamefont {Neugebauer}}]{grabowski_ab_2007}%
  \BibitemOpen
  \bibfield  {author} {\bibinfo {author} {\bibfnamefont {B.}~\bibnamefont {Grabowski}}, \bibinfo {author} {\bibfnamefont {T.}~\bibnamefont {Hickel}},\ and\ \bibinfo {author} {\bibfnamefont {J.}~\bibnamefont {Neugebauer}},\ }\href {https://doi.org/10.1103/PhysRevB.76.024309} {\bibfield  {journal} {\bibinfo  {journal} {Physical Review B}\ }\textbf {\bibinfo {volume} {76}},\ \bibinfo {pages} {024309} (\bibinfo {year} {2007})}\BibitemShut {NoStop}%
\bibitem [{\citenamefont {Woodgate}\ \emph {et~al.}(2024)\citenamefont {Woodgate}, \citenamefont {Marchant}, \citenamefont {Pártay},\ and\ \citenamefont {Staunton}}]{woodgate_structure_2024}%
  \BibitemOpen
  \bibfield  {author} {\bibinfo {author} {\bibfnamefont {C.~D.}\ \bibnamefont {Woodgate}}, \bibinfo {author} {\bibfnamefont {G.~A.}\ \bibnamefont {Marchant}}, \bibinfo {author} {\bibfnamefont {L.~B.}\ \bibnamefont {Pártay}},\ and\ \bibinfo {author} {\bibfnamefont {J.~B.}\ \bibnamefont {Staunton}},\ }\href {https://doi.org/10.1038/s41524-024-01445-w} {\bibfield  {journal} {\bibinfo  {journal} {npj Computational Materials}\ }\textbf {\bibinfo {volume} {10}},\ \bibinfo {pages} {271} (\bibinfo {year} {2024})}\BibitemShut {NoStop}%
\bibitem [{\citenamefont {Woodgate}\ \emph {et~al.}(2023)\citenamefont {Woodgate}, \citenamefont {Hedlund}, \citenamefont {Lewis},\ and\ \citenamefont {Staunton}}]{woodgate_interplay_2023}%
  \BibitemOpen
  \bibfield  {author} {\bibinfo {author} {\bibfnamefont {C.~D.}\ \bibnamefont {Woodgate}}, \bibinfo {author} {\bibfnamefont {D.}~\bibnamefont {Hedlund}}, \bibinfo {author} {\bibfnamefont {L.~H.}\ \bibnamefont {Lewis}},\ and\ \bibinfo {author} {\bibfnamefont {J.~B.}\ \bibnamefont {Staunton}},\ }\href {https://doi.org/10.1103/PhysRevMaterials.7.053801} {\bibfield  {journal} {\bibinfo  {journal} {Physical Review Materials}\ }\textbf {\bibinfo {volume} {7}},\ \bibinfo {pages} {053801} (\bibinfo {year} {2023})}\BibitemShut {NoStop}%
\bibitem [{\citenamefont {Woodgate}\ and\ \citenamefont {Staunton}(2024)}]{woodgate_competition_2024}%
  \BibitemOpen
  \bibfield  {author} {\bibinfo {author} {\bibfnamefont {C.~D.}\ \bibnamefont {Woodgate}}\ and\ \bibinfo {author} {\bibfnamefont {J.~B.}\ \bibnamefont {Staunton}},\ }\href {https://doi.org/10.1063/5.0200862} {\bibfield  {journal} {\bibinfo  {journal} {Journal of Applied Physics}\ }\textbf {\bibinfo {volume} {135}},\ \bibinfo {pages} {135106} (\bibinfo {year} {2024})}\BibitemShut {NoStop}%
\bibitem [{\citenamefont {Troparevsky}\ \emph {et~al.}(2015)\citenamefont {Troparevsky}, \citenamefont {Morris}, \citenamefont {Kent}, \citenamefont {Lupini},\ and\ \citenamefont {Stocks}}]{troparevsky_criteria_2015}%
  \BibitemOpen
  \bibfield  {author} {\bibinfo {author} {\bibfnamefont {M.~C.}\ \bibnamefont {Troparevsky}}, \bibinfo {author} {\bibfnamefont {J.~R.}\ \bibnamefont {Morris}}, \bibinfo {author} {\bibfnamefont {P.~R.}\ \bibnamefont {Kent}}, \bibinfo {author} {\bibfnamefont {A.~R.}\ \bibnamefont {Lupini}},\ and\ \bibinfo {author} {\bibfnamefont {G.~M.}\ \bibnamefont {Stocks}},\ }\href {https://doi.org/10.1103/PhysRevX.5.011041} {\bibfield  {journal} {\bibinfo  {journal} {Physical Review X}\ }\textbf {\bibinfo {volume} {5}},\ \bibinfo {pages} {011041} (\bibinfo {year} {2015})}\BibitemShut {NoStop}%
\bibitem [{\citenamefont {Santodonato}\ \emph {et~al.}(2018)\citenamefont {Santodonato}, \citenamefont {Liaw}, \citenamefont {Unocic}, \citenamefont {Bei},\ and\ \citenamefont {Morris}}]{santodonato_predictive_2018}%
  \BibitemOpen
  \bibfield  {author} {\bibinfo {author} {\bibfnamefont {L.~J.}\ \bibnamefont {Santodonato}}, \bibinfo {author} {\bibfnamefont {P.~K.}\ \bibnamefont {Liaw}}, \bibinfo {author} {\bibfnamefont {R.~R.}\ \bibnamefont {Unocic}}, \bibinfo {author} {\bibfnamefont {H.}~\bibnamefont {Bei}},\ and\ \bibinfo {author} {\bibfnamefont {J.~R.}\ \bibnamefont {Morris}},\ }\href {https://doi.org/10.1038/s41467-018-06757-2} {\bibfield  {journal} {\bibinfo  {journal} {Nature Communications}\ }\textbf {\bibinfo {volume} {9}},\ \bibinfo {pages} {4520} (\bibinfo {year} {2018})}\BibitemShut {NoStop}%
\bibitem [{sup()}]{supplemental}%
  \BibitemOpen
  \href@noop {} {}\bibinfo {note} {See Supplementary Material at \texttt{URL will be inserted by publisher} for tabulated partial lattice site occupancies as a function of $\eta$, additional site- and orbital-resolved electronic DoS plots, tabulated atom-atom effective pair interactions, and discussion of the electronic origins of the increased residual resistivity induced by the B2 ordering for the considered alloys.}\BibitemShut {Stop}%
\bibitem [{\citenamefont {Naguszewski}\ \emph {et~al.}()\citenamefont {Naguszewski}, \citenamefont {P\'artay}, \citenamefont {Quigley},\ and\ \citenamefont {Woodgate}}]{naguszewski_brawl_nodate}%
  \BibitemOpen
  \bibfield  {author} {\bibinfo {author} {\bibfnamefont {H.~J.}\ \bibnamefont {Naguszewski}}, \bibinfo {author} {\bibfnamefont {L.~B.}\ \bibnamefont {P\'artay}}, \bibinfo {author} {\bibfnamefont {D.}~\bibnamefont {Quigley}},\ and\ \bibinfo {author} {\bibfnamefont {C.~D.}\ \bibnamefont {Woodgate}},\ }\href@noop {} {\bibinfo {title} {The \texttt{BraWl} package}},\ \bibinfo {note} {\href{https://github.com/ChrisWoodgate/BraWl}{https://github.com/ChrisWoodgate/BraWl}}\BibitemShut {NoStop}%
\bibitem [{\citenamefont {Stukowski}(2010)}]{stukowski_visualization_2010}%
  \BibitemOpen
  \bibfield  {author} {\bibinfo {author} {\bibfnamefont {A.}~\bibnamefont {Stukowski}},\ }\href {https://doi.org/10.1088/0965-0393/18/1/015012} {\bibfield  {journal} {\bibinfo  {journal} {Modelling and Simulation in Materials Science and Engineering}\ }\textbf {\bibinfo {volume} {18}},\ \bibinfo {pages} {015012} (\bibinfo {year} {2010})}\BibitemShut {NoStop}%
\bibitem [{\citenamefont {Lowitzer}\ \emph {et~al.}(2010)\citenamefont {Lowitzer}, \citenamefont {Ködderitzsch}, \citenamefont {Ebert}, \citenamefont {Tulip}, \citenamefont {Marmodoro},\ and\ \citenamefont {Staunton}}]{lowitzer_ab_2010}%
  \BibitemOpen
  \bibfield  {author} {\bibinfo {author} {\bibfnamefont {S.}~\bibnamefont {Lowitzer}}, \bibinfo {author} {\bibfnamefont {D.}~\bibnamefont {Ködderitzsch}}, \bibinfo {author} {\bibfnamefont {H.}~\bibnamefont {Ebert}}, \bibinfo {author} {\bibfnamefont {P.~R.}\ \bibnamefont {Tulip}}, \bibinfo {author} {\bibfnamefont {A.}~\bibnamefont {Marmodoro}},\ and\ \bibinfo {author} {\bibfnamefont {J.~B.}\ \bibnamefont {Staunton}},\ }\href {https://doi.org/10.1209/0295-5075/92/37009} {\bibfield  {journal} {\bibinfo  {journal} {EPL (Europhysics Letters)}\ }\textbf {\bibinfo {volume} {92}},\ \bibinfo {pages} {37009} (\bibinfo {year} {2010})}\BibitemShut {NoStop}%
\bibitem [{\citenamefont {Thomas}(1951)}]{thomas_uber_1951}%
  \BibitemOpen
  \bibfield  {author} {\bibinfo {author} {\bibfnamefont {H.}~\bibnamefont {Thomas}},\ }\href {https://doi.org/10.1007/BF01333398} {\bibfield  {journal} {\bibinfo  {journal} {Zeitschrift für Physik}\ }\textbf {\bibinfo {volume} {129}},\ \bibinfo {pages} {219} (\bibinfo {year} {1951})}\BibitemShut {NoStop}%
\bibitem [{\citenamefont {Mahan}\ and\ \citenamefont {Sofo}(1996)}]{mahan1996}%
  \BibitemOpen
  \bibfield  {author} {\bibinfo {author} {\bibfnamefont {G.~D.}\ \bibnamefont {Mahan}}\ and\ \bibinfo {author} {\bibfnamefont {J.~O.}\ \bibnamefont {Sofo}},\ }\href {https://doi.org/10.1073/pnas.93.15.7436} {\bibfield  {journal} {\bibinfo  {journal} {Proceedings of the National Academy of Sciences}\ }\textbf {\bibinfo {volume} {93}},\ \bibinfo {pages} {7436} (\bibinfo {year} {1996})}\BibitemShut {NoStop}%
\bibitem [{\citenamefont {Scheidemantel}\ \emph {et~al.}(2003)\citenamefont {Scheidemantel}, \citenamefont {{Ambrosch-Draxl}}, \citenamefont {Thonhauser}, \citenamefont {Badding},\ and\ \citenamefont {Sofo}}]{scheidemantel2003}%
  \BibitemOpen
  \bibfield  {author} {\bibinfo {author} {\bibfnamefont {T.~J.}\ \bibnamefont {Scheidemantel}}, \bibinfo {author} {\bibfnamefont {C.}~\bibnamefont {{Ambrosch-Draxl}}}, \bibinfo {author} {\bibfnamefont {T.}~\bibnamefont {Thonhauser}}, \bibinfo {author} {\bibfnamefont {J.~V.}\ \bibnamefont {Badding}},\ and\ \bibinfo {author} {\bibfnamefont {J.~O.}\ \bibnamefont {Sofo}},\ }\href {https://doi.org/10.1103/PhysRevB.68.125210} {\bibfield  {journal} {\bibinfo  {journal} {Physical Review B}\ }\textbf {\bibinfo {volume} {68}},\ \bibinfo {pages} {125210} (\bibinfo {year} {2003})}\BibitemShut {NoStop}%
\end{thebibliography}

%

\end{document}